\documentclass[11pt,a4paper]{article}
\usepackage[top=2cm, bottom=2cm, left=2cm, right=2cm]{geometry}
\usepackage{amssymb}
\usepackage{amsbsy}
\usepackage{mathrsfs,amsmath}
\usepackage{slashed}
\usepackage{mathtools}
\usepackage{graphicx}
\usepackage{color}
\usepackage{colortbl}
\usepackage{tensor}
\usepackage{array}
\usepackage{graphbox}
\usepackage{multicol}
\usepackage{multirow}
\usepackage[nottoc]{tocbibind}
\usepackage{dsfont}
\usepackage{array}
\usepackage{supertabular}
\usepackage{longtable}
\usepackage{mdframed}
\usepackage[normalem]{ulem}
\graphicspath{{}}

\usepackage[force]{feynmp-auto}

\numberwithin{equation}{section}

\newcommand{\lol}{\lambda^{1\text{L}}}
\newcommand{\lori}{\lambda^{1\text{R}}}
\newcommand{\lbl}{\lambda^{\slashed{B}L}}

\newcommand{\lbr}{\lambda^{\slashed{B}R}}
\newcommand{\ltil}{\tilde{\lambda}}
\newcommand{\LsBq}{\Lambda_{q}^{\slashed{B}}}
\newcommand{\Ltil}{\tilde{\Lambda}_{\ell}}
\newcommand{\Ltild}{\tilde{\Lambda}_{d}}
\newcommand{\LsBu}{\Lambda_{u}^{\slashed{B}}}
\newcommand{\LsBd}{\Lambda_{d}^{\slashed{B}}}
\newcommand{\ltilone}{\tilde{\lambda}^1}
\newcommand{\lRL}{\lambda^{2RL}}
\newcommand{\lLR}{\lambda^{2LR}}
\newcommand{\lLLL}{\lambda^{3L}}
\newcommand{\ltilbsl}{\tilde{\lambda}^{1\slashed{B}}}
\newcommand{\lBBB}{\lambda^{3\slashed{B}}}
\newcommand{\XBLD}{Y^{\slashed{B}L}_{1D}}
\newcommand{\XBLU}{Y^{\slashed{B}L}_{1U}}

\newcommand{\XLU}{Y^{1L}_{1U}}
\newcommand{\XLE}{Y^{1L}_{1E}}
\newcommand{\XLDD}{Y^{1L}_{2D}}
\newcommand{\XLUU}{Y^{1L}_{2U}}
\newcommand{\XLEE}{Y^{1L}_{2E}}
\newcommand{\XRUU}{Y^{1R}_{2U}}
\newcommand{\XREE}{Y^{1R}_{2E}}
\newcommand{\XtilEE}{\tilde{Y}_{2E}}
\newcommand{\XtilDD}{\tilde{Y}_{2D}}
\newcommand{\XsBRDD}{Y^{\slashed{B}R}_{2D}}
\newcommand{\XsBRUU}{Y^{\slashed{B}R}_{2U}}
\newcommand{\XsBLUU}{Y^{\slashed{B}L}_{2U}}
\newcommand{\XsBLDD}{Y^{\slashed{B}L}_{2D}}
\newcommand{\XLEEE}{Y^{1L}_{3E}}
\newcommand{\XLUUU}{Y^{1L}_{3U}}
\newcommand{\XsBLDDD}{Y^{\slashed{B}L}_{3D}}
\newcommand{\XsBLUUU}{Y^{\slashed{B}L}_{3U}}
\newcommand{\cttilde}{\tilde{c}_{2}}
\newcommand{\coottilde}{c^{(1)}_{1\tilde{2}}}
\newcommand{\ctotilde}{c^{(2)}_{1\tilde{2}}}

\allowdisplaybreaks

\usepackage{jheppub}

\title{Universal Scalar Leptoquark Action for Matching}
\author{Athanasios Dedes,}
\author{Kostas Mantzaropoulos}
\affiliation{Division of Theoretical Physics, Department of Physics, University of Ioannina, GR 45110, Greece}

\emailAdd{adedes@uoi.gr}
\emailAdd{k.mantzaropoulos@uoi.gr}

\keywords{Effective Field Theory, Functional matching, Leptoquark models}

\abstract{
	In this study we present a universal effective action for one-loop matching  of all scalar leptoquarks. 
	We use both the Universal One-Loop Effective Action (UOLEA) and covariant diagrams to evaluate the 
	Wilson coefficients directly in the Green basis for	up-to dimension-6 operators.
	On the technical side, we use the newly developed method of evaluating supertraces, 
	to further validate the results stemming from the use of covariant diagrams.
	As an application, we perform a fully functional matching onto Standard Model Effective Field Theory (SMEFT) 
	of a model with two  scalar leptoquark fields: a weak isospin singlet and a doublet.  
	We demonstrate its use by  calculating several observables, such as lepton magnetic and electric dipole moments,
	neutrino masses, proton decay rate, while we comment upon fine tuning issues in this model.
	Apart from its phenomenological interest, 
	this model generates the majority of dimension-6 
	operators and provides an EFT benchmark 
	towards future matching automation.
	}

\begin{document}
	\maketitle

	\section{Introduction}
			Effective field theory (EFT)~\cite{Weinberg:1980wa,Callan:1969sn,Coleman:1969sm} is an important part of our understanding of nature, it constitutes a robust way of dealing with new physics phenomena for Beyond the Standard Model (BSM) physics. One can obtain a low energy EFT action by integrating out heavy degrees of freedom from a, more general than the SM, UV-theory.
	  At the end of the process we obtain an EFT Lagrangian with modified SM couplings and masses augmented by higher dimensional operators whose associated  Wilson coefficients (Wcs) encode the information about the UV-theory~\cite{Appelquist:1974tg}. The main technique to perform this kind of calculation has been Feynman diagrams. However, 
	 during the last decade, functional matching has seen a renewed interest.
	
	The first steps were taken with the application of the covariant derivative expansion (CDE) in~\cite{Gaillard:1985uh,Cheyette:1987qz,Chan:1986jq}, while the revival of these techniques and methods was recently made in~\cite{Henning:2014wua,Henning:2016lyp}. A first universal result named UOLEA (Universal One Loop Effective Action) was developed in~\cite{Drozd:2015rsp,Ellis:2016enq,Ellis:2017jns}. However, this result is not truly universal since it does not account for mixed statistics and open covariant derivatives, it can be used to decouple scalar particles only i.e. involving both heavy-heavy loops as well as heavy-light loops with the scalar particles running inside. Very recently the fermionic UOLEA was also constructed~\cite{Angelescu:2020yzf,Ellis:2020ivx} and the completion of the fermionic and scalar UOLEA was also developed~\cite{Kramer:2019fwz} taking also mixed statistics into account.
	
	Another approach to functional matching, which was used to derive the heavy-light part of the UOLEA, are the \emph{covariant diagrams}~\cite{Zhang:2016pja}, which mimic the usual Feynman diagrams but are at all steps gauge-covariant. This relatively new tool makes use of the expansion by regions~\cite{Beneke:1997zp,Jantzen:2011nz} and a simpler matching framework~\cite{Fuentes-Martin:2016uol}, which builds upon~\cite{Dittmaier:1995cr,Dittmaier:1995ee}, to further simplify the matching procedure. An example application can be found in~\cite{Zhang:2016pja,Wells:2017vla}. The logic of these diagrams was taken a step forward with the development of \emph{supertrace functional}-technique~\cite{Cohen:2020fcu} which establishes a cleaner way to make up diagrammatic traces. Soon after that, an automated application of the CDE  followed~\cite{Cohen:2020qvb,Fuentes-Martin:2020udw} easing further matching calculations. Although aiming at a different direction, similar effective actions based on supertrace and Grassmannian techniques were derived in ref.~\cite{Finn:2020nvn} using a field-space super-manifold.
	
	  Briefly, the idea behind functional matching is to equate the generating functionals,
	\begin{equation}
	\Gamma_{\rm EFT}[\phi] = \Gamma_{\rm L, \,UV}[\phi]\;,
	\end{equation}
	 for the EFT and UV-theory with light fields ($\phi$) respectively. If $S$ is a heavy field, say a leptoquark field, with mass $M_S \gg m_\phi$, then the 
	 matching conditions at tree and one-loop level read:
\begin{eqnarray}
\mathcal{L}_{\rm EFT}^{\rm (tree)}[\phi] \ &=& \ \mathcal{L}_{\rm UV} [ S, \phi ]\:
\Bigr |_{S=S_c[\phi]} \;, \label{eq:con1} \\[2mm]
\int d^d x \: \mathcal{L}_{\rm EFT}^{\rm (1-loop)}[\phi] \ &=& \ \Gamma_{\rm L, \, UV}[\phi] \: \Bigr |_{\rm hard} \label{eq:con2}\;.
\end{eqnarray}
Here	$S_c[\phi]$ is the classical heavy field which solves 
the classical equations of motion (EOMs), 
	\begin{equation}
	\frac{\delta \mathcal{S}_{\rm UV}[S,\phi]}{\delta S} \: \biggr |_{S=S_c[\phi]} \ = \ 0 \;.
	\label{eq:eoms}
\end{equation}	 
Moreover, the evaluation of the loop-integral in the rhs of eq.~\eqref{eq:con2}
is performed in the (hard) region assuming momenta $q \sim M_S \gg m_\phi$
and has the form
\begin{equation}
\int d^d x \: \mathcal{L}_{\rm EFT}^{\rm (1-loop)}[\phi] \ = \ \frac{i}{2} \: \mathrm{STr} \, \log \mathbf{K} \: \Bigr |_{\rm hard}  \ - \ \frac{i}{2} \sum_{n=1}^\infty \frac{1}{n}\: \mathrm{STr} \left [ (\mathbf{K}^{-1} \mathbf{X})^n \right ] \: \Bigr |_{\rm hard} \;.
\label{eq:master}
\end{equation}
Therefore, the EFT Lagrangian is a sum of functional  Supertraces through the
log-function of the propagator $\mathbf{K}$ in field space and a power expansion of
the operator $(\mathbf{K}^{-1} \mathbf{X})$, where $\mathbf{X}$ is a field operator - an interaction matrix - evaluated at $S=S_c[\phi]$. Basically, finding the $\mathbf{X}$-matrix, and evaluating the Supertrace functional at the desired order in the EFT 
Lagrangian is what is required for the master formula of eq.~\eqref{eq:master} to work. This is the  functional approach mainly of refs.~\cite{Cohen:2020fcu,Zhang:2016pja} that we use in our work here in order to
	\begin{itemize}
	\item[1.] derive a  universal one-loop effective action up-to dimension-6 operators for all scalar leptoquark (LQ) extensions of the Standard Model (SM)~\cite{Buchmuller:1986zs}. 
	\item[2.] apply the formalism in the decoupling of two heavy LQ fields, a coloured weak isospin singlet ($S_1$) and a coloured weak isospin doublet ($\tilde{S}_2$)  and derive the full set of $d\le 6$ operators, not resorting to Baryon or Lepton number conservation.
	\item[3.] support the usefulness and clarity of functional matching over traditional Feynman diagrammatic methods or within functional methods, by comparing both supertrace and covariant diagrammatic techniques.
	\end{itemize}
There are, various worked out examples functionally integrating out non-degenerate fields in refs.~\cite{Drozd:2015rsp,Angelescu:2020yzf,Kramer:2019fwz}, however, with an exception of ref.~\cite{Wells:2017vla} and to our knowledge, there is no other functional calculation  
with two-field decoupling and more general Yukawa interactions in the literature  as the one we present here.\footnote{A complete one-loop functional matching of the singlet scalar extension of the SM exists in ref. \cite{Cohen:2020fcu} and very recently, there have also been two complete one loop, but one-field-type, matching calculations using functional methods, where ref.~\cite{Zhang:2021jdf} matches the Type-I neutrino seesaw onto SMEFT, while ref.~\cite{Brivio:2021alv} matches the Higgs triplet extension of the electroweak gauge sector. Also recently, one-field heavy scalar decoupling has been classified in ref.~\cite{DasBakshi:2020pbf} by using the code of ref.~\cite{DasBakshi:2018vni}.}
The renormalization scheme in our calculation is a (modified) mass independent one
($\overline{\rm MS}$) and we regulate the integrals with dimensional regularization.
We match on to SMEFT operators within a redundant basis, referred to as  
{\em Green (or General) basis}, which consists of operators written 
before equations of motion for the light-fields are taken into account~\cite{Jiang:2018pbd,Gherardi:2020det}. Expressions for translating Wcs from Green to Warsaw basis~\cite{Grzadkowski:2010es} are given in ref.~\cite{Gherardi:2020det}.

However, before taking up the above analysis, we first validated calculations
performed with Feynman diagrammatic techniques. We  started from 
matching a single charged singlet, the model of ref.~\cite{Bilenky:1993bt} in SMEFT.
We found full agreement apart from a missing operator  \cite{Mantzaropoulos}.
Next, we applied functional covariant diagrams to a benchmark leptoquark model $S_1 + S_3 $ of ref.~\cite{Gherardi:2020det}, where we found perfect agreement with  v4 of ref.~\cite{Gherardi:2020det}. Part of our functional calculation in our paper here addresses this model too but now with the inclusion of Baryon number violating terms in the UV-Lagrangian. Finally, regarding the tree-level part of our calculation 
we found agreement with ref.~\cite{deBlas:2017xtg}.

We have chosen to study the decoupling of heavy scalar leptoquark 
fields for two main reasons: first, there is a plethora of 
interesting BSM phenomena associated to them, i.e. from neutrino masses and proton decay~\cite{Dorsner:2016wpm}, to possible interpretation of recent flavour anomalies and enhanced anomalous magnetic moment of the 
muon~\cite{Bauer:2015knc,Angelescu:2021lln,Crivellin:2020mjs,Gherardi:2020qhc,Alasfar:2020mne,Bordone:2020lnb}, and second,
leptoquark fields are naturally embedded  in Grand Unified 
Theories (GUTs) which may
be in turn linked to  even more fundamental theories.


		\section{Universal One Loop Functional Matching for Scalar Leptoquarks}

	Leptoquarks  ($S$) are hypothetical fields defined by their Yukawa interactions to both SM quarks and leptons via the Lagrangian,
	\begin{equation}
	\mathcal{L}_{\rm S-f} \ = \ \bar{F}^c\, \pmb{\lambda}_i^L \, F S_i \ + \ 
	\bar{f}^c\, \pmb{\lambda}^R_i \, f S_i \ +  \ \bar{f}\: \tilde{\pmb{\lambda}}_i\, F S_i \ + \ {\rm h.c.} \;, \label{eq:lagSf}
	\end{equation}
where  fermion $F = \{ q, \ell \}$ 	is a Left handed
quark or lepton weak doublet field, while $f = \{u,d,e \}$ is a Right-handed fermion weak singlet field. $F^c, f^c$ denote charge-conjugated fermion fields. Gauge and flavour indices are all suppressed in \eqref{eq:lagSf},
or otherwise encoded in the Yukawa couplings, $\pmb{\lambda}^L,\pmb{\lambda}^R$ and,
$ \tilde{\pmb{\lambda}}$.
	Therefore, there are five different scalar leptoquark field representations in weak isospin space: three singlets, two doublets and one triplet. Their gauge quantum numbers
	under the SM gauge group, are shown in Table \ref{tab:allcharges}.\footnote{In  
	notation of \eqref{eq:lagSf} some leptoquark fields from Table~\ref{tab:allcharges}
	may be their charge-conjugated fields.} The LQ-flavour index $i$ in \eqref{eq:lagSf} takes the values $i=\{1,\tilde{1},2,\tilde{2},3\}$.~\footnote{With apologies to the reader, 
	the LQ-flavour indices-$i,j,k$ used throughout this section should not be confused with 
	the colour indices-$i,j,k$ introduced in section~\ref{sec:3}.} 
	\begin{table}[h]
		\centering
		\begin{tabular}[c]{| c | c  c  c |}
			\hline
			LQ-fields ($S$) & SU(3) & SU(2) & U(1)\\
			\hline\hline
			$ S_{1} $ & $ \phantom{\displaystyle\frac{1}{1}}\bar{3} $ & $ 1 $ &  $ \frac{1}{3} $\\
			$ \tilde{S}_{1} $ & $ \phantom{\displaystyle\frac{1}{1}}\bar{3} $ & $ 1 $ &  $ \frac{4}{3} $\\
			$ S_{2} $ & $ \phantom{\displaystyle\frac{1}{1}} 3 $ & $ 2 $ &  $ \frac{7}{6} $\\
			$ \tilde{S}_{2} $ & $ \phantom{\displaystyle\frac{1}{1}} 3 $ & $ 2 $ &  $ \frac{1}{6} $\\
			$ S_{3} $ & $ \phantom{\displaystyle\frac{1}{1}}\bar{3} $ & $ 3 $ &  $ \frac{1}{3} $\\
			\hline
		\end{tabular}
		\caption{All possible representations of  leptoquark fields under the SM gauge group.}
		\label{tab:allcharges}
	\end{table}	
%
Obviously, by picking up only quarks from $F-$ (or from $f-$) fields we arrive at  Baryon ($B$) and Lepton ($L$) number non-conservation 
LQ-interactions.

All five scalar LQs can interact with the SM Higgs-field\footnote{The hypercharge of the SM Higgs doublet is defined so that $ Y_{H} = 1/2 $.} ($H$) through trilinear
and quadratic terms of the form:
\begin{equation}
\mathcal{L}_{\rm S-H} \ = \  (A_{ij} H^\dagger S_i S_j \ + \ {\rm h.c.} ) \ + \ \lambda_{Hi} (S^\dagger_i S_i) (H^\dagger H) \ + \ (\lambda_{3S} S_i S_j S_k H^\dagger + \mathrm{h.c.}) \ + \ \dots \;,
\label{eq:S-H}
\end{equation}
where ``\dots " mean other gauge invariant terms of the form $(SSHH)$ and 
$(SSSH)$. Their exact form is irrelevant for drawing the 
supertrace functional diagrams  since their explicit details  entered only 
 at the end in $\mathbf{X}$-matrices of \eqref{eq:master}.
Note that the $A$-term of \eqref{eq:S-H} has mass dimension one, there are
only two options $H S_1 \tilde{S}_2$ and  $H S_3 \tilde{S}_2$,  and it 
plays an important role in the effective Lagrangian at $d\ge 5$ level as
we shall see in the next section.

Furthermore, self-interactions among LQs read in general as
\begin{equation}
\mathcal{L}_{\rm S} \ =\   - M_i^2 |S_i|^2 \ + \  A^\prime_{ijk} (S_i^\dagger S_j S_k) \ +\ c_{ijkl} (S_i^\dagger S_j) (S_k^\dagger S_l) \ + \ \cdots \;,
\label{eq:S}
\end{equation}
where again ``$\cdots"$ refer to different internal gauge group invariant structure of terms not in our immediate interest in constructing the effective action. 
Again $A^\prime$ is a mass dimension one parameter but break baryon and lepton numbers. Among the fields arranged in Table~\ref{tab:allcharges},  there 
are three choices of $A'$-terms: $S_1^\dagger \tilde{S_2}\tilde{S_2}$, 
  $S_3^\dagger \tilde{S}_2 \tilde{S}_2$ and $\tilde{S}^\dagger_1S_2\tilde{S}_2$.
  Masses $M_i$ in  \eqref{eq:S} are assumed much heavier than the electroweak scale $m_W$ but the dimension-full parameters introduced above could in general range within
  \begin{equation}
  0 \ \le \ \left \{ A/M_i, \; A'/M_i \right \}\ \lesssim \ 1 \;.
  \label{eq:Aregion}
  \end{equation}
In total, the BSM Lagrangian is 
\begin{equation}
\mathcal{L}_{\rm BSM} = \mathcal{L}_{\rm S-f} + 
\mathcal{L}_{\rm S-H} + \mathcal{L}_{\rm S}\;.
\label{eq:BSM}
\end{equation}
This ``universal" way of writing down leptoquark interactions will be the stepping stone for the calculation of the effective action at tree and one-loop levels using eqs.~\eqref{eq:con1},\eqref{eq:con2}
and  \eqref{eq:master}, respectively, since these will determine the dimensionality of the $\mathbf{X}$-matrices that we will introduce shortly. Otherwise, the explicit form of $\mathcal{L}_{\rm BSM}$
is given in Appendix~\ref{app:X}.


		\subsection{Tree level EFT, $\mathcal{L}^{\rm (tree)}_{\rm EFT}$}
		
We start out with the UV-Lagrangian $\mathcal{L}_{\rm UV}[S,\phi] = \mathcal{L}_{\rm SM}[\phi] + \mathcal{L}_{\rm BSM}[S,\phi]$ and derive the EOMs \eqref{eq:eoms}  for the heavy fields $S_i$  in Table~\ref{tab:allcharges}. We 
solve EOMs and substitute the solutions for the classical fields $S_{i,c}[\phi]$
back into $\mathcal{L}_{UV}$ in order to obtain the tree-level EFT from eq.~\eqref{eq:con1}.
By expanding the classical field in inverse powers of heavy masses $M_i$
\begin{equation}
S_{i,c}[\phi] \ = \ S_{i,c}^{(3)} \ + \ S_{i,c}^{(4)}  \ + \  \dots \;,
\end{equation}
we find
\begin{eqnarray}
(S_{i,c}^{(3)})^\dagger \ &=& 
\ \frac{1}{M_i^2} \left ( \bar{F}^c\, \pmb{\lambda}_i^L \, F  \ + \ 
	\bar{f}^c\, \pmb{\lambda}^R_i \, f  \ +  \ \bar{f}\: \tilde{\pmb{\lambda}}_i\, F    \right )\;, \label{eq:Sc3} \\[2mm]	
	(S_{i,c}^{(4)})^\dagger \ &=& \frac{1}{M_i^2} \, A_{ij} \, H^\dagger \, S_{j,c}^{(3)} \;.
	\label{eq:Sc4}
\end{eqnarray}
The solutions $S_{i,c}^{(n)}$ contain operators with mass dimension $n$  which are 
suppressed by factors that scale like $M_i^{n-1}$.  Plugging in this back to eqs.~\eqref{eq:lagSf}-\eqref{eq:S} we obtain the tree-level effective Lagrangian containing $d\le 7$ operators 
\begin{equation}
\mathcal{L}^{\rm (tree)}_{\rm EFT}[\phi] \ = \ M_i^2\:  (S_{i,c}^{(3)})^\dagger\: (S_{i,c}^{(3)}) 
\ +  \  ( A_{ij} H^\dagger S_{i,c}^{(3)} S_{j,c}^{(3)} + \mathrm{h.c.}) \;,
\label{eq:treeEFT}
\end{equation}
where $S_{i,c}^{(3)}$ is the hermitian conjugate of \eqref{eq:Sc3}. Therefore, the tree-level EFT contains
\emph{only} four-fermion dimension-6 operators proportional to the product of couplings from the 
set $\pmb{ \{\lambda}^L,\pmb{\lambda}^R,\pmb{\tilde{\lambda} \} }$. On the other hand,
\emph{all} tree-level dimension-7 operators are proportional to the dimension-full
combination of parameters $A \times \pmb{\lambda}^2$. From eq.~\eqref{eq:S-H} we see that 
the parameters $\lambda_{Hi}$ and $\lambda_{3S}$ appear first to multiplying 
operators with dimensions-8 and 10, respectively, while from eq.~\eqref{eq:S}, the 
parameters $A'$ and $c$ are associated with dimension-9 and 12, respectively.
Although our main focus in this paper is on operators with dimensions less or equal to 
six it is obvious that dimension-7 operators at tree-level may become equally important  in the parameter region where $A \approx M_i$ and one other leptoquark mass is $M_j \approx \sqrt{v M_i}$,
with $v$ being the electroweak vev.


			\subsection{$\mathbf{K}$- and $\mathbf{X}$-matrices}
			 \label{sec:Setup}			
				The neccessary steps for one loop matching are neatly outlined in \cite{Cohen:2020fcu} and are followed closely here. In performing the matching, the method of functional \emph{supertraces} will be used as introduced in~\cite{Cohen:2020fcu}. After collecting and constructing the supertraces, the application of the CDE (Covariant Derivative Expansion) is carried out automatically through two recently developed packages~\cite{Cohen:2020qvb,Fuentes-Martin:2020udw}. We will be using mainly the package, {\tt STrEAM} of~\cite{Cohen:2020qvb}.

	The rationale of these diagrams comes from an earlier diagrammatic approach to matching which uses the so called covariant diagrams~\cite{Zhang:2016pja}. 
	In Appendix~\ref{app:a} we make this comparison more explicit by presenting the equivalent covariant diagrams that match to the diagrammatic supertraces.
	
	We begin by creating field multiplets, where we denote the five (heavy) leptoquark  fields, listed in Table~\ref{tab:allcharges}, as  $ \{S_{i}\} = \left\{S_1, \tilde{S}_1,  S_2, \tilde{S}_2, S_3  \right\} $. Additionally, to treat chiral fermions we introduce fictitious fields promoting Weyl fermions into Dirac and properly inserting projections operators to single-out the correct chirality of the fields in the end.
	For simplicity these projection operators are left implicit throughout the text. The field multiplets then read,
		\begin{align}
			\varphi_{S} &= \left\{\varphi_{S_{i}}\right\} = 
			\left\{
			\begin{pmatrix}
				S_{i}\\ S_{i}^{\ast}
			\end{pmatrix}
			\right\}\;,\\
			\varphi_{H} &=
			\left\{
			\begin{pmatrix}
				H \\ H^{\ast}
			\end{pmatrix}
			\right\}\;,\\
			\varphi_{f} &= \left\{\varphi_{\ell},\,\varphi_{q},\,\varphi_{u},\varphi_{e},\,\varphi_{d}\right\} = 
			\left\{
			\begin{pmatrix}
			\ell \\ \ell^{c}
			\end{pmatrix},\,
			\begin{pmatrix}
			q \\ q^{c}
			\end{pmatrix},\,
			\begin{pmatrix}
			u \\ u^{c}
			\end{pmatrix},\,
			\begin{pmatrix}
			e \\ e^{c}
			\end{pmatrix},\,
			\begin{pmatrix}
			d \\ d^{c}
			\end{pmatrix}
			\right\}\;,\\
			\varphi_{V} &= \left\{B,\,W,\,G\right\}\;.
		\end{align}
	We also introduce the conjugate field multiplets,
		\begin{align}
			\bar{\varphi}_{S} &= \left\{\bar{\varphi}_{S_{i}}\right\} = 
			\left\{
			\begin{pmatrix}
				S^{\dagger}_i & S_{i}^{T}
			\end{pmatrix}
			\right\}\;,\\
			\bar{\varphi}_{H} &=
			\left\{
			\begin{pmatrix}
				H^{\dagger} & H^{T}
			\end{pmatrix}
			\right\}\;,\\
			\bar{\varphi}_{f} &= \left\{\bar{\varphi}_{\ell},\,\bar{\varphi}_{q},\,\bar{\varphi}_{u},\bar{\varphi}_{e},\,\bar{\varphi}_{d}\right\} = 
			\left\{
			\begin{pmatrix}
			\bar{\ell} & \bar{\ell}^{c}
			\end{pmatrix},\,
			\begin{pmatrix}
			\bar{q} & \bar{q}^{c}
			\end{pmatrix},\,
			\begin{pmatrix}
			\bar{u} & \bar{u}^{c}
			\end{pmatrix},\,
			\begin{pmatrix}
			\bar{e} & \bar{e}^{c}
			\end{pmatrix},\,
			\begin{pmatrix}
			\bar{d} & \bar{d}^{c}
			\end{pmatrix}
			\right\}\;,\\
			\bar{\varphi}_{V} &= \left\{B,\,W,\,G\right\}\;.
		\end{align}
	These field multiplets are connected to the inverse propagator matrix-$\mathbf{K}$
	and to interaction matrix-$\mathbf{X}$, both needed for master formula \eqref{eq:master}, via the second variation of the action as
		\begin{equation}
			+\frac{1}{2}\,
			\delta\bar{\varphi}\,
			\mathbf{K}\,
			\delta\varphi
			\ -\  \frac{1}{2}
			\begin{pmatrix}
				\delta\bar{\varphi}_{S} & \delta\bar{\varphi}_{L}
			\end{pmatrix}
			\begin{bmatrix}
				X_{SS} & X_{SL}\\
				X_{LS} & X_{LL}
			\end{bmatrix}
			\begin{pmatrix}
				\delta\varphi_{S} \\ \delta\varphi_{L}
			\end{pmatrix}
			\;.
			\label{eq:varsec}
		\end{equation}
	We have gathered all \emph{light} multiplets in $ \varphi_{L} $ and we denote the whole field multiplet with $ \varphi $ for brevity.  Matrix-$\mathbf{K}$ is block-diagonal with 
	$(P^2 -m^2)$ for spin-0
	fields, $(\slashed P -m)$ for spin-1/2  and $-\eta^{\mu\nu} (P^2 -m^2)$ for spin-1 fields in Feynman gauge. Here $P_\mu \equiv i D_\mu$ is basically the (Hermitian) covariant derivative.
	The $ \mathbf{X} $-matrix may contain potential-only interactions and/or terms with open covariant derivatives as well. It is evaluated with $S_i = S_{i,c}[\phi]$. Moreover, in the most general case,
	\begin{equation}
		\mathbf{X} \ = \ \mathbf{U} \ +\ P^{\kappa}\mathbf{Z}_{\kappa} \ +\  \mathbf{\bar{Z}}_{\kappa}P^{\kappa} \ + \ \ldots
		\label{eq:X}
	\end{equation}
	where the dots contain terms with two or more open covariant derivatives, however these higher derivative terms do not appear in any renormalizable UV-model, such as the LQ-models under consideration and can be ignored.
	
	The $\mathbf{X}$-interaction matrix structure in \eqref{eq:varsec} is split into \emph{heavy-heavy}, $ \mathbf{X}_{SS} $, \emph{light-light}, $ \mathbf{X}_{LL} $ and \emph{heavy-light (light-heavy)}, $ \mathbf{X}_{SL} $ $ (\mathbf{X}_{LS}) $ sub-blocks. 
In terms of the expansion matrices of 	\eqref{eq:X} these sub-blocks  are organized in the following way,
		\begin{align}
			\left(\mathbf{X}_{SS}\right)_{10\times 10} &=
			\begin{pmatrix}
				\left(U_{S_{i}S_{j}}\right)_{10\times 10}
			\end{pmatrix}\;, \label{eq:XSS}\\
				\left(\mathbf{X}_{SL}\right)_{10\times 15} &=
			\begin{pmatrix}
				\left(U_{S_{i}f}\right)_{10\times 10} &
				\left(U_{S_{i}H}\right)_{10\times 2} &
				\left(X_{S_{i}V}\right)_{10\times 3}
			\end{pmatrix}\;, \label{eq:XSL}\\
				\left(\mathbf{X}_{LS}\right)_{15\times 10} &=
			\begin{pmatrix}
				\left(U_{fS_{i}}\right)_{10\times 10} \\
				\left(U_{HS_{i}}\right)_{2\times 10} \\
				\left(X_{VS_{i}}\right)_{3\times 10} 
			\end{pmatrix}\;,\\
				\left(\mathbf{X}_{LL}\right)_{15\times 15} &=
			\begin{pmatrix}
				\left(U_{ff}\right)_{10\times 10} & \left(U_{fH}\right)_{10\times 2} & \left(U_{fV}\right)_{10\times 3}\\
				\left(U_{Hf}\right)_{2\times 10} & \left(U_{HH}\right)_{2\times 2} & \left(U_{HV}\right)_{2\times 3}\\
				\left(U_{Vf}\right)_{3\times 10} & \left(U_{VH}\right)_{3\times 2} & \left(U_{VV}\right)_{3\times 3}
			\end{pmatrix}\;, \label{eq:XLL}
		\end{align}
	where with subscript we denote the respective matrix dimensionality
	for each generation of light fermions.
	
	From the general interactions, eqs.~\eqref{eq:lagSf},\eqref{eq:S-H} and \eqref{eq:S}, of the scalar leptoquarks outlined in the previous subsection, we can now read the mass dimensions of the corresponding elements of $\mathbf{U}$-matrices. By schematically performing a second variation, for example on $ S(\bar{f}f) $-terms,
		\begin{align}
			\delta^2 \left(S\,\bar{f}f\right) &\propto (\delta S) (\delta\bar{f}) f + (\delta S) \bar{f} (\delta f) + S (\delta \bar{f}) (\delta f)\nonumber\\ 
			&= (\delta \bar{f}) U_{\bar{f}S} (\delta S) + (\delta S^T)U_{S^Tf}(\delta f) + (\delta\bar{f}) U_{\bar{f}f} (\delta f)\;,
		\end{align}
 we can obtain the mass dimensions of the $ \mathbf{X} $-matrices.	
	Adding the h.c. of this interaction and doing again the exact calculation for the conjugate fermion fields, namely $ S \bar{f}^c f^c + \text{h.c.} $, we can get the mass dimension of the matrices $ [\mathbf{U}_{ff}] = 3 $, $ [\mathbf{U}_{S_{i}f}] = 3/2 $ and $ [\mathbf{U}_{fS_{i}}] = 3/2 $. Consequently, we arrive at the following mass dimensions for all involved matrices in notation of eqs.~\eqref{eq:XSS}-\eqref{eq:XLL},
		\begin{align}
					\left[\mathbf{X}_{SS}\right] &=
		\begin{bmatrix}
			(1,2,3,4,6)
		\end{bmatrix}\;, \\[2mm]
		\left[\mathbf{X}_{SL}\right] &= \left[\mathbf{X}_{LS}\right] =
		\begin{bmatrix}
			3/2 & (3,4,6) & (3,4)
		\end{bmatrix}\;,\\[2mm]
		\left[\mathbf{X}_{LL}\right] &= \left[\mathbf{X}_{LL}\right]_{\text{SM}} \ + \ \left[\mathbf{X}_{LL}\right]_{\text{BSM}} =
		\begin{bmatrix}
			1 & 3/2 & 3/2\\
			3/2 & 2 & 2\\
			3/2 &  2  & 2
		\end{bmatrix} +
		\begin{bmatrix}
		3 & 0 & 0\\
		0 & 6 & 0\\
		0 &  0  & 6
		\end{bmatrix}\;,
		\end{align}
	 where in  parenthesis we denote all possible mass 
	 dimensions with $d\le 6$  (starting from the lowest) of the $\mathbf{X}$-matrices. 
	 
	 For all scalar leptoquark interactions we need to calculate the explicit form of the $\mathbf{X}$-matrices in eqs.~\eqref{eq:XSS}-\eqref{eq:XLL}. This is done in Appendix~\ref{app:X}.
	 As an application, we shall deploy those matrices in section~\ref{sec:3},
	  for a detailed functional matching procedure in
	 a particular model for decoupling together two heavy LQ fields, the $S_1$ 
	 and the $\tilde{S}_2$.


			\subsection{Enumerating: UOLEA and supertraces}
				\label{sec:1loop}
				
There are two contributions in the rhs of one-loop effective action [eq.~\eqref{eq:master}]: the log-type term, $\mathrm{STr}\log \mathbf{K}$, and the power-type, 
$\mathrm{STr} \left [ (\mathbf{K}^{-1} \mathbf{X})^n \right ]$.
However, a great deal of contributions  in eq.~\eqref{eq:master} are encoded in 19-UOLEA-terms [\textit{c.f.} eq.~\eqref{uolea}]  for only-heavy scalars. These UOLEA terms include the full expressions of log-type terms and all power-type diagrams with \emph{only heavy} scalars in the loop. 
  
 What remains to be added is all \emph{heavy-light} diagrams.
 For those we use the technique of functional supertraces of ref.~\cite{Cohen:2020fcu} and, as a cross check, the technique of covariant diagrams of ref.~\cite{Zhang:2016pja}.
  In fact, a detailed diagrammatic comparison of both techniques is given in Appendix~\ref{app:a}.

	The $\mathbf{X}$-matrices  are the building blocks for the functional supertraces. In  most of the cases only the $\mathbf{U}$-matrices appear in the expansion  \eqref{eq:X}.
	Different combinations of these matrices are inserted into diagrams and make up operators of up-to mass 
	dimension-6. In what follows we list all diagrammatic supertraces along with the equivalent expressions that arise through this process (see~\cite{Cohen:2020fcu} for details). Our notation in 
	functional diagrams below is the following:  heavy leptoquark fields $S_{i}$ with masses $M_i$ (double-dashed lines), 
	$f,f',f^{''},f^{'''}={q,u,d,\ell, e}$ are  the SM fermion fields (solid lines), $H$ is the SM Higgs-doublet (single dashed-lines), and $V={G,W,B}$ are the SM gauge fields (wavy lines).  
	Every circle indicates an insertion from $U$(or in general $X$)-matrices and $P_\mu$ is the covariant derivative.
	Furthermore,  all SM fields are taken to be massless and $\eta^{\mu\nu}=(1,-1,-1,-1)$ is the metric tensor. We these definitions we obtain:
	\vspace{1cm}
	\begin{align}
		\parbox{20mm}{\begin{fmffile}{OU2}
	\begin{fmfgraph*}(10,10)
		\fmfset{dash_len}{2mm}
		\fmfcurved
		\fmfsurroundn{v}{2}
		\fmfv{d.shape=circle, d.filled=empty,d.size=5}{v1,v2}
		\fmf{dashes,right=1,label=$ H $,label.side=right,label.dist=1.5mm}{v1,v2}
		\fmf{dbl_dashes,right=1,label=$ S_i $,label.side=right,label.dist=1.5mm}{v2,v1}
	\end{fmfgraph*}
\end{fmffile}} &=\quad -\frac{i}{2}\,\text{STr}\left.\left[\frac{1}{P^2 - M_i^2}U_{S_iH}\frac{1}{P^2}U_{HS_i}\right]\right|_{\text{hard}}\;,\label{diag1}\\
		\nonumber\\
		\nonumber\\
		\parbox{20mm}{\begin{fmffile}{OU2PStr}
	\begin{fmfgraph*}(10,10)
		\fmfset{dash_len}{2mm}
		\fmfcurved
		\fmfsurroundn{v}{2}
		\fmfv{d.shape=circle, d.filled=empty,d.size=5}{v1,v2}
		\fmf{dbl_dashes,right=1,label=$ S_i $,label.side=right,label.dist=1.5mm}{v1,v2}
		\fmf{plain,right=1,label=$ f $,label.side=right,label.dist=1.5mm}{v2,v1}
	\end{fmfgraph*}
\end{fmffile}} &=\quad -\frac{i}{2}\,\text{STr}\left.\left[\frac{1}{P^2 - M_i^2}U_{S_if}\frac{1}{\slashed{P}}U_{fS_i}\right]\right|_{\text{hard}}\;,\\
		\nonumber\\
		\nonumber\\
		\parbox{20mm}{\begin{fmffile}{OU3PStr}
	\begin{fmfgraph*}(10,10)
		\fmfset{dash_len}{2mm}
		\fmfcurved
		\fmfsurroundn{v}{8}
		\fmfv{d.shape=circle, d.filled=empty,d.size=5}{v1,v3,v5}
		\fmf{dbl_dashes,right=0.5,label=$ S_j $,label.side=right,label.dist=1.5mm}{v1,v3}
		\fmf{dbl_dashes,right=0.5,label=$ S_i $,label.side=right,label.dist=1.5mm}{v3,v5}
		\fmf{plain,right=1,label=$ f$,label.side=right,label.dist=.5mm}{v5,v1}
	\end{fmfgraph*}
\end{fmffile}} &=\quad -\frac{i}{2}\,\text{STr}\left.\left[\frac{1}{P^2 - M_i^2}U_{S_iS_j}\frac{1}{P^2-M_j^2}U_{S_jf}\frac{1}{\slashed{P}}U_{fS_i}\right]\right|_{\text{hard}}\;,\label{Strg-2a}\\
		\nonumber\\
		\nonumber\\
		\parbox{20mm}{\begin{fmffile}{OU3Str}
	\begin{fmfgraph*}(10,10)
		\fmfset{dash_len}{2mm}
		\fmfcurved
		\fmfsurroundn{v}{8}
		\fmfv{d.shape=circle, d.filled=empty,d.size=5}{v1,v7,v5}
		\fmf{plain,right=1,label=$ f $,label.side=right,label.dist=1.5mm}{v1,v5}
		\fmf{dbl_dashes,left=0.5,label=$ S_i $,label.side=left,label.dist=1.5mm}{v7,v5}
		\fmf{plain,right=0.5,label=$ f^\prime $,label.side=right,label.dist=.5mm}{v7,v1}
	\end{fmfgraph*}
\end{fmffile}} &=\quad -\frac{i}{2}\,\text{STr}\left.\left[\frac{1}{P^2-M_i^2}U_{S_if}\frac{1}{\slashed{P}}U_{ff^\prime}\frac{1}{\slashed{P}}U_{f^\prime S_i}\right]\right|_{\text{hard}}\;,\label{Strg-2b}\\
		\nonumber\\
		\nonumber\\
		\parbox{20mm}{\begin{fmffile}{OU4HStr}
	\begin{fmfgraph*}(10,10)
		\fmfset{dash_len}{2mm}
		\fmfcurved
		\fmfsurroundn{v}{8}
		\fmfv{d.shape=circle, d.filled=empty,d.size=5}{v1,v3,v5,v7}
		\fmf{dashes,right=0.5,label=$ H $,label.side=right,label.dist=1mm}{v1,v3}
		\fmf{plain,right=0.5,label=$ f $,label.side=right,label.dist=.5mm}{v3,v5}
		\fmf{dbl_dashes,right=0.5,label=$ S_i $,label.side=right,label.dist=1.5mm}{v5,v7}
		\fmf{plain,right=0.5,label=$ f^\prime $,label.side=right,label.dist=.5mm}{v7,v1}
	\end{fmfgraph*}
\end{fmffile}} &=\quad 	-\frac{i}{2}\,\text{STr}\left.\left[\frac{1}{P^2-M_i^2}U_{S_if}\frac{1}{\slashed{P}}U_{fH}\frac{1}{P^2}U_{Hf^\prime}\frac{1}{\slashed{P}}U_{f^\prime S_i}\right]\right|_{\text{hard}}\;,\\
		\nonumber\\
		\nonumber\\
		\parbox{20mm}{\begin{fmffile}{OU4sfStr}
	\begin{fmfgraph*}(10,10)
		\fmfset{dash_len}{2mm}
		\fmfcurved
		\fmfsurroundn{v}{8}
		\fmfv{d.shape=circle, d.filled=empty,d.size=5}{v1,v3,v5,v7}
		\fmf{dbl_dashes,right=0.5,label=$ S_j $,label.side=right,label.dist=1mm}{v1,v3}
		\fmf{plain,right=0.5,label=$ f $,label.side=right,label.dist=.5mm}{v3,v5}
		\fmf{dbl_dashes,right=0.5,label=$ S_i $,label.side=right,label.dist=1.5mm}{v5,v7}
		\fmf{plain,right=0.5,label=$ f^\prime $,label.side=right,label.dist=.5mm}{v7,v1}
	\end{fmfgraph*}
\end{fmffile}} &=\quad 	-\frac{i}{4}\,\text{STr}\left.\left[\frac{1}{P^2-M_i^2}U_{S_if}\frac{1}{\slashed{P}}U_{fS_j}\frac{1}{P^2-M_j^2}U_{S_jf^\prime}\frac{1}{\slashed{P}}U_{f^\prime S_i}\right]\right|_{\text{hard}}\;,\\
		\nonumber\\
		\nonumber\\
		\parbox{20mm}{\begin{fmffile}{OU4SSStr}
	\begin{fmfgraph*}(10,10)
		\fmfset{dash_len}{2mm}
		\fmfcurved
		\fmfsurroundn{v}{8}
		\fmfv{d.shape=circle, d.filled=empty,d.size=5}{v1,v3,v5,v7}
		\fmf{plain,right=0.5,label=$ f $,label.side=right,label.dist=1mm}{v1,v3}
		\fmf{dbl_dashes,right=0.5,label=$ S_j $,label.side=right,label.dist=.5mm}{v3,v5}
		\fmf{dbl_dashes,right=0.5,label=$ S_i $,label.side=right,label.dist=1.5mm}{v5,v7}
		\fmf{plain,right=0.5,label=$ f^\prime $,label.side=right,label.dist=.5mm}{v7,v1}
	\end{fmfgraph*}
\end{fmffile}} &=\quad -\frac{i}{2}\,\text{STr}\left.\left[\frac{1}{P^2-M_i^2}U_{S_iS_j}\frac{1}{P^2-M_j^2}U_{S_jf}\frac{1}{\slashed{P}}U_{ff^\prime}\frac{1}{\slashed{P}}U_{f^\prime S_i}\right]\right|_{\text{hard}}\;,\label{eq:2.34}\\
		\nonumber\\
		\nonumber\\
		\parbox{20mm}{\begin{fmffile}{OU4XStr}
	\begin{fmfgraph*}(10,10)
		\fmfset{dash_len}{2mm}
		\fmfcurved
		\fmfsurroundn{v}{8}
		\fmfv{d.shape=circle, d.filled=empty,d.size=5}{v1,v3,v5,v7}
		\fmf{boson,right=0.5,label=$ V $,label.side=right,label.dist=1mm}{v1,v3}
		\fmf{plain,right=0.5,label=$ f $,label.side=right,label.dist=.5mm}{v3,v5}
		\fmf{dbl_dashes,right=0.5,label=$ S_i $,label.side=right,label.dist=1.5mm}{v5,v7}
		\fmf{plain,right=0.5,label=$ f^\prime $,label.side=right,label.dist=.5mm}{v7,v1}
	\end{fmfgraph*}
\end{fmffile}} &=\quad -\frac{i}{2}\,\text{STr}\left.\left[\frac{1}{P^2-M_i^2}U_{S_if}\frac{1}{\slashed{P}}U^{\mu}_{fV}\frac{-\eta _{\mu\nu}}{P^2}U^{\nu}_{Vf^\prime}\frac{1}{\slashed{P}}U_{f^\prime S_i}\right]\right|_{\text{hard}}\;,\label{eq:2.35}\\
		\nonumber\\
		\nonumber\\
		\parbox{20mm}{\begin{fmffile}{OU4PStr}
	\begin{fmfgraph*}(10,10)
		\fmfset{dash_len}{2mm}
		\fmfcurved
		\fmfsurroundn{v}{8}
		\fmfv{d.shape=circle, d.filled=empty,d.size=5}{v1,v3,v5,v7}
		\fmf{plain,right=0.5,label=$ f^\prime $,label.side=right,label.dist=1mm}{v1,v3}
		\fmf{plain,right=0.5,label=$ f $,label.side=right,label.dist=.5mm}{v3,v5}
		\fmf{dbl_dashes,right=0.5,label=$ S_i $,label.side=right,label.dist=1.5mm}{v5,v7}
		\fmf{plain,right=0.5,label=$ f^{\prime\prime} $,label.side=right,label.dist=.5mm}{v7,v1}
	\end{fmfgraph*}
\end{fmffile}} &=\quad -\frac{i}{2}\,\text{STr}\left.\left[\frac{1}{P^2 - M_i^2}U_{S_if}\frac{1}{\slashed{P}}U_{ff^\prime}\frac{1}{\slashed{P}}U_{f^\prime 	f^{\prime\prime}}\frac{1}{\slashed{P}}U_{f^{\prime\prime}S_i}\right]\right|_{\text{hard}}\;,\\
		\nonumber\\
		\nonumber\\
		\parbox{20mm}{\begin{fmffile}{OU4PStrnew}
	\begin{fmfgraph*}(10,10)
		\fmfset{dash_len}{2mm}
		\fmfcurved
		\fmfsurroundn{v}{16}
		\fmfv{d.shape=circle, d.filled=empty,d.size=5}{v1,v5,v9,v13}
		\fmf{plain,right=.5,label=$ f $,label.side=right,label.dist=2mm}{v1,v5}
		\fmf{dbl_dashes,right=.5,label=$ S_k $,label.side=right,label.dist=2mm}{v5,v9}
		\fmf{dbl_dashes,right=.5,label=$ S_i $,label.side=right,label.dist=2mm}{v13,v1}
		\fmf{dbl_dashes,right=.5,label=$ S_j $,label.side=right,label.dist=2mm}{v9,v13}
	\end{fmfgraph*}
\end{fmffile}} &=\quad -\frac{i}{2}\,\text{STr}\left.\left[\frac{1}{P^2 - M_i^2}U_{S_iS_j}\frac{1}{P^2-M_j^2}U_{S_jS_k}\frac{1}{P^2-M^2_k}U_{S_kf }\frac{1}{\slashed{P}}U_{fS_i}\right]\right|_{\text{hard}}\;,\\
		\nonumber\\
		\nonumber\\
		\parbox{20mm}{\begin{fmffile}{OU5Str}
	\begin{fmfgraph*}(10,10)
		\fmfset{dash_len}{2mm}
		\fmfcurved
		\fmfsurroundn{v}{10}
		\fmfv{d.shape=circle, d.filled=empty,d.size=5}{v1,v3,v5,v7,v9}
		\fmf{plain,right=0.4,label=$ f^{\prime\prime} $,label.side=right,label.dist=1mm}{v1,v3}
		\fmf{plain,right=0.4,label=$ f^\prime $,label.side=right,label.dist=1mm}{v3,v5}
		\fmf{plain,right=0.4,label=$ f $,label.side=right,label.dist=1mm}{v5,v7}
		\fmf{dbl_dashes,right=0.4,label=$ S_i $,label.side=right,label.dist=1.5mm}{v7,v9}
		\fmf{plain,right=0.4,label=$ f^{\prime\prime\prime} $,label.side=right,label.dist=1mm}{v9,v1}
	\end{fmfgraph*}
\end{fmffile}} &=\quad -\frac{i}{2}\,\text{STr}\left.\left[\frac{1}{P^2 - M_i^2}U_{S_if}\frac{1}{\slashed{P}}U_{ff^\prime}\frac{1}{\slashed{P}}U_{f^\prime 	f^{\prime\prime}}\frac{1}{\slashed{P}}U_{f^{\prime\prime}f^{\prime\prime\prime}}\frac{1}{\slashed{P}}U_{f^{\prime\prime\prime}S_i}\right]\right|_{\text{hard}}\;,\\
		\nonumber\\
		\nonumber\\
		\parbox{20mm}{\begin{fmffile}{OU5Strnew}
	\begin{fmfgraph*}(10,10)
		\fmfset{dash_len}{2mm}
		\fmfcurved
		\fmfsurroundn{v}{10}
		\fmfv{d.shape=circle, d.filled=empty,d.size=5}{v1,v3,v5,v7,v9}
		\fmf{dbl_dashes,right=0.4,label=$ S_i $,label.side=right,label.dist=1mm}{v1,v3}
		\fmf{plain,right=0.4,label=$ f^\prime $,label.side=right,label.dist=1mm}{v3,v5}
		\fmf{plain,right=0.4,label=$ f $,label.side=right,label.dist=1mm}{v5,v7}
		\fmf{dbl_dashes,right=0.4,label=$ S_k $,label.side=right,label.dist=1.5mm}{v7,v9}
		\fmf{dbl_dashes,right=0.4,label=$ S_j $,label.side=right,label.dist=1mm}{v9,v1}
	\end{fmfgraph*}
\end{fmffile}} &=\quad -\frac{i}{2}\,\text{STr}\left.\left[\frac{1}{P^2 - M_i^2}U_{S_iS_j}\frac{1}{P^2-M_j^2}U_{S_jS_k}\frac{1}{P^2-M_k^2}U_{S_kf}\frac{1}{\slashed{P}}U_{ff^\prime}\frac{1}{\slashed{P}}U_{f^\prime S_i}\right]\right|_{\text{hard}}\;,\\
		\nonumber\\
		\nonumber\\
		\parbox{20mm}{\begin{fmffile}{OZ11Str}
	\begin{fmfgraph*}(10,10)
		\fmfset{dash_len}{2mm}
		\fmfcurved
		\fmfsurroundn{v}{8}
		\fmfv{d.shape=circle, d.filled=empty,d.size=5}{v1,v3,v5}
		\fmf{plain,right=.5,label=$ f $,label.side=right,label.dist=2mm}{v1,v3}
		\fmf{boson,right=.5,label=$ V $,label.side=right,label.dist=2mm}{v3,v5}
		\fmf{dbl_dashes,right=1,label=$ S_i $,label.side=right,label.dist=2mm}{v5,v1}
	\end{fmfgraph*}
\end{fmffile}} &=\quad -\frac{i}{2}\,\text{STr}\left.\left[\frac{1}{P^2-M_i^2}X^{\mu}_{S_iV}\frac{-\eta_{\mu\nu}}{P^2}X^{\nu}_{Vf}\frac{1}{\slashed{P}}U_{fS_i}\right]\right|_{\text{hard}}\;,\\
		\nonumber\\
		\nonumber\\
		\parbox{20mm}{\begin{fmffile}{OZ12Str}
	\begin{fmfgraph*}(10,10)
		\fmfset{dash_len}{2mm}
		\fmfcurved
		\fmfsurroundn{v}{8}
		\fmfv{d.shape=circle, d.filled=empty,d.size=5}{v1,v3,v5}
		\fmf{boson,right=.5,label=$ V $,label.side=right,label.dist=2mm}{v1,v3}
		\fmf{plain,right=.5,label=$ f $,label.side=right,label.dist=2mm}{v3,v5}
		\fmf{dbl_dashes,right=1,label=$ S_i $,label.side=right,label.dist=2mm}{v5,v1}
	\end{fmfgraph*}
\end{fmffile}} &=\quad -\frac{i}{2}\,\text{STr}\left.\left[\frac{1}{P^2-M_i^2}U_{S_if}\frac{1}{\slashed{P}}X^{\mu}_{fV}\frac{-\eta_{\mu\nu}}{P^2}X^{\nu}_{VS_i}\right]\right|_{\text{hard}}\;,\\
		\nonumber\\
		\nonumber\\
		\parbox{20mm}{\begin{fmffile}{OZ2Str}
	\begin{fmfgraph*}(10,10)
		\fmfset{dash_len}{2mm}
		\fmfcurved
		\fmfsurroundn{v}{2}
		\fmfv{d.shape=circle, d.filled=empty,d.size=5}{v1,v2}
		\fmf{dbl_dashes,right=1,label=$ S_i $,label.side=right,label.dist=1.5mm}{v1,v2}
		\fmf{wiggly,right=1,label=$ V $,label.side=right,label.dist=1.5mm}{v2,v1}
	\end{fmfgraph*}
\end{fmffile}} &=\quad -\frac{i}{2}\,\text{STr}\left.\left[\frac{1}{P^2-M_i^2}X^{\mu}_{S_iV}\frac{-\eta_{\mu\nu}}{P^2}X^{\nu}_{VS_i}\right]\right|_{\text{hard}}\;.
		\label{diag15}
	\end{align}
	\vspace{1cm}
	
	\noindent The total amount of \emph{heavy-light} supertrace diagrams adds up to number 15. 
	
		In Appendix~\ref{app:a} one can find the explicit comparison between the number of covariant diagrams that match to a single Supertrace diagram. There the advantage of Supertraces is more evident.

\subsection{Evaluating $\mathcal{L}_{\rm EFT}^{(\rm 1-loop)}$}	

The full 1-loop effective action is the sum of UOLEA for \emph{heavy-heavy} loops and functional supertrace diagrams for \emph{heavy-light} loops
\begin{equation}
\mathcal{L}_{\rm EFT}^{(\rm 1-loop)} \ = \ \mathcal{L}_{\rm UOLEA} \ + \ \mathcal{L}_{\rm STr} \;,
\label{eq:LEFT1L}
\end{equation}
respectively.
The UOLEA for \emph{only-heavy} particles circulating in the loop, derived in Ref.~\cite{Drozd:2015rsp} and then re-derived in Ref.~\cite{Zhang:2016pja}, reads:
				\begin{align}
					\mathcal{L}_{\text{UOLEA}} &= -ic_{s}\,\text{tr}\,\left\{f^{S_i}_{2}\,U_{S_iS_i} + f^{S_i}_{3}\,G^{\prime\mu\nu}_{S_i}G^{\prime}_{\mu\nu,S_i}\right. + f^{S_iS_j}_{4}\,U_{S_iS_j}U_{S_jS_i} + f^{S_i}_{5}\,\left(P^{\mu}G^{\prime}_{\mu\nu,S_i}\right)\left(P_{\rho}G^{\prime\rho\nu}_{S_i}\right)\nonumber\\
					&+ f^{S_i}_{6}\,G^{\prime\mu}_{\nu,S_i}G^{\prime\nu}_{\rho,S_i}G^{\prime\rho}_{\mu,S_i} + f^{S_iS_j}_{7}\,(P^{\mu}U_{S_iS_j})(P_{\mu}U_{S_jS_i}) + f^{S_iS_jS_k}_{8}\,U_{S_iS_j}U_{S_jS_k}U_{S_kS_i}\nonumber\\
					&+ f^{S_i}_{9}\,U_{S_iS_i}G^{\prime\mu\nu}_{S_i}G^{\prime}_{\mu\nu,S_i} + f^{S_iS_jS_kS_l}_{10}\,U_{S_iS_j}U_{S_jS_k}U_{S_kS_l}S_{S_lS_i} \nonumber \\
					&+ f^{S_iS_jS_k}_{11}\,U_{S_iS_j}(P_{\mu}U_{S_jS_k})(P^{\mu}U_{S_kS_i})\nonumber\\
					&+ f^{S_iS_j}_{12}\,(P^{2}U_{S_iS_j})(P^{2}U_{S_jS_i}) + f^{S_iS_j}_{13}\,U_{S_iS_j}U_{S_jS_i}G^{\prime\mu\nu}_{S_i}G^{\prime}_{\mu\nu,S_i} \nonumber \\
					&+ f^{S_iS_j}_{14}\,(P_{\mu}U_{S_iS_j})(P_{\nu}U_{S_jS_i})G^{\prime\nu\mu}_{S_i}\nonumber\\
					&+ f^{S_iS_j}_{15}\,\left(U_{S_iS_j}(P^{\mu}U_{S_jS_i}) - (P^{\mu}U_{S_iS_j})U_{S_jS_i}\right)(P^{\nu}G^{\prime}_{\nu\mu,S_i})\nonumber\\
					&+ f^{S_iS_jS_kS_lS_m}_{16}\,U_{S_iS_j}U_{S_jS_k}U_{S_kS_l}U_{S_lS_m}U_{S_mS_i}  \nonumber \\
					&+f^{S_iS_jS_kS_l}_{17}\,U_{S_iS_j}U_{S_jS_k}(P^{\mu}U_{S_kS_l})(P_{\mu}U_{S_lS_i})\nonumber\\
					&+ f^{S_iS_jS_kS_l}_{18}\,U_{S_iS_j}(P^{\mu}U_{S_jS_k})U_{S_kS_l}(P_{\mu}U_{S_lS_i})\nonumber\\
					&+ \left.f^{S_iS_jS_kS_lS_mS_n}_{19}\,U_{S_iS_j}U_{S_jS_k}U_{S_kS_l}U_{S_lS_m}U_{S_mS_n}U_{S_nS_i}\right\}\;,
					\label{uolea}
				\end{align}
where $ G^{\prime} = -[P_{\mu},P_{\nu}] = -igG_{\mu\nu} $ and $ g $ is the coupling of the corresponding field strength tensor $ G_{\mu\nu} $.  $P_\mu$ is the covariant derivative
that act to the right  in every parenthesis.
To get the correct contribution for Wilson coefficients we multiply these terms with $ (-ic_{s}) $. Since we separate each complex scalar into a two component field multiplet, each component counts as a real degree of freedom, thus the correct value is, $ c_{s} = 1/2 $.
			We should note that after the term $ f_{9} $ only the non-diagonal terms in $ U_{SS} $ contribute and exclusively the mass dimension-$ 1 $ terms of the Lagrangian. 
			In this formula  a summation over leptoquark fields $ S_{i} $ from Table~\ref{tab:allcharges} is implied. The expressions for the coefficients, $f_1,\dots, f_{19}$, can be found in ref.~\cite{Zhang:2016pja}. Appropriate limits of these expressions must be taken
			in case of degeneracies (i.e. more than one single $S$-field in the loop). Note that only $\textbf{U}$-matrices, calculated with $S_{i} = S_{i,c}[\phi]$, appear in \eqref{uolea}.

We now need to calculate  $\mathcal{L}_{\rm STr}$ in \eqref{eq:LEFT1L}. For this purpose we use the package {\tt STrEAM}~\cite{Cohen:2020qvb} in order to calculate the relevant supertraces in \eqref{diag1}-\eqref{diag15}.
  The main function of this package is the automation of the CDE application. As a result it computes local traces for further calculation inserting the explicit expressions of the $ \mathbf{X} $-matrices. We note that the option $ {\tt No\gamma inU} $ removes \emph{all} spinor indices from all matrices. However, in some instances the outer matrices contain spinor indices while the internal ones do not, or if they do, these matrices (anti)commute with $ \gamma $-matrices. Therefore some manual intervention is necessary to obtain the final result.
	
	The result of this procedure is given below. The traces are categorized depending on the number of $ U $'s and $ Z $'s involved in the respective diagram. The single term from the heavy-light UOLEA [\eqref{diag1}], derived in Refs.~\cite{Ellis:2016enq,Ellis:2017jns}, is also included here. The prefactor $ -ic_s $ is omitted, $ c_s = 1/2 $ for complex scalars, and note that the matrix $ U_{ff} $ contains only chirality projection operators $ P_{(L,R)} $, which have been taken into account while anti-commuting $ \gamma $-matrices. At the end, $\mathcal{L}_{\rm STr}$ in \eqref{eq:LEFT1L} is obtained from the equation, 
	\begin{equation}
	\mathcal{L}_{\rm STr} = (-ic_s) \times {\rm (sum ~of ~all ~contributions ~below [\textit{c.f.}~ \eqref{Str1}-\eqref{Str2}])} \;. 
	\label{eq:LStr}
	\end{equation}
	\vspace{1cm}
	\begin{itemize}
		\item[-]$\underline{\mathcal{O}\left(U^{2}\right)}$
		\begin{align}
		\left(1+\log\frac{\mu^{2}}{M^{2}_{i}}\right)\,&\text{tr}\left\{U_{S_iH}U_{HS_i}\right\}\;,\label{Str1}\\
		\frac{1}{2}\left(\frac{1}{2}+\log\frac{\mu^{2}}{M_i^2}\right)\,&\text{tr}\left\{U_{S_if}\gamma_{\mu}(P^{\mu}U_{fS_i})\right\}\;,\\
		\frac{1}{12M_i^2}\,&\text{tr}\left\{U_{S_if}\gamma^{\mu}(P^{2}P_{\mu}U_{fS_i}) + U_{S_if}\gamma^{\mu}(P_{\mu}P^{2}U_{fS_i})\right\}\;,\\
		-\frac{1}{18M_i^2}\,&\text{tr}\left\{U_{S_if}\gamma_{\nu}U_{fS_i}(P_{\mu}G^{\prime\mu\nu}_{S_i})\right\}\;,\\
		-\frac{1}{3M_i^2}\left(\frac{7}{12} + \log\frac{\mu^2}{M_i^2}\right)\,&\text{tr}\left\{U_{S_if}\gamma_{\nu}(P^{\mu}G^{\prime\,\mu\nu}_{f})U_{fS_i}\right\}\;,\\
		\frac{i}{2M_i^2}\,&\text{tr}\left\{(P^{\mu}U_{S_if})\tilde{G}^{\prime}_{\mu\nu}\gamma^{\nu}\gamma_{5}U_{fS_i} - U_{S_if}\tilde{G}^{\prime}_{\mu\nu}\gamma^{\nu}\gamma_{5}(P^{\mu}U_{fS_i})\right\}\;.
		\end{align}
		\item[-]$\underline{\mathcal{O}\left(U^{3}\right)}$
		\begin{align}
		\left(1+\log\frac{\mu^{2}}{M^{2}_{i}}\right)\,&\text{tr}\left\{U_{S_if}U_{ff^\prime}U_{f^\prime S_i}\right\}\;,\\
		\frac{2\Delta^2_{ij} + (M_i^2 + M_j^2)\log M_j^2/M_i^2}{4(\Delta^2_{ij})^2}\,&\text{tr}\left\{(P^{\mu}U_{S_iS_j})U_{S_jf}\gamma_{\mu}U_{fS_i}\right\}\;,\\
		\frac{1}{4\Delta_{ij}^{2}}\log\frac{M_{j}^{2}}{M_i^2}\,&\text{tr}\left\{U_{S_iS_j}U_{S_jf}\gamma_{\mu}(P^{\mu}U_{fS_i}) - U_{S_iS_j}(P^{\mu}U_{S_jf})\gamma_{\mu}U_{fS_i}\right\}\;,\\
		-\frac{1}{2M_i^2}\left(\frac{1}{2} + \log\frac{\mu^2}{M_i^2}\right)\,&\text{tr}\left\{U_{S_if}(P^{2}U_{ff^\prime})U_{f^\prime S_i}\right\}\;,\\
		\frac{1}{2M_i^2}\,&\text{tr}\left\{U_{S_if}U_{ff^\prime}(P^{2}U_{f^\prime S_i})\right\}\;,\\
		\frac{1}{2M_i^2}\,&\text{tr}\left\{U_{S_if}(P_{\mu}U_{ff^\prime})(P^{\mu}U_{f^\prime S_i})\right\}\;,\\
		-\frac{1}{4M_i^2}\left(\frac{3}{2} + \log\frac{\mu^2}{M_i^2}\right)\,&\text{tr}\left\{U_{S_if}U_{ff^\prime}i\sigma_{\mu\nu}G^{\prime\,\mu\nu}_{f^\prime}U_{f^\prime S_i}\right\}\;,\label{g-2a}\\
		-\frac{1}{4M_i^2}\left(\frac{1}{2} + \log\frac{\mu^2}{M_i^2}\right)\,&\text{tr}\left\{U_{S_if}i\sigma_{\mu\nu}G^{\prime\,\mu\nu}_{f}U_{ff^\prime}U_{fS_i}\right\}\;,\label{g-2b}\\
		\frac{1}{2M_i^2}\,&\text{tr}\left\{U_{S_if}i\sigma_{\mu\nu}(P^{\mu}U_{ff^\prime})(P^{\nu}U_{f^\prime S_i})\right\}\;.
		\end{align}
		\item[-]$\underline{\mathcal{O}\left(U^{4}\right)}$
		\begin{align}
		-\frac{1}{4M^2_i}\left(\frac{3}{2} + \log\frac{\mu^2}{M^2_i}\right)\,&\text{tr}\left\{U_{S_if}\gamma^{\mu}U_{fH}U_{Hf^\prime}\gamma_{\mu}U_{f^\prime S_i}\right\}\;,\\
		-\frac{\log{M^2_i/M^2_j}}{4\Delta^{2}_{ij}}\,&\text{tr}\left\{U_{S_if}\gamma^{\mu}U_{fS_j}U_{S_jf^\prime}\gamma_{\mu}U_{f^\prime S_i}\right\}\;,\\
		-\frac{\log{M^2_i/M^2_j}}{\Delta^{2}_{ij}}\,&\text{tr}\left\{U_{S_iS_j}U_{S_jf}U_{ff^\prime}U_{f^\prime S_i}\right\}\;,\label{eq:2.62}\\
		-\frac{1}{4M^2_i}\left(\frac{3}{2} + \log\frac{\mu^2}{M^2_i}\right)\,&\text{tr}\left\{U_{S_if}\gamma^{\mu}U_{fV}U_{Vf^\prime}\gamma_{\mu}U_{f^\prime S_i}\right\}\;,\\
		-\frac{1}{4M_i^2}\,&\text{tr}\left\{U_{S_if}U_{ff^\prime}U_{f^\prime f^{\prime\prime}}\gamma_{\mu}(P^{\mu}U_{f^{\prime\prime}S_i})\right.\nonumber\\
		&\left.- (P^{\mu}U_{fS_i})U_{ff^\prime}U_{f^\prime f^{\prime\prime}}\gamma_{\mu}U_{f^{\prime\prime}S_i	}\right\}\;,\\
		-\frac{1}{2M_i^2}\left(1 + \log\frac{\mu^{2}}{M_i^2}\right)\,&\text{tr}\left\{U_{fS_i}U_{ff^\prime}(P^{\mu}U_{f^\prime f^{\prime\prime}})\gamma_{\mu}U_{f^{\prime\prime}S_i}\right.\nonumber\\
		&\left.- U_{fS_i}(P^{\mu}U_{ff^\prime})U_{f^\prime f^{\prime\prime}}\gamma_{\mu}U_{f^{\prime\prime}S_i}\right\}\;,\\
		\mathcal{I}[q^2]^{1112}_{ijk0}\,&\text{tr}\left\{U_{S_iS_j}U_{S_jS_k}U_{S_kf}\gamma^{\mu}(P_{\mu}U_{fS_i})\right.\nonumber\\
		&\left.-U_{S_iS_j}U_{S_jS_k}(P^{\mu}U_{S_kf})\gamma_{\mu}U_{fS_i}\right\}\;,\\
		-\mathcal{I}[q^2]^{1211}_{ijk0}\,&\text{tr}\left\{U_{S_iS_j}(P^\mu U_{S_jS_k})U_{S_kf}\gamma_{\mu}U_{fS_i}\right.\nonumber\\
		&\left.- (P^\mu U_{S_iS_j})U_{S_jS_k}U_{S_kf}\gamma_{\mu}U_{S_if}\right\}\;.
		\end{align}
		\item[-]$\underline{\mathcal{O}\left(U^{5}\right)}$
		\begin{align}
		\frac{1}{M^2_i}\left(1 + \log\frac{\mu^2}{M^2_i}\right)\,&\text{tr}\left\{U_{S_if}U_{ff^\prime}U_{f^\prime f^{\prime\prime}}U_{f^{\prime\prime}f^{\prime\prime\prime}}U_{f^{\prime\prime\prime}S_i}\right\}\;,\\
		4\mathcal{I}[q^2]^{1112}_{ijk0}\,&\text{tr}\left\{U_{S_iS_j}U_{S_jS_k}U_{S_kf}U_{ff^\prime}U_{f^\prime S_i}\right\}\;.
		\end{align} 
		\item[-]$ \underline{\mathcal{O}\left(Z^1\right)} $
		\begin{equation}
		\frac{1}{4}\left(\frac{3}{2} + \log\frac{\mu^2}{M^2_{i}}\right)\,\text{tr}\left\{Z^{\mu}_{S_iV}U_{Vf}\gamma_{\mu}U_{fS_i} + U_{S_if}\gamma_{\mu}U_{fV}\bar{Z}^{\mu}_{VS_i}\right\}\;.
		\end{equation} 
		\item[-]$ \underline{\mathcal{O}\left(Z^2\right)} $
		\begin{equation}
		\frac{M_{i}^{2}}{4}\left(\frac{3}{2}+\log\frac{\mu^{2}}{M_{i}^2}\right)\,\text{tr}\left\{Z^{\mu}_{S_iV}\bar{Z}_{VS_i,\mu}\right\}\;.
		\label{Str2}
		\end{equation}
	\end{itemize}
In eqs.~\eqref{Str1}-\eqref{Str2} above, the trace $(\text{tr})$ stands for a normal trace over the product of matrices-$(U,Z)$ that are direct products of spinor, gauge or flavour matrices. Also, with $ \widetilde{G} $ we denote the usual dual tensor $ \widetilde{G}^{\prime}_{\mu\nu} = \frac{1}{2}\epsilon^{\mu\nu\rho\sigma}G^{\prime}_{\rho\sigma}.  $
The $\mu$-parameter denotes the renormalization scale and  the $\overline{MS}$-renormalization scheme with dimensional regularization is used throughout.
	Finally, the expressions for few integrals appearing before traces are 
	$(\Delta_{ij}^2 \equiv M_i^2-M_j^2)$,
	\begin{align}
		\mathcal{I}[q^2]^{1112}_{ijk0} &= \frac{\log M_k^2/M_i^2}{4\Delta_{ij}^2 \Delta_{ik}^2} - \frac{\log M_k^2/M_j^2}{4 \Delta_{ij}^2 \Delta_{jk}}\;,\\
		\mathcal{I}[q^2]^{1211}_{ijk0} &= \frac{1}{4 \Delta_{ij}^2}\left(\frac{3}{2} + \frac{M_i^2 \log \mu^2/M_i^2 + M_k^2 \log \mu^2/M_k^2}{\Delta_{ik}^2}\right)\nonumber\\
		&+\frac{\log \mu^2/M_j^2 (M_i^2 M_k^2 - M_j^4) + M_k^2 \log\mu^2/M_k^2 (M_i^2 - 2M_j^2 + M_k^2)}{4(\Delta_{ij}^2)^2 (\Delta_{jk}^2)^2}\nonumber\\
		&+\frac{2M_i^2 - 5M_j^2 + 3M_k^2}{8(\Delta_{ij}^2)^2 \Delta_{jk}^2}\;.
	\end{align}

	Following this  general procedure, the operators extracted from $\mathcal{L}_{\rm EFT}^{(\rm 1-loop)}$ in \eqref{eq:LEFT1L} are given in a general operator
	basis, usually referred to as Green basis, which does not involve field EOMs in reducing the number operators, but only integration-by-parts.  There is however, one more
	complication in writing down the Wilson coefficients even in Green basis. This is the appearance of the so-called \emph{evanescent operators}~\cite{Dugan:1990df,Herrlich:1994kh} that vanish 
	in $d=4$ but do not vanish in general for $d\ne 4$ in certain four-fermion interactions. The effect of evanescent operators in SM EFT~\cite{Dekens:2019ept, Aebischer:2020dsw,Gherardi:2020det}
	is taken into account in the application we present 	in  the next section.

	
	It is easy to make the connection between eqs.~\eqref{Str1}-\eqref{Str2} and the functional supertrace diagrams in eqs.~\eqref{diag1}-\eqref{diag15}.
	For example, say we want to find a diagram candidate for  neutrino-mass generation operator $\ell\ell H H$. We need four interaction vertices, \emph{i.e.} four $U$-matrices but no
	derivative ($P$) operators or $\gamma$-matrices. Looking at $\mathcal{O}(U^4)$ terms we see that only~\eqref{eq:2.62} satisfies this condition. Following the subscripts of 
	$U$-matrices, in this case $S_i - S_j - f -f' - S_i$, we trivially see the corresponding diagram is that of~\eqref{eq:2.34}. What is very nice in this approach is the
	fact that strict correlations between observables are now obvious, i.e., the operator resulting from~\eqref{eq:2.62} may be correlated with those containing the insertions
	$U_{S_iS_j}$, $U_{S_i f}$ and $U_{f f'}$.
	
	\subsection{Summary}

Our main formulae for the full 1-loop matching up-to dimension-6 order in EFT expansion for \emph{all} scalar leptoquarks are: 
\begin{enumerate}

\item Tree level matching: eq.~\eqref{eq:treeEFT}, 

\item One loop matching: eq.~\eqref{eq:LEFT1L} = eq.~\eqref{uolea} + eq.~\eqref{eq:LStr}.

\end{enumerate}	
 From now on we have to choose a specific model with heavy particles taken from Table~\ref{tab:allcharges}, plug in the explicit $\mathrm{X}(U,Z)$-matrices 
 and operators will pop-out of the traces. The general form of interaction $\mathrm{X}(U,Z)$-matrices is presented in Appendix~\ref{app:X}.


	\section{Application: The leptoquark model $ S_1 + \tilde{S}_2 $}
\label{sec:3}
		In this section, we apply the machinery of functional matching onto a particular scalar leptoquark model. Consequently,
		we consider an extension of the SM consisting of two scalar colored leptoquarks, an isospin singlet and a doublet,  $ S_1 $ and $ \tilde{S}_2 $, with masses 
		$ M_1$ and $\tilde{M}_2$, respectively. 
		Their charges under the SM gauge group are shown in Table~\ref{tab:charges}.
			\begin{table}[h]
				\centering
				\begin{tabular}[c]{| c | c  c  c |}
					\hline
					Field/Group & SU(3) & SU(2) & U(1)\\
					\hline\hline
					$ S_{1} $ & $ \phantom{\displaystyle\frac{1}{1}}\bar{3} $ & $ 1 $ &  $ \frac{1}{3} $\\
					$ \tilde{S}_{2} $ & $ \phantom{\displaystyle\frac{1}{1}} 3 $ & $ 2 $ &  $ \frac{1}{6} $\\
					\hline
				\end{tabular}
				\caption{\label{sqnums}Leptoquark charges under the SM gauge group for the $ S_1 + \tilde{S}_2 $ model.}
				\label{tab:charges}
			\end{table}

Interestingly, $S_1$ and $\tilde{S}_2$ belong to irreducible representations
of an $SU(5)$ [or $SO(10)$] Grand Unified Theory (GUT). For example,
$S_1$ may belong to $\bar{\mathbf{5}}, \overline{\mathbf{45}}, \overline{\mathbf{50}}$ and $\tilde{S}_2$ to $ \mathbf{10}, \mathbf{15}$ irreps  of minimal $SU(5)$, respectively. 
From the SM EFT operator content we derive below, we see that this model predicts fast proton decay and neutrino masses  (and related $B$- and $L$-violating phenomena). 
Then it is natural for the two fields to have heavy masses $M_i \approx 10^{12}~\mathrm{GeV}$ which at the same time 
control the proton decay rate and generate neutrino masses consistent with experimental constraints~\cite{Dorsner:2012uz, Dorsner:2017wwn,Dorsner:2012nq,Heeck:2019kgr}. 

On the other hand, one may apply a baryon parity where lepton, quark and therefore leptoquark fields, transform differently under a symmetry
 in order to protect the model from  proton decay (although such a symmetry is not in general natural in GUTs as we argue below). In this
 case, $M_1$ and $\tilde{M}_2$   may be within the few-TeV range. Again, one may be able to account 
 for radiative neutrino masses~\cite{Mahanta:1999xd,Zhang:2021dgl}, or current anomalous events such as
 the muon anomalous magentic moment~\cite{Bauer:2015knc,Crivellin:2021rbq,Athron:2021iuf,Zhang:2021dgl}  and certain 
 $B$-meson decays~\cite{Bauer:2015knc,Angelescu:2021lln}.
 
 In either cases, this $S_1 + \tilde{S}_2$-model seems to attract a certain phenomenological interest which motivates us  
 for studying its effective operators and their matching onto the SM EFT Lagrangian. However, 
 further than a functional matching demonstration, such as a detailed phenomenological consideration, are beyond the scope of this paper.

	\subsection{Lagrangian and  symmetries}

		We split the leptoquark Lagrangian into three parts,
			\begin{equation}
				\mathcal{L}_{\text{BSM}} \ = \ \mathcal{L}_{\text{S-f}} \ +\
				 \mathcal{L}_{\text{S-H}} \ + \ \mathcal{L}_{\text{S}}\;.
				\label{eq:bsmshort}
			\end{equation}
		The first part refers to leptoquark-fermion interactions, the second one to leptoquark-Higgs interactions, while the last part contains self and mixed terms between the two leptoquarks. Explicitly the first part reads~\cite{Gherardi:2020det,deBlas:2017xtg},
			\begin{align}
				\mathcal{L}_{\text{S-f}} = 
				&\left[
				\left(\lol_{pr}\right) \bar{q}^{c}_{pi}\cdot\epsilon\cdot\ell_{r} + \left(\lori_{pr}\right) \bar{u}_{i}^{c}\,e_{r} 
				\right]S_{1i} +
				\text{h.c.}\nonumber\\[2mm]
				&+ (\lbl_{pr})\,S_{1i}\,\epsilon^{ijk}\,\bar{q}_{pj}\cdot\epsilon\cdot q^{c}_{rk} +(\lbr_{pr})\,S_{1i}\,\epsilon^{ijk}\bar{d}_{pj}\,u^{c}_{rk} + \text{h.c.}\nonumber\\[2mm]
				&+ (\ltil_{pr})\,\bar{d}_{pi}\tilde{S}^{T}_{2i}\cdot\epsilon\cdot\ell_{r} + \text{h.c.}\;, 
				\label{eq:sferm}
			\end{align} 
		where $ \ell $ and $ q $ are the lepton and quark field $SU(2)_L$-doublets while the singlets are denoted by $ u,d,e $ in gauge basis\footnote{Field redefinitions and flavour rotations to mass-basis 
		are performed  following ref.~\cite{Dedes:2017zog} after running the SM EFT parameters down to the EW scale.} and $\epsilon$ is the antisymmetric tensor with $SU(2)_L$ indices. The matrix $ \lbl_{pr} $ is complex symmetric in flavor space while all other matrices in \eqref{eq:sferm} are in general complex ones. 
		From now on, we use the indices $ p,r,s,t $ to denote \textit{flavor} without making any distinction between quark and lepton flavors. We also use $ i,j,k,l $ to label $SU(3)$ \textit{fundamental} indices, while the dot-product denotes $SU(2)$ contractions in the fundamental representation. Later we will also use the letters $ \alpha,\beta,\gamma,\delta $ for $SU(2)$ \textit{fundamental} and $ I,J,K,L $ for $SU(2)$ \textit{adjoint} representation. Lastly, we use $ A,B,C,D $ for the $SU(3)$ \textit{adjoint} representation and suppress spinor-indices throughout. 
		
		The next part of the Lagrangian, namely leptoquark-Higgs interactions, reads~\cite{Crivellin:2020mjs},
			\begin{align}
				\mathcal{L}_{\text{S-H}} = &-\left(M_{1}^{2} + \lambda_{H1}|H|^{2}\right)|S_{1}|^{2} - \left(\tilde{M}_{2}^{2} + \tilde{\lambda}_{H2}|H|^{2}\right)|\tilde{S}_{2}|^{2} +
				 \lambda_{\tilde{2}\tilde{2}}\,(\tilde{S}^{\dagger}_{2i}\cdot H)\,(H^{\dagger}\cdot\tilde{S}_{2i})\nonumber\\
				&-A_{\tilde{2}1}\left(\tilde{S}_{2i}^{\dagger}\cdot H\right)S^{\dagger}_{1i} +\frac{1}{3}\lambda_{3}\epsilon^{ijk}\left(\tilde{S}_{2i}^{T}\cdot\epsilon\cdot\tilde{S}_{2j}\right)\left(H^\dagger\cdot\tilde{S}_{2k}\right) + \text{h.c.}
				\label{eq:shiggs}
			\end{align}
			
					The last part containing leptoquark self-interactions is,
			\begin{align}
				-\mathcal{L}_{\text{S}} &= \frac{c_{1}}{2}\,(S_{1}^{\dagger}S_{1})^{2} + \frac{\tilde{c}_{2}}{2}\,(\tilde{S}^{\dagger}_{2}\cdot\tilde{S}_{2})^{2} + c_{1\tilde{2}}^{(1)}\,(S_{1}^{\dagger}S_{1})(\tilde{S}_{2}^{\dagger}\cdot\tilde{S}_{2}) + c^{(2)}_{1\tilde{2}} (\tilde{S}_{2\alpha}^{\dagger}S_{1})(S_{1}^{\dagger}\tilde{S}_{2\alpha})\nonumber\\
				&+ c_{\tilde{2}}^{(8)} (\tilde{S}^{\dagger}_{2i}\cdot\tilde{S}_{2j})\,(\tilde{S}^{\dagger}_{2j}\cdot\tilde{S}_{2i}) + \left[A^\prime S_{1i}^{\dagger}\,\epsilon^{ijk}\left(\tilde{S}_{2j}^{T}\cdot\epsilon\cdot \tilde{S}_{2k}\right) + \text{h.c.}\right]\;.
				\label{eq:sonly}				
			\end{align}
		Our convention for the covariant derivative is,
			\begin{equation}
				D_{\mu} = \partial_{\mu} - ig^{\prime}YB_{\mu} - igT^{I}W_{\mu}^{I} - ig_{s}T^{A}G^{A}_{\mu}\;,
				\label{eq:covder}
			\end{equation}
		where each $ T $ represents the gauge-group generators of the corresponding representation of a generic field and $Y$ is its hypercharge. The field strength tensors for $U(1)_Y$, $SU(2)_L$ and
		$SU(3)_c$ gauge-fields are respectively, 
			\begin{align}
				B_{\mu\nu} &= \partial_{\mu}B_{\nu} - \partial_{\nu}B_{\mu}\;,\\
				W^{I}_{\mu\nu} &= \partial_{\mu}W_{\nu} - \partial_{\nu}W_{\mu} + g\epsilon^{IJK}W_{\mu}^{J}W_{\nu}^{K}\;,\\
				G^{A}_{\mu\nu} &= \partial_{\mu}G_{\nu} - \partial_{\nu}G_{\mu} + g_{s}f^{ABC}G_{\mu}^{B}G_{\nu}^{C}\;.
				\label{eq:fstrenght}
			\end{align}
		Finally,   the SM Yukawa couplings are defined as,
			\begin{align}
				-\mathcal{L}_{\text{Y}} &= (y_{E})_{pr}\bar{\ell}_{p}\cdot H\,e_{r} + (y_{U})_{pr}\,\bar{q}_{pi}\cdot\epsilon\cdot H^{\ast}\,u_{ri} + (y_{D})_{pr}\,\bar{q}_{pi}\cdot H\,d_{ri} + \text{h.c.}\;,
				\label{eq:yuksm}
			\end{align}
		while the Higgs potential  is,
			\begin{align}
				V_{H} = -m^{2}(H^{\dagger}\cdot H) + \frac{\lambda}{2} (H^{\dagger}H)^{2}\;.
				\label{eq:higgspot}
			\end{align}

	It is always instructive in a given Lagrangian to check upon global symmetries such as Baryon ($B$) and Lepton ($L$) number,  
	which may be broken by certain interaction parameters. For the $S_1+\tilde{S}_2$ model these are given in Table~\ref{tab:BL}.	
\begin{table}[th!]
\centering
\begin{tabular}{|c|c|c|c|}
\hline 
LQ-fields & $B$ & $L$ & $B-L$\\[2mm] 
\hline
$S_1$ & -1/3 & -1 & 2/3\\[2mm] 
\hline 
$\tilde{S}_2$ & +1/3 & -1 &4/3 \\[2mm]
\hline 
Parameters & & & \\[2mm]
\hline
$ \lbl $ & +1 & +1 & 0\\[2mm]
\hline
$ \lbr $  & +1 & +1 & 0\\[2mm]
\hline
$A_{\tilde{2}1}$ & 0 & -2 & 2 \\[2mm]
\hline
$A'$ & -1 & 1 & -2 \\[2mm]
\hline
$\lambda_3$ & -1 & 3 & -4 \\[2mm]
\hline
\end{tabular} 
\caption{$B$, $L$ and $B-L$ quantum numbers for the fields $S_1$ and $\tilde{S}_2$  and parameters (promoted to fields). 
For normalization we take $B(q)=1/3$ and $L(\ell)=1$. All other parameters in eqs.~\eqref{eq:sferm},\eqref{eq:shiggs} and \eqref{eq:sonly}  which are not quoted here,
 have zero $B$ and $L$ quantum numbers.}
\label{tab:BL}
\end{table}

Obviously, by assuming baryon and/or lepton 
number conservation we can eliminate all terms proportional to couplings $(\lbl, \lbr, A', \lambda_3 )$ and/or $A_{\tilde{2}1}$, respectively.
 However, baryon and lepton symmetries cannot be well-defined gauge 
 symmetries of a  GUT model since they lead to chiral anomalies. 
On the other hand however,  in SO(10)-GUTs for example (and also in the SM), $B-L$  is an anomaly free
 gauge  symmetry and therefore we search for linear combinations of $B-L$-quantum numbers~\cite{Arnold:2013cva}. In this case,
 a $\mathbb{Z}_2$-symmetry $\exp(2\pi i (B-L)/2)$, does not exclude any of couplings above, 
  a $\mathbb{Z}_3$-symmetry $\exp(2\pi i (B-L)/3)$, excludes  $A_{\tilde{2}1},A',\lambda_3$,  a $\mathbb{Z}_4$-symmetry $\exp(2\pi i (B-L)/4)$,
  excludes  $A_{\tilde{2}1},A',$
  which is not desirable if we want to generate neutrino masses by loop-corrections.
 Therefore, there is no $B-L$ discrete symmetry that rejects
 terms proportional to $\lbl$ (and $\lbr$) which 
 lead to proton decay. We conclude that in general, an extra (and possibly ad-hoc) symmetry should be in order
 if we are about to pick a combination of leptoquark fields with masses nearby the TeV scale. For a recent discussion 
 the reader is referred to ref.~\cite{Murgui:2021bdy}.


		\subsection{Tree Level Matching}
							As is the case in EFTs, we assume that both leptoquarks masses, $ M_i = \{M_{1},\,\tilde{M}_{2}\} $, are heavier than any other scale in the theory. 
			Moreover, the parameters
			$A_{\tilde{2}1}$, and $A'$ should lay in region \eqref{eq:Aregion}. 			
			To match at \textit{tree level} we need the equations of motion \eqref{eq:eoms}, in order to derive the classical fields,
				\begin{align}
					S_{1i,c} &=\frac{1}{M_1^2}\left[ -(\lol_{pr})^{\dagger}\bar{\ell}_{p}\cdot\epsilon\cdot q^{c}_{ri} + (\lori_{pr})^{\dagger}\bar{e}_{p}u^{c}_{ri} + (\lbl_{pr})^{\dagger}\epsilon^{ijk} \bar{q}^{c}_{pj}\cdot\epsilon\cdot q_{ri} - (\lbr_{pr})^{\dagger}\epsilon^{ijk}\bar{u}^{c}_{pj}d_{rk}\right]\;,\\
					\tilde{S}_{2i\alpha ,c} &=\frac{1}{\tilde{M}_{2}^{2}} \left[ -(\ltil_{pr})^{\dagger}\left(\bar{\ell}_{p}\cdot\epsilon\right)_{\alpha}d_{ri}\right]\;.
					\label{Seoms}
				\end{align}
			Substituting back into the Lagrangian [see \eqref{eq:treeEFT}]  we obtain the following Wilson coefficients that accompany $d=6$-operators.
			 In what follows, we split the Wilson coefficients to \emph{tree} and \emph{loop} contributions as $ G = G^{(0)} + \frac{1}{(4\pi)^2}G^{(1)} $. The symbol $G$ denotes Wilson coefficients in Green basis, 
			and we use exactly the same naming for operators as in  ref.~\cite{Gherardi:2020det} that we append in Appendix~\ref{app:GreenOps} for complementarity purposes.

			 In summary, we find the following twelve $B$-number conserving tree-level coefficients
				\begin{align}
					&\left[G_{\ell q}^{(1)}\right]^{(0)}_{prst} = \frac{(\lol_{sp})^{\ast}(\lol_{tr})}{4M_{1}^{2}}\;, &&\left[G_{\ell q}^{(3)}\right]^{(0)}_{prst} = -\frac{(\lol_{sp})^{\ast}(\lol_{tr})}{4M_{1}^{2}}\;,\\
					&\left[G_{lequ}^{(1)}\right]^{(0)}_{prst} = \frac{(\lol_{sp})^{\ast}(\lori_{tr})}{2M_{1}^{2}}\;,
					&&\left[G_{\ell equ}^{(3)}\right]^{(0)}_{prst} = -\frac{(\lol_{sp})^{\ast}(\lori_{tr})}{8M_{1}^{2}}\;,
					\label{eq:3.14}\\
					&\left[G_{eu}\right]^{(0)}_{prst} = \frac{(\lori_{sp})^{\ast}(\lori_{tr})}{2M_{1}^{2}}\;,
					&&\left[G_{\ell d}\right]^{(0)}_{prst} = -\frac{(\ltil_{tp})^{\ast}(\ltil_{sr})}{2\tilde{M}_{2}^{2}}\;,\label{eq:Gldtree}\\
					&\left[G_{qq}^{(1)}\right]^{(0)}_{prst} = \frac{(\lbl_{rt})^{\ast}(\lbl_{sp})}{2M_{1}^{2}}\label{eq:Gqqtree}\;,
					&&\left[G_{qq}^{(3)}\right]^{(0)}_{prst} =- \frac{(\lbl_{rt})^{\ast}(\lbl_{sp})}{2M_{1}^{2}}\;,\\
					&\left[G_{ud}^{(1)}\right]^{(0)}_{prst} = \frac{(\lbr_{tr})^{\ast}(\lbr_{sp})}{3M_{1}^{2}}\;,
					&&\left[G_{ud}^{(8)}\right]^{(0)}_{prst} = -\frac{(\lbr_{tr})^{\ast}(\lbr_{sp})}{M_{1}^{2}}\;,\\
					&\left[G_{quqd}^{(1)}\right]^{(0)}_{prst} = \frac{4}{3}\frac{(\lbr_{ts})^{\ast}(\lbl_{pr})}{M_{1}^{2}}\;,
					&&\left[G_{quqd}^{(8)}\right]^{(0)}_{prst} = -4\frac{(\lbr_{ts})^{\ast}(\lbl_{pr})}{M_{1}^{2}}\;,
					\label{eq:dim6BNC}
				\end{align}
			and four $B$-number violating ones
				\begin{align}
					&\left[G_{qqq}\right]^{(0)}_{prst} = -2\frac{(\lbl_{pr})^{\ast}(\lol_{st})}{M_{1}^{2}}\;,
					&&\left[G_{qqu}\right]^{(0)}_{prst} = \frac{(\lbl_{pr})^{\ast}(\lori_{st})}{M_{1}^{2}}\;,\label{eq:dim6BNV1}\\
					&\left[G_{duq}\right]^{(0)}_{prst} = \frac{(\lbr_{pr})^{\ast}(\lol_{st})}{M_{1}^{2}}\;,
					&&\left[G_{duu}\right]^{(0)}_{prst} = \frac{(\lbr_{pr})^{\ast}(\lori_{st})}{M_{1}^{2}}\;.
					\label{eq:dim6BNV2}
				\end{align}
As we see all tree-level dimension-six operators are four-fermion operators in the effective Lagrangian. 
Besides $M_i$, their strength is governed by products of leptoquark Yukawa couplings.   			
			
			Finally for completeness, we present the five tree-level $d=7$ Wilson coefficients in the basis of ref.~\cite{Lehman:2014jma}. They are,
				\begin{align}
				\label{dim7s}
					&\left[G^{(1)}_{LLQ\bar{d}H}\right]^{(0)}_{prst} =
					-\frac{A_{\tilde{2}1}}{M_{1}^{2}\tilde{M}_{2}^{2}}(\lol_{st})(\ltil_{pr})\;,
					&&\left[G^{(2)}_{LLQ\bar{d}H}\right]^{(0)}_{prst} =
					\frac{A_{\tilde{2}1}}{M_{1}^{2}\tilde{M}_{2}^{2}}(\lol_{st})(\ltil_{pr})\;,\\
					&\left[G_{Leu\bar{d}H}\right]^{(0)}_{prst} =
					\frac{A_{\tilde{2}1}}{2M_{1}^{2}\tilde{M}_{2}^{2}}(\lori_{tr})(\ltil_{sp})\;,
					&&\left[G_{LuddH}\right]^{(0)}_{prst} =
					\frac{A^{\ast}_{\tilde{2}1}}{M_{1}^{2}\tilde{M}_{2}^{2}}(\lbr_{st})^{\dagger}(\ltil_{pr})^{\dagger}\;,\\
					&\left[G_{\bar{L}QQdH}\right]^{(0)}_{prst} =
					-2\frac{A^{\ast}_{\tilde{2}1}}{M_{1}^{2}\tilde{M}_{2}^{2}}(\lbl_{st})^{\ast}(\ltil_{pr})^{\dagger}\;.
					\label{dim7e}
				\end{align}
				As noted in the paragraph below eq.~\eqref{eq:treeEFT},  coefficients associated with  
				$d=7$ operators in eqs.~\eqref{dim7s}-\eqref{dim7s}
			 can be competitive to $d=6$ ones in \eqref{eq:dim6BNV1}-\eqref{eq:dim6BNV2} if there is a certain hierarchy 
				between the two scales involved, e.g., $M_1 \gg \tilde{M}_2 $ and $A_{\tilde{2}1} v/\tilde{M}_2^2  \simeq 1$.
				
				The appearance of products only  $\lambda\cdot \lambda^\prime$ with $|\Delta(B-L)|=0$ and  $A_{\tilde{2}1} \lambda \cdot \lambda^\prime$ with $|\Delta(B-L)|=2$
				 in coefficients for $d=6$ and $d=7$ tree-level EFT operators respectively,
				is not accidental. It follows from the $B,L$-numbers for the parameters quoted in Table~\ref{tab:BL},  
				and an interesting connection~\cite{Kobach:2016ami,Helset:2019eyc} between $\Delta B$ and $\Delta L$  with the minimum and possible dimensionality of operators 
				\begin{eqnarray}
				d_{\rm min} \ & \ge & \ \frac{9}{2} |\Delta B| + \frac{3}{2} |\Delta L| \;, \label{eq:dmin} \\[2mm]
				|\Delta (B-L)| \ &=& \ 0,4,8,12,16, \dots \qquad (d-{\rm even}) \;, \label{eq:deven} \\[2mm]
				|\Delta (B-L)| \ &=& \ 2,6,10,14,18,\dots \qquad (d-{\rm odd}) \;. \label{eq:dodd} 	 
				\end{eqnarray}
              For example, a coefficient proportional to $A_{\tilde{2}1} \times (|\Delta(B-L)|=0~{\rm couplings}) $ must necessarily be associated 
              to odd-dimensional $d=3,5,7,\dots$ case which is confirmed here at tree and below at one-loop level. 
              Similarly, the dimension-full parameter $A' \times (|\Delta(B-L)|=0~{\rm couplings})$ 
              will be associated for the first time with $d=7$-operators in EFT  at 1-loop and at $d=9$ at tree level; the dimensionless parameter $\lambda_3\times (|\Delta(B-L)|=0~{\rm couplings})$  
              will appear first at $d=10$ and so on.
				

			\subsection{One Loop Matching in the Green basis}			
				As we explained in section~\ref{sec:1loop},	one loop matching is carried out in two steps. First, the original heavy-only UOLEA in  eq.~\eqref{uolea} is used to derive operators with heavy leptoquark
fields (for this model $S_1$ and $\tilde{S}_2$)  circulating in the loop. 
Second, we use the general results from evaluating functional Supertraces in \eqref{eq:LStr} to calculate Wilson coefficients involving both heavy and light fields in the loop.
		
		 We list all one-loop Wilson coefficients produced both from the UOLEA and the Supertraces. As before, we split them in tree and loop level coefficients as $ G_i = G_i^{(0)} + \frac{1}{(4\pi)^2}G_{i}^{(1)} $. Furthermore, for the quantities that renormalize $d=4$ operators we write, $ \lambda^{\prime} = \lambda + \frac{1}{(4\pi)^2}\delta\lambda $, $ m^{\prime 2} = m^2 + \frac{1}{(4\pi)^2}\delta m^2 $, $ y^{\prime}_{n} = y_{n} + \frac{1}{(4\pi)^2}\delta y_n  $, where $ n=E,U,D $ and the wave function renormalization is $ Z_k = 1 + \frac{1}{(4\pi)^2}\delta Z_k $ with $k=q,u,d,\ell,e$. For gauge bosons, we factor out of the resulting trace calculation the whole canonical kinetic term $ -1/4 F_{\mu\nu}F^{\mu\nu} $, with $ F_{\mu\nu} $ being a generic field strength tensor.	Following an analogous naming scheme as in ref.~\cite{Gherardi:2020det}  we define,
		\begin{align}
			&L_i = \log\frac{\mu^{2}}{M_i^2}\;,\qquad \text{with}\quad i=1,2\;,
		\end{align}
and the general $3\times 3$ matrices,
		\begin{align}
			&\Lambda_{\ell} = (\lol)^{\dagger} \lol\;,
			&&\Lambda_{q} = (\lol)^{\ast} (\lol)^{T}\;,
			&&\Lambda_{u} = (\lori)^\ast(\lori)^{T}\;,\label{eq:Lambdaell}\\
			&\Lambda_{e}=(\lori)^\dagger\lori\;,
			&&\Ltil = \ltil^{\dagger}\ltil\;,
			&&\Ltild = \ltil\ltil^{\dagger}\;,\label{eq:Lambdae}\\
			&\LsBq = \lbl(\lbl)^{\dagger}\;,
			&&\LsBu = (\lbr)^{T}(\lbr)^{\ast}\;,
			&&\LsBd = \lbr(\lbr)^{\dagger}\;,
		\end{align}
%
		\begin{align}
			&\XLU = (\lol)^{\dagger}y_U^\ast\lori\;,
			&&\XLE=\left(\lori y_E^\dagger(\lol)^\dagger\right)^{T}\;,\label{eq:Y1U1L}\\
			&\XLUU = (\lol)^\dagger y_U^\ast y_U^T \lol\;,
			&&\XLDD = (\lol)^\dagger y_D^\ast y_D^T \lol\;,\\
			&\XRUU = (\lori)^{\dagger} y_U^T y_U^\ast \lori\;,
			&&\XREE = (\lori)^{\ast} y_E^T y_E^\ast (\lori)^{T}\;,\\
			&\XLEEE = \left(\lori y_E^\dagger y_E y_E^\dagger(\lol)^\dagger\right)^{T}\,
			&&\XLUUU = (\lol)^{\dagger}y_U^\ast y_U^T y_U^\ast\lori\;,\\
			&\XtilEE = \tilde{\lambda}y_E y_E^\dagger\tilde{\lambda}^{\dagger}\;,
			&&\XtilDD = \tilde{\lambda}^{\dagger}y_D^\dagger y_D\tilde{\lambda}\;,\\
			&\XBLD = \lbl y^{\ast}_{D}(\lbr)^{\ast}\;, 
			&&\XBLU = \lbl y^{\ast}_{U}(\lbr)^{\dagger}\;,\\
			&\XsBRUU = \left((\lbr)^{\dagger}y_U^{\dagger}y_U\lbr\right)^{T} \;,
			&&\XsBRDD = \left((\lbr)^{\dagger}y_D^{\dagger}y_D\lbr\right)^{T}\;,\\
			&\XsBLUU = \left((\lbl)^{\dagger} y_Uy_U^\dagger\lbl\right)^{T}\;,
			&&\XsBLDD = \left((\lbl)^{\dagger} y_Dy_D^\dagger\lbl\right)^{T}\;,\\
			&\XsBLDDD = \left((\lbr)^\dagger y_D^\dagger y_D y_D^\dagger\lbl\right)^{T}\;,
			&&\XsBLUUU = (\lbr)^\ast y_U^\dagger y_U y_U^\dagger \lbl\;.
		\end{align}
Armed with those definitions we can now write the, lengthy but complete, one-loop Wilson coefficients  associated to $d=4$ (renormalizable operators) 
 all the way up-to $d=6$ operators in Green basis. All $d=5~{\rm and}~6$ operators and the categories they belong to are given in Appendix~\ref{app:GreenOps}.
The hypercharges of $S_1$ and $\tilde{S}_2$ leptoquark fields are denoted as $Y_{S_1}$ and $Y_{\tilde{S}_2}$ respectively, and can be read from Table~\ref{tab:allcharges}.
$N_c$ is the number of colours and $ C_{F}^{G} $ is the quadratic Casimir of fundamental representation of group $ G $. Finally, we define the squared mass difference quantity 
$\Delta_{12}^2 \equiv M_1^2 - \tilde{M}_2^2$.

		\subsubsection{Renormalizable Operators}
		\begin{align}
			\left(\delta Z_{B}\right) &= \frac{N_{c}}{3}g^{\prime 2}\left[Y^2_{S_1}L_1 + 2Y^2_{\tilde{S}_2}L_2\right]\;,\\
			\left(\delta Z_{W}\right) &= \frac{N_c}{6}g^{2}L_2\;,\\
			\left(\delta Z_{G}\right) &= \frac{N_c}{6}g_s^2\left[L_1 + 2L_2\right]\;,\\
			\left(\delta Z_{\ell}\right)_{pr} &= \frac{N_c}{2}\left[\left(\frac{1}{2} + L_1\right)(\Lambda_{\ell})_{pr} + \left(\frac{1}{2} + L_2\right)(\tilde{\Lambda}_{\ell})_{pr}\right]\;,\label{eq:3.43}\\
			\left(\delta Z_{e}\right)_{pr} &=\frac{N_c}{2}\left(\frac{1}{2}+L_1\right)(\Lambda_{e})_{pr}\;,\\
			\left(\delta Z_{q}\right)_{pr} &= \frac{1}{2}\left(\frac{1}{2}+L_1\right)(\Lambda_{q} -8\LsBq)_{pr}\;,\label{eq:3.45}\\
			\left(\delta Z_{u}\right)_{pr} &=\frac{1}{2}\left(\frac{1}{2}+L_1\right)(\Lambda_{u}+2\LsBu)_{pr}\;,\\
			\left(\delta Z_{d}\right)_{pr} &=\left(\frac{1}{2}+L_2\right)(\Ltild)_{pr}+\left(\frac{1}{2}+L_1\right)(\LsBd)_{pr}\;,\\
			\left(\delta Z_{H}\right) &= N_c\,\left|A_{\tilde{2}1}\right|^{2} \left[\frac{M_1^2+\tilde{M}_2^2}{2(\Delta^2_{12})^2}+\frac{M_1^2\tilde{M}_2^2\log M_1^2/\tilde{M}_2^2}{(\Delta^2_{12})^3}\right]\;,
			\end{align}
			\begin{align}
			\left(\delta y_{E}\right)_{pr} &= -N_c(1+L_1)\,(\XLU)_{pr}\;,\label{eq:Ymu} \\
			\left(\delta y_{U}\right)_{pr} &= -N_c(1+L_1)\,(\XLE)_{pr} - 4(\XBLD)_{pr}(1+L_1)\;,\\
			\left(\delta y_{D}\right)_{pr} &= -4(\XBLU)_{pr}(1+L_1)\;,
		\end{align}
		\begin{align}
			(\delta \lambda) &= N_c\left[\lambda^2_{H1}L_1 + \left(\tilde{\lambda}^2_{H2} +  (\tilde{\lambda}_{H2} - \lambda_{\tilde{2}\tilde{2}})^2\right)L_2 -|A_{\tilde{2}1}|^2\lambda_{H1}\frac{\Delta^2_{12}+\tilde{M}^2_2\log\tilde{M}^2_2/M^2_1}{(\Delta^2_{12})^2}\right.\nonumber\\
			&\left.+|A_{\tilde{2}1}|^2(\tilde{\lambda}_{H2} - \lambda_{\tilde{2}\tilde{2}})\frac{\Delta^2_{12}+M^2_1\log\tilde{M}^2_2/M^2_1}{(\Delta^2_{12})^2}-\frac{1}{2}|A_{\tilde{2}1}|^4\frac{2\Delta_{12}^2+(M_1^2+\tilde{M}_2^2)\log \tilde{M}_2^2/M_1^2}{(\Delta^2_{12})^3}\right]\;,\\
			\left(\delta m^2\right) &= N_c\left[\lambda_{H1}M_1^2(1+L_1)+(2\tilde{\lambda}_{H2}-\lambda_{\tilde{2}\tilde{2}})\tilde{M}_2^2(1+L_2)\right.\nonumber\\
			&\left.+|A_{\tilde{2}1}|^{2}\left(1+\frac{M_1^2\log\mu^2/M_1^2-\tilde{M}_2^2\log\mu^2/\tilde{M}_2^2}{\Delta^2_{12}}\right)\right]\;.\label{eq:dm2}
		\end{align}
		
		\subsubsection{Dimension-5 Operator}

		\begin{align}
		\left[G_{\nu\nu}\right]^{(1)}_{pr} = N_c\,A_{\tilde{2}1}\left((\lol)^Ty_D\ltil\right)_{pr}\,\frac{\log M_1^2/\tilde{M}_2^2}{M_1^2 - \tilde{M}_2^2}\;.
		\label{eq:WeinbergOp}
		\end{align}
		
		\subsubsection{Vector Bosons-Scalar Operators}

		$ X^{3} $
		\begin{align}
			G_{3G}^{(1)} &= \frac{g_s^3}{360}\left(\frac{1}{M_1^2}+\frac{2}{\tilde{M}_2^2}\right)\;,\\
			G_{3W}^{(1)} &= \frac{g^3 N_c}{360}\,\frac{1}{\tilde{M}_2^2}\;.
		\end{align}
		$ X^2D^2 $
		\begin{align}
			G_{2B}^{(1)} &=\frac{N_c}{30}g^{\prime 2}\left(\frac{Y^2_{S_1}}{M_1^2}+2\frac{Y^2_{\tilde{S}_2}}{\tilde{M}_2^2}\right)\;,\\
			G_{2W}^{(1)} &=\frac{N_c}{60}g^2\frac{1}{\tilde{M}_2^2}\;,\\
			G_{2G}^{(1)} &=\frac{1}{60}g_s^2\left(\frac{1}{M_1^2}+\frac{2}{\tilde{M}_2^2}\right)\;.
		\end{align}
		$ X^2H^2 $
		\begin{align}
			G^{(1)}_{HB} &=N_cg^{\prime 2}\left[Y_{S_1}^2\left(\frac{\lambda_{H1}}{12M_1^2}-|A_{\tilde{2}1}|^2\tilde{f}^{S_1\tilde{S}_2}_{13}\right) + Y^2_{\tilde{S}_2}\left(\frac{(2\tilde{\lambda}_{H2}-\lambda_{\tilde{2}\tilde{2}})}{12\tilde{M}_2^2}-|A_{\tilde{2}1}|^2\tilde{f}^{\tilde{S}_2S_1}_{13}\right)\right]\;,\\
			G^{(1)}_{HW} &=N_c\frac{g^2}{4}\left(\frac{(\tilde{\lambda}_{H2} - \lambda_{\tilde{2}\tilde{2}})}{6\tilde{M}_2^2}-|A_{\tilde{2}1}|^2\tilde{f}^{\tilde{S}_2S_1}_{13}\right)\;,\\
			G^{(1)}_{HG} &=\frac{g_s^2}{2}\left[\frac{\lambda_{H1}}{12M_1^2}+\frac{(2\tilde{\lambda}_{H2} - \lambda_{\tilde{2}\tilde{2}})}{12\tilde{M}_2^2}-|A_{\tilde{2}1}|^2\left(\tilde{f}_{13}^{S_1\tilde{S}_2}+\tilde{f}_{13}^{\tilde{S}_2S_1}\right)\right]\;,\\
			G^{(1)}_{HWB} &=-N_c\,gg^{\prime}\,\left[|A_{\tilde{2}1}|^{2}\tilde{f}_{13}^{\tilde{S}_2S_1} + Y_{\tilde{S}_2}\frac{\lambda_{\tilde{2}\tilde{2}}}{12\tilde{M}_2^2}\right]\;. 
		\end{align}
		\\
		$ H^2X^2D^2 $
		\begin{align}
			G^{(1)}_{BDH} &= N_c|A_{\tilde{2}1}|^{2}g^{\prime}\left(Y_{S_1}+Y_{\tilde{S}_2}\right)\tilde{f}_{14}^{S_1\tilde{S}_2}\;,\\
			G^{(1)}_{WDH} &= N_{c}|A_{\tilde{2}1}|^{2}\frac{g}{2}\tilde{f}_{14}^{\tilde{S}_2S_1}\;.
		\end{align}
		$ H^2D^4 $
		\begin{align}
			G_{DH}^{(1)} &=2N_c|A_{\tilde{2}1}|^{2}\tilde{f}_{12}^{S_1\tilde{S}_2}\;.
		\end{align}
		$ H^4D^2 $
		\begin{align}
			G_{H\square}^{(1)} &=-\frac{N_c}{12}\left(\frac{\lambda_{H1}^2}{M_1^2} + \frac{\tilde{\lambda}^2_{H2}}{\tilde{M}_2^2}\right)\;,\\
			G_{HD}^{(1)} &=-N_c\frac{\lambda_{\tilde{2}\tilde{2}}}{6\tilde{M}_2^2} +N_c\lambda_{\tilde{2}\tilde{2}}|A_{\tilde{2}1}|^2\,(\tilde{f}_{11}^{\tilde{S}_2\tilde{S}_2S_1} - 2\tilde{f}_{11}^{S_1\tilde{S}_2\tilde{S}_2})  -N_c|A_{\tilde{2}1}|^{4}\left(\tilde{f}_{17}^{\tilde{S}_2S_1}-2\tilde{f}^{S_1\tilde{S}_2}_{18}\right)\;,\\
			G_{HD}^{\prime (1)} &=-N_c|A_{\tilde{2}1}|^{2}\left(\lambda_{H1}\tilde{f}_{11}^{S_1S_1\tilde{S}_2} + \tilde{\lambda}_{H2}\tilde{f}_{11}^{\tilde{S}_2\tilde{S}_2S_1} - 2\lambda_{\tilde{2}\tilde{2}}\tilde{f}_{11}^{S_1\tilde{S}_2\tilde{S}_2}\right) - N_c|A_{\tilde{2}1}|^{4}\tilde{f}_{17}^{S_1\tilde{S}_2}\;,\\
			G_{HD}^{\prime\prime (1)} &=-iN_c|A_{\tilde{2}1}|^{2}\left(\lambda_{H1}\tilde{f}_{11}^{S_1\tilde{S}_2S_1} + \tilde{\lambda}_{H2}\tilde{f}_{11}^{\tilde{S}_2S_1\tilde{S}_2} -\lambda_{\tilde{2}\tilde{2}}\tilde{f}_{11}^{S_1\tilde{S}_2\tilde{S}_2}\right)-iN_c|A_{\tilde{2}1}|^4\tilde{f}_{18}^{S_1\tilde{S}_2}\nonumber\\
			&-\frac{iN_c}{12\tilde{M}_2^2}\left(2\tilde{\lambda}_{H2}\lambda_{\tilde{2}\tilde{2}} - \lambda_{\tilde{2}\tilde{2}}^{2}\right)\;.
		\end{align}
		$ H^6 $
		\begin{align}
			G_{H}^{(1)} &=-\frac{N_c}{6}\,\left(\frac{\lambda_{H1}^3}{M_1^2} + 2\frac{\tilde{\lambda}_{H2}^3}{\tilde{M}_2^2}\right)+ 2N_c|A_{\tilde{2}1}|^6\,\tilde{f}^{S_1\tilde{S}_2}_{19}\;.
		\end{align}
		
		\subsubsection{Two Fermion Operators}

		$ \psi^2 D^3 $
		\begin{align}
			\left[G_{\ell D}\right]_{pr}^{(1)} &=-\frac{N_c}{6}\left[\frac{(\Lambda_{\ell})_{pr}}{M_1^2} + \frac{(\tilde{\Lambda}_{\ell})_{pr}}{\tilde{M}_2^2}\right]\;,\\
			\left[G_{eD}\right]_{pr}^{(1)} &=-\frac{N_c}{6}\,\frac{(\Lambda_{e})_{pr}}{M_1^2}\;,\\
			\left[G_{qD}\right]_{pr}^{(1)} &=-\frac{1}{6M_1^2}\left[(\Lambda_{q})_{pr} - 8(\LsBq)_{pr}\right]\;,\\
			\left[G_{uD}\right]_{pr}^{(1)} &=-\frac{1}{6M_1^2}\left[2(\LsBu)_{pr} + (\Lambda_{u})_{pr}\right]\;,\\
			\left[G_{dD}\right]_{pr}^{(1)} &=\frac{(\tilde{\Lambda}_{d})_{pr}}{3\tilde{M}_2^2}-\frac{(\LsBd)_{pr}}{3M_1^2}\;.
		\end{align}
		$ \psi^2 X D $
		\begin{align}
			\left[G_{W\ell}\right]_{pr}^{(1)} &=-\frac{N_c}{6}\,g\,\left[\left(\frac{7}{12} + L_1\right)\frac{(\Lambda_{\ell})_{pr}}{M_1^2}+ \frac{(\tilde{\Lambda}_{\ell})_{pr}}{6\tilde{M}_2^2}\right]\;,\\
			\left[G^{\prime}_{\widetilde{W}\ell}\right]_{pr}^{(1)} &=\frac{N_c}{4}\,g\,\frac{(\Lambda_{\ell})_{pr}}{M_1^2}\;,\\
			\left[G_{B\ell}\right]_{pr}^{(1)} &=\frac{N_c}{3}\,g^{\prime}\,\left[\left(\frac{7Y_q - 2Y_{S_1}}{12} + Y_qL_1\right)\frac{(\Lambda_{\ell})_{pr}}{M_1^2} - \left(\frac{7Y_d + 2Y_{\tilde{S}_{2}}}{12} + Y_dL_2\right)\frac{(\tilde{\Lambda}_{\ell})_{pr}}{\tilde{M}_2^2} \right]\;,\\
			\left[G^{\prime}_{\widetilde{B}\ell}\right]_{pr}^{(1)} &=-\frac{N_c}{2}\,g^{\prime}\,\left[Y_{q}\frac{(\Lambda_{\ell})_{pr}}{M_1^2} - Y_{d}\frac{(\tilde{\Lambda}_{\ell})_{pr}}{\tilde{M}_2^2}\right]\;,\\
			\left[G_{Be}\right]_{pr}^{(1)} &=\frac{N_c}{3}\,g^\prime\,\left(\frac{7Y_u - 2Y_{S_1}}{12} + Y_u L_1 \right)\frac{(\Lambda_{e})_{pr}}{M_1^2}\;,\\
			\left[G^{\prime}_{\widetilde{B}e}\right]_{pr}^{(1)} &=\frac{N_c}{2}\,g^{\prime}\,Y_u\,\frac{(\Lambda_{e})_{pr}}{M_1^2}\;,\\
			\left[G_{Gq}\right]_{pr}^{(1)} &=\frac{1}{18}\,g_s\,\frac{(\Lambda_{q})_{pr}}{M_1^2} - \frac{4}{3}\,g_s\,\frac{(\LsBq)_{pr}}{M_1^2}\left(\frac{3}{4} + L_1\right)\;,\\
			\left[G^{\prime}_{\widetilde{G}q}\right]_{pr}^{(1)} &=-2\,g_s\,\frac{(\LsBq)_{pr}}{M_1^2}\;,\\
			\left[G_{Wq}\right]_{pr}^{(1)} &=-\frac{1}{6}\,g\,\left(\frac{7}{12} + L_1\right)\frac{(\Lambda_{q})_{pr} + 8(\LsBq)_{pr}}{M_1^2}\;,\\
			\left[G^{\prime}_{\widetilde{W}q}\right]_{pr}^{(1)} &=\frac{1}{4}\,g\,\frac{(\Lambda_{q})_{pr} - 8(\LsBq)_{pr}}{M_1^2}\;,\\
			\left[G_{Bq}\right]_{pr}^{(1)} &=\frac{1}{3}\,g^{\prime}\,\left[\left(\frac{7Y_\ell - 2Y_{S_1}}{12} + Y_{\ell}L_1\right)\frac{(\Lambda_{q})_{pr}}{M_1^2} + \left(\frac{7Y_q + 2Y_{S_1}}{12}+ Y_qL_1\right)\frac{8(\LsBq)_{pr}}{M_1^2}\right]\;,\\
			\left[G^{\prime}_{\widetilde{B}q}\right]_{pr}^{(1)} &=-\frac{1}{2}\,g^{\prime}\,\left[Y_\ell\frac{(\Lambda_{\ell})_{pr}}{M_1^2}+ Y_q\frac{8(\LsBq)_{pr}}{M_1^2}\right]\;,
		\end{align}
		\begin{align}
			\left[G_{Gu}\right]_{pr}^{(1)} &=\frac{1}{18}g_s\frac{(\Lambda_{u})_{pr}}{M_1^2} - \frac{1}{3}\,g_s\,\left(\frac{3}{4} + L_1\right)\frac{(\LsBu)_{pr}}{M_1^2}\;,\\
			\left[G^{\prime}_{\widetilde{G}u}\right]_{pr}^{(1)} &=-\frac{1}{2}\,g_s\,\frac{(\LsBu)_{pr}}{M_1^2}\;,\\
			\left[G_{Bu}\right]_{pr}^{(1)} &=\frac{1}{3}\,g^{\prime}\left[\left(\frac{7Y_e - 2Y_{S_1}}{12} + Y_eL_1\right)\frac{(\Lambda_{u})_{pr}}{M_1^2} + \left(\frac{7Y_d + 2Y_{S_1}}{12} + Y_dL_1\right)\frac{2(\LsBu)_{pr}}{M_1^2}\right]\,\;,\\
			\left[G^{\prime}_{\widetilde{B}u}\right]_{pr}^{(1)} &=\frac{1}{2}\,g^{\prime}\,\left[Y_{e}\frac{(\Lambda_{u})_{pr}}{M_1^2} + Y_{d}\frac{2(\LsBu)_{pr}}{M_1^2}\right]\;,\\
			\left[G_{Gd}\right]_{pr}^{(1)} &=\frac{1}{9}\,g_s\,\frac{(\tilde{\Lambda}_{d})_{pr}}{\tilde{M}_2^2} - \frac{1}{3}\,g_s\,\left(\frac{3}{4} + L_1\right)\frac{(\LsBd)_{pr}}{M_1^2}\;,\\
			\left[G^{\prime}_{\widetilde{G}d}\right]_{pr}^{(1)} &=-\frac{1}{2}\,g_s\,\frac{(\LsBd)_{pr}}{M_1^2}\;,\\
			\left[G_{Bd}\right]_{pr}^{(1)} &=-\frac{2}{3}\,g^{\prime}\left[\left(\frac{7Y_\ell - 2Y_{S_1}}{12} + Y_\ell L_1\right)\frac{(\tilde{\Lambda}_{d})_{pr}}{\tilde{M}_2^2} - \left(\frac{7Y_u + 2Y_{S_1}}{12} + Y_uL_1\right)\frac{(\LsBd)_{pr}}{M_1^2}\right]\,\;,\\
			\left[G^{\prime}_{\widetilde{B}d}\right]_{pr}^{(1)} &=\,g^{\prime}\,\left[2Y_{\ell}\frac{(\tilde{\Lambda}_{d})_{pr}}{\tilde{M}_2^2} + Y_{u}\frac{(\LsBd)_{pr}}{M_1^2}\right]\;,
		\end{align}
		\begin{align}
			&\left[G^{\prime}_{Gq}\right]_{pr}^{(1)} = \left[G^{\prime}_{Wq}\right]_{pr}^{(1)}=\left[G^{\prime}_{Bq}\right]_{pr}^{(1)}=0\;,\\
			&\left[G^{\prime}_{Gu}\right]_{pr}^{(1)}=\left[G^{\prime}_{Bu}\right]_{pr}^{(1)}=0\;,\\
			&\left[G^{\prime}_{Gd}\right]_{pr}^{(1)}=\left[G^{\prime}_{Bd}\right]_{pr}^{(1)}=0\;,\\
			&\left[G^{\prime}_{W\ell}\right]_{pr}^{(1)}=\left[G^{\prime}_{B\ell}\right]_{pr}^{(1)}=\left[G^{\prime}_{Be}\right]_{pr}^{(1)}=0\;.
		\end{align}
		$ \psi^2 H D^2 $
		\begin{align}
			\left[G_{eHD1}\right]_{pr}^{(1)} &=\frac{N_c}{2}\left(\frac{1}{2} + L_1\right)\frac{(\XLU)_{pr}}{M_1^2}\;,\\
			\left[G_{eHD2}\right]_{pr}^{(1)} &=+\frac{N_c}{2}\frac{(\XLU)_{pr}}{M_1^2}\;,\\
			\left[G_{eHD3}\right]_{pr}^{(1)} &=-\frac{N_c}{2}\frac{(\XLU)_{pr}}{M_1^2}\;,\\
			\left[G_{eHD4}\right]_{pr}^{(1)} &=-\frac{N_c}{2}\frac{(\XLU)_{pr}}{M_1^2}\;,\\
			\left[G_{uHD1}\right]_{pr}^{(1)} &=\frac{1}{2}\left(\frac{1}{2} + L_1\right)\frac{(\XLE)_{pr} - 4(\XBLD)_{pr}}{M_1^2}\;,\\
			\left[G_{uHD2}\right]_{pr}^{(1)} &=\frac{(\XLE)_{pr} - 4(\XBLD)_{pr}}{2M_1^2}\;,\\
			\left[G_{uHD3}\right]_{pr}^{(1)} &=-\frac{(\XLE)_{pr} - 4(\XBLD)_{pr}}{2M_1^2}\;,\\
			\left[G_{uHD4}\right]_{pr}^{(1)} &=-\frac{(\XLE)_{pr} - 4(\XBLD)_{pr}}{2M_1^2}\;,
			\end{align}
			\begin{align}
			\left[G_{dHD1}\right]_{pr}^{(1)} &=-\frac{2}{M_1^2}\left(\frac{1}{2} + L_1\right)(\XBLU)_{pr}\;,\\
			\left[G_{dHD2}\right]_{pr}^{(1)} &=-\frac{2}{M_1^2}(\XBLU)_{pr}\;,\\
			\left[G_{dHD3}\right]_{pr}^{(1)} &=\frac{2}{M_1^2}(\XBLU)_{pr}\;,\\
			\left[G_{dHD4}\right]_{pr}^{(1)} &=\frac{2}{M_1^2}(\XBLU)_{pr}\;.
		\end{align}
		$ \psi^2 X H $
		\begin{align}
			\left[G_{eW}\right]_{pr}^{(1)} &=-\frac{N_c}{8}\,g\,\left(\frac{1}{2} + L_1\right)\frac{(\XLU)_{pr}}{M_1^2}\;,\\
			\left[G_{eB}\right]_{pr}^{(1)} &=\frac{N_c}{4}\,g^{\prime}\,\left[(Y_q + Y_u)L_1 +\frac{1}{2}Y_q + \frac{3}{2}Y_u\right]\frac{(\XLU)_{pr}}{M_1^2}\;,\\
			\left[G_{uG}\right]_{pr}^{(1)} &=g_s\,\left(1 + L_1\right)\frac{(\XBLD)_{pr}}{M_1^2}\;,\\
			\left[G_{uW}\right]_{pr}^{(1)} &=-\frac{1}{8}\,g\,\left[\left(\frac{1}{2} + L_1\right)\frac{(\XLE)_{pr} - 4(\XBLD)_{pr}}{M_1^2}\right]\;,\\
			\left[G_{uB}\right]_{pr}^{(1)} &=\frac{1}{4}\,g^{\prime}\,\left[(Y_\ell + Y_e)L_1 + \frac{1}{2}Y_\ell + \frac{3}{2}Y_e\right]\frac{(\XLE)_{pr}}{M_1^2}\\ &-g^{\prime}\,\left[(Y_q + Y_d)L_1+\frac{1}{2}Y_q+\frac{3}{2}Y_d\right]\frac{(\XBLD)_{pr}}{M_1^2}\;,\\
			\left[G_{dG}\right]_{pr}^{(1)} &=+g_s\left(1+L_1\right)\frac{(\XBLU)_{pr}}{M_1^2}\;,\\
			\left[G_{dW}\right]_{pr}^{(1)} &=\frac{1}{2}\,g\,\left(\frac{1}{2} + L_1\right)\frac{(\XBLU)_{pr}}{M_1^2}\;\\
			\left[G_{dB}\right]_{pr}^{(1)} &=-g^{\prime}\,\left[(Y_u + Y_q)L_1 + \frac{1}{2}Y_q + \frac{3}{2}Y_u\right]\frac{(\XBLU)_{pr}}{M_1^2}\;.
		\end{align}
		$ \psi^2 D H^2 $
		\begin{align}
			\left[G^{(1)}_{H\ell}\right]_{pr}^{(1)} &= -\frac{N_c}{4}\left[\left(1 + L_1\right)\frac{(\XLDD)_{pr}-(\XLUU)_{pr}}{M_1^2}+ \left(1 + L_2\right)\frac{2(\XtilDD)_{pr}}{\tilde{M}_2^2}\right]\nonumber\\
			&-\frac{N_c}{2}|A_{\tilde{2}1}|^{2}\left[\frac{1}{(\Delta_{12}^2)} + \frac{M_1^2 + \tilde{M}_2^2}{2(\Delta_{12}^2)^{2}}\log\frac{\tilde{M}_2^2}{M_1^2}\right]\frac{(\Lambda_{\ell})_{pr} + (\Ltil)_{pr}}{\Delta_{12}^2}\;,\\
			\left[G^{\prime(1)}_{H\ell}\right]_{pr}^{(1)} &= -\frac{N_c}{8}\left[\frac{(\XLUU)_{pr} + (\XLDD)_{pr} + 2\lambda_{H1}(\Lambda_{\ell})_{pr}}{M_1^2} + \frac{(2\tilde{\lambda}_{H2} - \lambda_{\tilde{2}\tilde{2}})(\tilde{\Lambda}_{\ell})_{pr} + (\XtilDD)_{pr}}{\tilde{M}_2^2}\right]\nonumber\\
			&-\frac{1}{4}\frac{|A_{\tilde{2}1}|^{2}}{\Delta_{12}^2}\left[\frac{\log \tilde{M}_2^2/M_1^2}{\Delta_{12}^2}\left((\Lambda_{\ell})_{pr} + (\Ltil)_{pr}\right) + \left(\frac{(\Lambda_{\ell})_{pr}}{M_1^2} + \frac{(\Ltil)_{pr}}{\tilde{M}_2^2}\right)\right]\;,\\
			\left[G^{(3)}_{H\ell}\right]_{pr}^{(1)} &= \frac{N_c}{4}\left[\left(1 + L_1\right)\frac{(\XLDD)_{pr} + (\XLUU)_{pr}}{M_1^2}\right]\;,\\
			\left[G^{\prime(3)}_{H\ell}\right]_{pr}^{(1)} &= \frac{N_c}{8}\left[\frac{(\XLDD)_{pr} - (\XLUU)_{pr}}{M_1^2}\right]\;,\\
			\left[G_{He}\right]_{pr}^{(1)} &= -\frac{N_c}{2}\left(1 + L_1\right)\frac{(\XRUU)_{pr}}{M_1^2} -\frac{N_c}{2}|A_{\tilde{2}1}|^{2}\left[\frac{1}{(\Delta_{12}^2)} + \frac{M_1^2 + \tilde{M}_2^2}{2(\Delta_{12}^2)^{2}}\log\frac{\tilde{M}_2^2}{M_1^2}\right]\frac{(\Lambda_{e})_{pr}}{\Delta_{12}^2}\;,\\
			\left[G^{\prime}_{He}\right]_{pr}^{(1)} &= -\frac{N_c}{4}\frac{(\XRUU)_{pr} + \lambda_{H1}(\Lambda_{e})_{pr}}{M_1^2}
			-\frac{1}{4}|A_{\tilde{2}1}|^{2}\left[\frac{1}{M_1^2} + \frac{\log \tilde{M}_2^2/M_1^2}{\Delta_{12}^2}\right]\frac{(\Lambda_{e})_{pr}}{\Delta_{12}^2}\;,\\
			\left[G^{(1)}_{Hq}\right]_{pr}^{(1)} &= -\frac{1}{4}\left[\left(1 + L_1\right)\frac{(\XLEE)_{pr} + 8(\XsBLDD)_{pr} - 8(\XsBLUU)_{pr}}{M_1^2}\right]\nonumber\\
			&-\frac{1}{2}|A_{\tilde{2}1}|^{2}\left[\frac{1}{(\Delta_{12}^2)} + \frac{M_1^2 + \tilde{M}_2^2}{2(\Delta_{12}^2)^{2}}\log\frac{\tilde{M}_2^2}{M_1^2}\right]\frac{(\Lambda_{q})_{pr} - 8(\LsBq)_{pr}}{\Delta_{12}^2}\;,\\
			\left[G^{\prime(1)}_{Hq}\right]_{pr}^{(1)} &= -\frac{1}{8}\left[\frac{(\XLEE)_{pr} + 2\lambda_{H1}(\Lambda_{q})_{pr} +16\lambda_{H1}(\LsBq)_{pr} + 8(\XsBLDD)_{pr} - 8(\XsBLUU)_{pr}}{M_1^2}\right]\nonumber\\
			&-\frac{1}{4}|A_{\tilde{2}1}|^{2}\left[\frac{1}{M_1^2} + \frac{\log \tilde{M}_2^2/M_1^2}{\Delta_{12}^2}\right]\frac{(\Lambda_{q})_{pr} + 8(\LsBq)_{pr}}{\Delta_{12}^2}\;,\\
			\left[G^{(3)}_{Hq}\right]_{pr}^{(1)} &= \frac{1}{4}\left[\left(1 + L_1\right)\frac{(\XLEE)_{pr} + 8(\XsBLDD)_{pr} - 8(\XsBLUU)_{pr}}{M_1^2}\right]\;,\\
			\left[G^{\prime(3)}_{Hq}\right]_{pr}^{(1)} &= \frac{1}{8}\left[\frac{(\XLEE)_{pr} + 8(\XsBLDD)_{pr} - 8(\XsBLUU)_{pr}}{M_1^2}\right]\;,\\
			\left[G_{Hu}\right]_{pr}^{(1)} &=\frac{1}{2}\left(1 + L_1\right)\frac{(\XREE)_{pr} + 2(\XsBRDD)_{pr}}{M_1^2}\nonumber\\
			&-\frac{1}{2}|A_{\tilde{2}1}|^{2}\left[\frac{1}{(\Delta_{12}^2)} + \frac{M_1^2 + \tilde{M}_2^2}{2(\Delta_{12}^2)^{2}}\log\frac{\tilde{M}_2^2}{M_1^2}\right]\frac{(\Lambda_{u})_{pr} - 2(\LsBu)_{pr}}{\Delta_{12}^2}\;,\\
			\left[G^{\prime}_{Hu}\right]_{pr}^{(1)} &=-\frac{1}{4}\left[\frac{(\XREE)_{pr} + \lambda_{H1}(\Lambda_{u})_{pr} + 2\lambda_{H1}(\LsBu)_{pr} + 2(\XsBRDD)_{pr}}{M_1^2}\right]\nonumber\\
			&-\frac{1}{4}|A_{\tilde{2}1}|^{2}\left[\frac{1}{M_1^2} + \frac{\log \tilde{M}_2^2/M_1^2}{\Delta_{12}^2}\right]\frac{(\Lambda_{u})_{pr} + 2(\LsBu)_{pr}}{\Delta_{12}^2}\;,\\
			\left[G_{Hd}\right]_{pr}^{(1)} &=\frac{1}{2}\left(1 + L_2\right)\frac{(\XtilEE)_{pr}}{\tilde{M}_2^2} - \left(1 + L_1\right)\frac{(\XsBRUU)_{pr}}{M_1^2}\nonumber\\
			&-\frac{1}{2}|A_{\tilde{2}1}|^{2}\left[\frac{1}{(\Delta_{12}^2)} + \frac{M_1^2 + \tilde{M}_2^2}{2(\Delta_{12}^2)^{2}}\log\frac{\tilde{M}_2^2}{M_1^2}\right]\frac{(\Ltild)_{pr} - 2(\LsBd)_{pr}}{\Delta_{12}^2}\;,\\
			\left[G^{\prime}_{Hd}\right]_{pr}^{(1)} &=-\frac{1}{2}\frac{(\XsBRUU)_{pr}+\lambda_{H1}(\LsBd)_{pr}}{M_1^2}-\frac{1}{4}\frac{(\XtilEE)_{pr} + (2\lambda_{H2} - \lambda_{\tilde{2}\tilde{2}})(\tilde{\Lambda}_{d})_{pr}}{\tilde{M}_2^2}\nonumber\\
			&-\frac{1}{4}\frac{|A_{\tilde{2}1}|^{2}}{\Delta_{12}^2}\left[\frac{\log \tilde{M}_2^2/M_1^2}{\Delta_{12}^2}\left((\Ltild)_{pr} + 2(\LsBd)_{pr}\right) + \left(\frac{(\Ltild)_{pr}}{M_1^2} + \frac{2(\LsBd)_{pr}}{\tilde{M}_2^2}\right)\right]\;,\\
			\left[G^{\prime\prime}_{H(d,u,e)}\right]_{pr}^{(1)} &= \left[G^{\prime\prime(1,3)}_{H\ell}\right]_{pr}^{(1)} = \left[G^{\prime\prime(1,3)}_{Hq}\right]_{pr}^{(1)} = 0\;.
		\end{align}
		$ \psi^{2} H^3 $
		\begin{align}
			\left[G_{eH}\right]_{pr}^{(1)} &=N_c\frac{\left(1 + L_1\right)(\XLUUU)_{pr}-\lambda_{H1}(\XLU)_{pr}}{M_1^2}-N_c|A_{\tilde{2}1}|^{2}\left(\frac{1}{M_1^2} + \frac{\log \tilde{M}_2^2/M_1^2}{\Delta_{12}^2}\right)\frac{(\XLU)_{pr}}{\Delta_{12}^2}\;,\\
			\left[G_{uH}\right]_{pr}^{(1)} &=\frac{\left(1 + L_1\right)\left[(\XLEEE)_{pr} - 4(\XsBLDDD)_{pr}\right]-\lambda_{H1}\left[(\XLE)_{pr} - 4(\XBLD)_{pr}\right]}{M_1^2}\nonumber\\
			&-|A_{\tilde{2}1}|^{2}\left(\frac{1}{M_1^2} + \frac{\log \tilde{M}_2^2/M_1^2}{\Delta_{12}^2}\right)\frac{(\XLE)_{pr} + 4(\XBLD)_{pr}}{\Delta_{12}^2}\;,\\
			\left[G_{dH}\right]_{pr}^{(1)} &=-\frac{4\left(1 + L_1\right)(\XsBLUUU)_{pr} - 4\lambda_{H1}(\XBLU)_{pr}}{M_1^2}-4|A_{\tilde{2}1}|^{2}\left(\frac{1}{M_1^2} + \frac{\log \tilde{M}_2^2/M_1^2}{\Delta_{12}^2}\right)\frac{(\XBLU)_{pr}}{\Delta_{12}^2}\;.
		\end{align}
		
		\subsubsection{Four Fermion Operators}
		\textbf{Four Quarks}

		\begin{align}
			\left[G^{(1)}_{qq}\right]^{(1)}_{prst} &=-\frac{1}{16}\frac{(\Lambda_{q})_{pt}(\Lambda_{q})_{sr}}{M_1^2}-\frac{4(\LsBq)_{pr} (\LsBq)_{st} + 2(\LsBq)_{pt} (\LsBq)_{sr} +2(\LsBq)_{pr} (\Lambda_{q})_{st} - 2(\LsBq)_{pt} (\Lambda_{q})_{sr}}{M_1^2}\nonumber\label{eq:Gqqloop}\\
			&+\left[\left(1+N_c\right)\left(1 + L_1\right)c_1 + \frac{2\tilde{M}_2^2}{M_1^2}\left(1 + L_2\right) \left(N_c\coottilde + \ctotilde\right) + 8|A^\prime|^2 \ \frac{1}{M_1^2}\,L_2\right] \frac{(\lbl_{ps})(\lbl_{rt})^\ast}{2M_1^2}\nonumber\\
			&-\frac{2(\lbl_{ps})(\lbl_{rt})^\ast}{M_1^2}\left[\left(g^{\prime 2}Y_q^2 - \frac{g_s^2}{12} - \frac{3g^2}{4}\right)\left(\frac{1}{2} + a_{\text{ev}}\right) + \left(\frac{g_s^2}{6} + \frac{3g^2}{4}\right) (1 + L_1)\right]\;,\\
			\left[G^{(3)}_{qq}\right]^{(1)}_{prst} &=-\frac{1}{16}\frac{(\Lambda_{q})_{pt}(\Lambda_{q})_{sr}}{M_1^2}-2\frac{(\LsBq)_{pt} (\LsBq)_{sr}}{M_1^2}\nonumber\\
			&-\left[\left(1+N_c\right)\left(1 + L_1\right)c_1 + \frac{2\tilde{M}_2^2}{M_1^2}\left(1 + L_2\right) \left(N_c\coottilde + \ctotilde\right) + 8|A^\prime|^2 \ \frac{1}{M_1^2}\,L_2\right] \frac{(\lbl_{ps})(\lbl_{rt})^\ast}{2M_1^2}\nonumber\\
			&+\frac{2(\lbl_{ps})(\lbl_{rt})^\ast}{M_1^2}\left[\left(g^{\prime 2}Y_q^2 - \frac{g_s^2}{12} - \frac{g^2}{4}\right)\left(\frac{1}{2} + a_{\text{ev}}\right) + \left(\frac{g_s^2}{6} + \frac{g^2}{4}\right) (1 + L_1)\right]\;,\\
			\left[G_{uu}\right]^{(1)}_{prst} &=-\frac{1}{8M_1^2}\left[(\Lambda_{u})_{pt}(\Lambda_{u})_{sr} + 2(\LsBu)_{pr} (\LsBu)_{st} + 2(\LsBu)_{sr} (\LsBu)_{pt}\right.\nonumber\\
			&\left.+ 4(\Lambda_{u})_{pr}(\LsBu)_{st}- 4(\Lambda_{u})_{pt}(\LsBu)_{sr}\right]\;,\\
			\left[G_{dd}\right]^{(1)}_{prst} &=\frac{1}{4}\frac{(\tilde{\Lambda}_{d})_{pt}(\tilde{\Lambda}_{d})_{sr} - 2(\LsBd)_{pr} (\LsBd)_{st} -2 (\LsBd)_{sr} (\LsBd)_{pt}}{\tilde{M}_2^2}\;,\\
			\left[G^{(1)}_{ud}\right]^{(1)}_{prst} &=\frac{(\Lambda_{u})_{pr} (\LsBd)_{st}}{3M_1^2} - \frac{3}{4}\frac{(\LsBu)_{pr} (\LsBd)_{st}}{M_1^1}\nonumber\\
			&+\left[\left(1+N_c\right)\left(1 + L_1\right)c_1 + \frac{2\tilde{M}_2^2}{M_1^2}\left(1 + L_2\right) \left(N_c\coottilde + \ctotilde\right) + 8|A^\prime|^2 \ \frac{1}{M_1^2}\,L_2\right] \frac{(\lbr_{ps})^T(\lbr_{rt})^\dagger}{3M_1^2}\nonumber\\
			&+g^{\prime 2}\frac{(\lbr_{ps})^T (\lbr_{tr})^\ast}{3M_1^2}\left[(Y_u - Y_d)^{2}\left(\frac{1}{2} + a_{\text{ev}}\right) - (Y_u + Y_d + Y_{S_1})^2 \left(1 + L_1\right)\right]\nonumber\\
			&+g_s^2\left[\frac{4}{3}\left(\frac{1}{2} + a_{ev}\right) + \frac{5}{9} (1 + L_1)\right]\frac{(\lbr_{tr})^\ast (\lbr_{sp})}{M_1^2}\;,\\
			\left[G^{(8)}_{ud}\right]^{(1)}_{prst} &= -\frac{(\Lambda_{u})_{pr} (\LsBd)_{st}}{M_1^2}-\frac{(\LsBu)_{pr} (\LsBd)_{st}}{M_1^1}\nonumber\\
			&-\left[\left(1+N_c\right)\left(1 + L_1\right)c_1 + \frac{2\tilde{M}_2^2}{M_1^2}\left(1 + L_2\right) \left(N_c\coottilde + \ctotilde\right) + 8|A^\prime|^2 \ \frac{1}{M_1^2}\,L_2\right] \frac{(\lbr_{ps})^T(\lbr_{rt})^\dagger}{M_1^2}\nonumber\\
			&-g^{\prime 2}\frac{(\lbr_{ps})^T (\lbr_{tr})^\ast}{M_1^2}\left[(Y_u - Y_d)^{2}\left(\frac{1}{2} + a_{\text{ev}}\right) - (Y_u + Y_d + Y_{S_1})^2 \left(1 + L_1\right)\right]\nonumber\\
			&+g_s^2\,\left[\frac{7}{6}\left(\frac{1}{2} + a_{ev}\right) + \frac{13}{6} (1 + L_1)\right]\frac{(\lbr_{tr})^\ast (\lbr_{sp})}{M_1^2};,\\
			\left[G^{(1)}_{qu}\right]^{(1)}_{prst} &=-\frac{1}{12}\frac{(\Lambda_{q})_{pr}(\Lambda_{u})_{st}}{M_1^2}-\frac{3(\LsBq)_{pr} (\LsBu)_{st}}{M_1^2} + \frac{(\Lambda_{q})_{pr} (\LsBu)_{st} -4(\LsBq)_{pr} (\Lambda_{u})_{st}}{3M_1^2}\nonumber\\
			&+\left(\frac{3}{2} + L_1\right)\frac{(y_D\lbr + 2\lbl y_U^\ast)_{ps} (y_D^\ast (\lbr)^\ast + 2(\lbl)^\ast y_U)_{rt}}{6M_1^2}\;,\\
			\left[G^{(8)}_{qu}\right]^{(1)}_{prst} &=-\frac{1}{2}\frac{(\Lambda_{q})_{pr}(\Lambda_{u})_{st}}{M_1^2}-\frac{4(\LsBq)_{pr} (\LsBu)_{st}}{M_1^2} - \frac{(\Lambda_{q})_{pr} (\LsBu)_{st} -4(\LsBq)_{pr} (\Lambda_{u})_{st}}{M_1^2}\nonumber\\
			&-\left(\frac{3}{2} + L_1\right)\frac{(y_D\lbr + 2\lbl y_U^\ast)_{ps} (y_D^\ast (\lbr)^\ast + 2(\lbl)^\ast y_U)_{rt}}{2M_1^2}\;,\\
			\left[G^{(1)}_{qd}\right]^{(1)}_{prst} &= \frac{(\Lambda_{q})_{pr} (\LsBd)_{st}}{3M_1^2} -\frac{3(\LsBq)_{pr} (\LsBd)_{st}}{M_1^2}-\frac{1}{8}\frac{((\lol)^\ast \ltil^T)_{ps} (\ltil^\ast (\lol)^T)_{tr}}{M_1^2 - \tilde{M}_2^2}\log\frac{M_1^2}{\tilde{M}_2^2}\nonumber\\
			&-\left(\frac{3}{2} + L_1\right)\frac{(y_U(\lbr)^T + 2\lbl y_D^\ast)_{ps} (y_U^\ast (\lbr)^\dagger + 2(\lbl)^\ast y_D)_{rt}}{6M_1^2}\;,\\
			\left[G^{(8)}_{qd}\right]^{(1)}_{prst} &= -\frac{(\Lambda_{q})_{pr} (\LsBd)_{st}}{M_1^2} -\frac{4(\LsBq)_{pr} (\LsBd)_{st}}{M_1^2}\nonumber\\
			&+\left(\frac{3}{2} + L_1\right)\frac{(y_U(\lbr)^T + 2\lbl y_D^\ast)_{ps} (y_U^\ast (\lbr)^\dagger + 2(\lbl)^\ast y_D)_{rt}}{2M_1^2}\;,\\
			\left[G^{(1)}_{quqd}\right]^{(1)}_{prst} &=\frac{4}{3}\left[\left(1+N_c\right)\left(1 + L_1\right)c_1 + \frac{2\tilde{M}_2^2}{M_1^2}\left(1 + L_2\right) \left(N_c\coottilde + \ctotilde\right) + 8|A^\prime|^2 \ \frac{1}{M_1^2}\,L_2\right] \frac{(\lbl_{pr})(\lbr_{st})^\dagger}{M_1^2}\nonumber\\
			&-\frac{4}{3}g^{\prime2}\frac{(\lbl_{ps}) (\lbr_{tr})^\ast}{M_1^2}(1 + L_1)\left[Y_{q}(Y_{S_1} + 2Y_{q}) + Y_{d}(Y_{d} - Y_{S_1})\right]\nonumber\\
			&+\frac{44}{9}g_s^2\frac{(\lbr_{tr})^\ast (\lbl_{sp})}{M_1^2}\left(\frac{8}{11} + L_1\right)\;,\\
			\left[G^{(8)}_{quqd}\right]^{(1)}_{prst} &=-4\left[\left(1+N_c\right)\left(1 + L_1\right)c_1 + \frac{2\tilde{M}_2^2}{M_1^2}\left(1 + L_2\right) \left(N_c\coottilde + \ctotilde\right) + 8|A^\prime|^2 \ \frac{1}{M_1^2}\,L_2\right] \frac{(\lbl_{pr})(\lbr_{st})^\dagger}{M_1^2}\nonumber\\
			&+4g^{\prime2}\frac{(\lbl_{ps}) (\lbr_{tr})^\ast}{M_1^2}(1 + L_1)\left[Y_{q}(Y_{S_1} + 2Y_{q}) + Y_{d}(Y_{d} - Y_{S_1})\right]\nonumber\\
			&+\frac{2}{3}g_s^2\frac{(\lbr_{tr})^\ast (\lbl_{sp})}{M_1^2}\;.
		\end{align}

		\textbf{Four Leptons}
		\begin{align}
			\left[G^{(1)}_{\ell\ell}\right]^{(1)}_{prst} &=-\frac{N_c}{8}\left[\frac{(\Lambda_{\ell})_{pr}(\Lambda_{\ell})_{st}}{M_1^2} + \frac{(\tilde{\Lambda}_{\ell})_{pt} (\tilde{\Lambda}_{\ell	})_{sr}}{\tilde{M}_2^2}\right]\;,\\
			\left[G_{ee}\right]^{(1)}_{prst} &=-\frac{N_c}{8}\frac{(\Lambda_{e})_{pr}(\Lambda_{e})_{st}}{M_1^2}\;,\\
			\left[G_{\ell e}\right]^{(1)}_{prst} &=-\frac{N_c}{4}\frac{(\Lambda_{\ell})_{pr}(\Lambda_{e})_{st}}{M_1^2}\;.
		\end{align}

		\textbf{Semileptonic}
		\begin{align}
			\left[G^{(1)}_{\ell q}\right]^{(1)}_{prst} &=-\frac{1}{4}\frac{(\Lambda_{\ell})_{pr}(\Lambda_{q})_{st}}{M_1^2} -\frac{4(\Lambda_{\ell})_{pr} (\LsBq)_{st}}{M_1^2}  + \frac{4(\lbl\lol)^\ast_{tp} ((\lol)^T\lbl)_{rs}}{M_1^2}\nonumber\\
			&+\frac{1}{4}\left(\frac{3}{2} + L_2\right)\frac{(\ltil^\dagger y_D^\dagger)_{pt} (y_D\ltil)_{sr}}{\tilde{M}_2^2}\nonumber\\
			&+\left[\frac{1+N_c}{4}\left(1 + L_1\right)c_1 + \frac{\tilde{M}_2^2}{2M_1^2}\left(1 + L_2\right) \left(N_c\coottilde + \ctotilde\right) + 8|A^\prime|^2 \ \frac{1}{M_1^2}\,L_2\right] \frac{(\lol_{ps})^\dagger(\lol_{tr})}{M_1^2}\nonumber\\
			&+\frac{1}{4}\left(\frac{1}{2} + a_{\text{ev}}\right)\left[g_s^2C_F^{\text{SU(3)}} + g^{\prime 2}(Y_q - Y_\ell)^{2} + 3g^2\right]\frac{(\lol_{ps})^{\dagger} \lol_{tr}}{M_1^2}\;,\\
			\left[G^{(3)}_{\ell q}\right]^{(1)}_{prst} 
			&=\frac{4(\lbl\lol)^\ast_{tp} ((\lol)^T\lbl)_{rs}}{M_1^2}\nonumber\\
			&-\left[\frac{1+N_c}{4}\left(1 + L_1\right)c_1 + \frac{\tilde{M}_2^2}{2M_1^2}\left(1 + L_2\right) \left(N_c\coottilde + \ctotilde\right) + 8|A^\prime|^2 \ \frac{1}{M_1^2}\,L_2\right] \frac{(\lol_{ps})^\dagger(\lol_{tr})}{M_1^2}\nonumber\\
			&-\frac{1}{4}\left(\frac{1}{2} + a_{\text{ev}}\right)\left[g_s^2C_F^{\text{SU(3)}} + g^{\prime 2}(Y_q - Y_\ell)^{2} - 3g^2\right]\frac{(\lol_{ps})^{\dagger} \lol_{tr}}{M_1^2}\;,\\
			\left[G_{eu}\right]^{(1)}_{prst} &=
			\left[\frac{1+N_c}{2}\left(1 + L_1\right)c_1 + \frac{\tilde{M}_2^2}{M_1^2}\left(1 + L_2\right) \left(N_c\coottilde + \ctotilde\right) + 8|A^\prime|^2 \ \frac{1}{M_1^2}\,L_2\right] \frac{(\lori_{ps})^\dagger(\lori_{tr})}{M_1^2}\nonumber\\
			&+\frac{1}{2}\left(\frac{1}{2} + a_{\text{ev}}\right)\left[g_s^2C_F^{\text{SU(3)}} + g^{\prime 2}(Y_u - Y_e)^{2} + 3g^2\right]\frac{(\lori_{ps})^{\dagger} \lori_{tr}}{M_1^2}
			+ \frac{(\Lambda_{e})_{pr} (\LsBu)_{st}}{M_1^2}\;,\\
			\left[G_{ed}\right]^{(1)}_{prst} &= \frac{(\Lambda_{e})_{pr} (\LsBd)_{st}}{M_1^2} -\frac{2(\lbr\lori)^\ast_{tp} (\lbr\lori)_{sr}}{M_1^2}-\frac{1}{2}\left(\frac{3}{2} - L_2\right)\frac{(y_E^\dagger \ltil^\dagger)_{pt} (\ltil y_E)_{sr}}{\tilde{M}_2^2}\;,\\
			\left[G_{qe}\right]^{(1)}_{prst} &=-\frac{1}{4}\frac{(\Lambda_{q})_{pr}(\Lambda_{e})_{st}}{M_1^2} - \frac{4(\LsBq)_{pr} (\Lambda_{e})_{st}}{M_1^2}\nonumber\\
			&-\frac{1}{4}\left(\frac{3}{2} + L_1\right)\frac{((\lol)^\ast y_E^\ast-y_U(\lori)^\ast)_{ps}(\lol y_E - y_U^\ast \lori)_{rt}}{M_1^2}\;,\\
			\left[G_{\ell u}\right]^{(1)}_{prst} &=-\frac{1}{4}\frac{(\Lambda_{\ell})_{pr}(\Lambda_{u})_{st}}{M_1^2} + \frac{(\Lambda_{\ell})_{pr} (\LsBu)_{st}}{M_1^2}\nonumber\\
			&-\frac{1}{4}\left(\frac{3}{2} + L_1\right)\frac{((\lol)^\dagger y_U^\ast-y_E(\lori)^\dagger)_{ps}((\lol)^T y_U - y_E^\ast (\lori)^T)_{rt}}{M_1^2}\nonumber\\
			&+\frac{1}{2}\frac{(\ltil^\dagger\lbr)_{ps} ((\lbr)^\dagger\ltil)_{tr}}{M_1^2 - \tilde{M}_2^2}\log\frac{M_1^2}{\tilde{M}_2^2}\;,\\
			\left[G_{\ell d}\right]^{(1)}_{prst} &= \frac{(\Lambda_{\ell})_{pr} (\LsBd)_{st}}{M_1^2} + \frac{1}{4}\frac{(\tilde{\Lambda}_{\ell})_{pr} (\tilde{\Lambda}_{d})_{st}}{M_1^2 - \tilde{M}_2^2}\log\frac{M_1^2}{\tilde{M}_2^2}-\frac{1}{4}\left(\frac{3}{2} + L_1\right)\frac{((\lol)^\dagger y_D^\ast)_{ps}((\lol)^T y_D)_{rt}}{M_1^2}\nonumber\label{eq:Gldloop}\\
			&-\left[\frac{M_1^2}{\tilde{M}_2^2}\left(1 + L_1\right)\left(N_c\coottilde + \ctotilde\right) + \left(\left(1+2N_c\right)\cttilde + (2+N_c)c_{\tilde{2}}^{(8)}\right)\left(1 + L_2\right)\right]\frac{\ltil_{pt}^{\dagger}\ltil_{sr}}{2\tilde{M}_2^2}\nonumber\\
			&-\frac{1}{2}\left(\frac{1}{2} + a_{\text{ev}}\right)\left[g_2^2 C^{\text{SU(3)}}_F + g^{2}C_F^{\text{SU(2)}}+ g^{\prime 2}(Y_{\ell} + Y_d)^{2}\right]\frac{(\ltil^\ast_{pt}) (\ltil_{sr})}{\tilde{M}_2^2}\nonumber\\
			&-2|A^\prime|^2\frac{(\ltil_{pt})^\dagger (\ltil_{sr})}{\tilde{M}_2^4}\left(1 + \frac{M_1^2 L_1 - \tilde{M}_2^2 L_2}{\Delta^2_{12}}\right)\;,\\
			\left[G_{\ell edq}\right]^{(1)}_{prst} & = -\frac{6(\lbl\lol)^\ast_{tp} (\lbr\lori)_{sr}}{M_1^2}\nonumber\\
			&-\frac{1}{2}\left(\frac{3}{2} + L_1\right)\frac{((\lol)^\dagger y_D^\ast)_{ps}(\lol y_E - y_U^\ast\lori	)_{tr}}{M_1^2} + \frac{1}{2}\left(\frac{3}{2} + L_2\right)\frac{(\ltil^\dagger y_D^\dagger)_{pt} (\ltil y_E)_{sr}}{\tilde{M}_2^2}\;,\\
			\left[G^{(1)}_{\ell equ}\right]^{(1)}_{prst} &=\left[\frac{1+N_c}{2}\left(1 + L_1\right)c_1 + \frac{\tilde{M}_2^2}{M_1^2}\left(1 + L_2\right) \left(N_c\coottilde + \ctotilde\right) + 8|A^\prime|^2 \ \frac{1}{M_1^2}\,L_2\right] \frac{(\lol_{pt})^\dagger(\lori_{sr})}{M_1^2}\nonumber\\
			&-\frac{3}{2}\left(\frac{3}{2} + L_1\right)\left[g_s^2C_F^{\text{SU(3)}} + g^{\prime 2}(Y_q-Y_\ell)(Y_u-Y_e)\right]\frac{(\lol_{ps})^\dagger\lori_{tr}}{M_1^2}\;,\\
			\left[G^{(3)}_{\ell equ}\right]^{(1)}_{prst} &= -\left[\frac{1+N_c}{8}\left(1 + L_1\right)c_1 + \frac{\tilde{M}_2^2}{4M_1^2}\left(1 + L_2\right) \left(N_c\coottilde + \ctotilde\right) + 8|A^\prime|^2 \ \frac{1}{M_1^2}\,L_2\right] \frac{(\lol_{pt})^\dagger(\lori_{sr})}{M_1^2}\nonumber\\
			&-\frac{1}{8}\left(\frac{3}{2} + L_1\right)\left[g_s^2C_F^{\text{SU(3)}} + g^{\prime 2}(Y_q - Y_\ell)(Y_u - Y_e)\right]\frac{(\lol_{ps})^\dagger\lori_{tr}}{M_1^2}\;.
		\end{align}

		\textbf{B-violating}
		\begin{align}
			\left[G_{duq}\right]^{(1)}_{prst} &=\left[\left(1+N_c\right)\left(1 + L_1\right)c_1 + \frac{2\tilde{M}_2^2}{M_1^2}\left(1 + L_2\right) \left(N_c\coottilde + \ctotilde\right) + 8|A^\prime|^2 \ \frac{1}{M_1^2}\,L_2\right] \frac{(\lbr_{pr})^\ast(\lol_{st})}{M_1^2}\nonumber\\
			&+\left(1 + L_1\right)\left[g^{\prime 2}\left(Y_d(Y_q - Y_{S_1}) + Y_u(Y_{\ell} - Y_{S_1}) + Y_d(Y_u - Y_{S_1})\right)\right]\frac{(\lbr_{pr})^\ast (\lol_{st})}{M_1^2}
			\label{eq:bnv1}\nonumber\\
			&-\left(\frac{1}{2} + a_{ev}\right)(Y_{\ell} + Y_u)(Y_u + Y_d)\frac{(\lbr_{pr})^\ast (\lol_{st})}{M_1^2}\nonumber\\
			&+ g_s^2\left[\frac{4}{3}\left(\frac{1}{2} + a_{ev}\right) -\frac{13}{6}(1 + L_1)\right]\frac{(\lbr_{pr	})^\ast (\lol_{st})}{M_1^2}\nonumber\\
			&+\frac{( (\ltil)^\ast (\lol)^T)_{ps} ((\lbr)^\dagger \ltil)_{rt}}{M_1^2 - \tilde{M}_2^2}\log\frac{M_1^2}{\tilde{M}_2^2}\nonumber\\
			&+\frac{(\lol y_E + y_U^\ast \lori)_{pt} ((\lbr)^\dagger y_D^\dagger + 2y_U^T (\lbl)^\ast)_{sr}}{2M_1^2}\left(\frac{3}{2} + L_1\right)\;,\\
			\left[G_{qqu}\right]^{(1)}_{prst} &=\left[\left(1+N_c\right)\left(1 + L_1\right)c_1 + \frac{2\tilde{M}_2^2}{M_1^2}\left(1 + L_2\right) \left(N_c\coottilde + \ctotilde\right) + 8|A^\prime|^2 \ \frac{1}{M_1^2}\,L_2\right] \frac{(\lbl_{pr})^\ast(\lori_{st})}{M_1^2}\nonumber\\
			&-\left(1 + L_1\right)\left[4 g^{\prime 2}Y_{S_1}(2Y_q + Y_{S_1})\right]\frac{(\lbl_{pr})^\ast (\lori_{st})}{4M_1^2}\nonumber\\
			&+2g^{\prime2}\left(\frac{1}{2} + a_{ev}\right)Y_q(Y_e - Y_u)\frac{(\lbl_{pr})^\ast (\lori_{st})}{M_1^2}\nonumber\\
			&+ g_s^2\frac{(\lbl_{pr})^\ast (\lori_{st})}{M_1^2}\left[\frac{2}{3}\left(\frac{1}{2} + a_{ev}\right)-\frac{5}{2}(1+L_1)\right]\nonumber\\
			&+\frac{(2(\lbl)^\ast y_D - y_U^\ast(\lbr)^\dagger)_{sp} (y_E^\ast(\lori)^T -(\lol)^T y_U)_{tr}}{2M_1^2}\left(\frac{3}{2} + L_1\right)\nonumber\\
			&-\frac{(y_D^T \lol)_{pt} (2(\lbl)^\ast y_U + y_D^\ast (\lbr	)^\ast)_{sr}}{2M_1^2}\left(\frac{3}{2} + L_1\right)\;,\\
			\left[G_{qqq}\right]^{(1)}_{prst} &=-\left[\left(1+N_c\right)\left(1 + L_1\right)c_1 + \frac{2\tilde{M}_2^2}{M_1^2}\left(1 + L_2\right) \left(N_c\coottilde + \ctotilde\right) + 8|A^\prime|^2 \ \frac{1}{M_1^2}\,L_2\right] \frac{2(\lbl_{pr})^\ast(\lol_{st})}{M_1^2}\nonumber\\
			&+\left(1 + L_1\right)\left[2g^{\prime 2}Y_{S_1}(6Y_q + Y_{\ell})\right] \frac{(\lbl_{pr})^\ast (\lol_{st})}{2M_1^2}\nonumber\\
			&+ g_s^2\left(\frac{3}{2} + L_1\right)\frac{(\lbl_{ps})^\ast (\lol_{rt})}{12M_1^2} - \frac{5}{3}g_s^2\frac{(\lbl_{pr})^\ast (\lol_{st})}{M_1^2}\left(-\frac{3}{5} + L_1\right)\nonumber\\
			&+\frac{2g^2}{M_1^2}\left(\frac{3}{2} + L_1\right)\left[2(\lbl_{ps})^\ast (\lol_{rt}) + (\lbl_{rs})^\ast (\lol_{tp}) - (\lbl_{pr})^\ast (\lol_{st})\right]\;,\\
			\left[G_{duu}\right]^{(1)}_{prst} &=\left[\left(1+N_c\right)\left(1 + L_1\right)c_1 + \frac{2\tilde{M}_2^2}{M_1^2}\left(1 + L_2\right) \left(N_c\coottilde + \ctotilde\right) + 8|A^\prime|^2 \ \frac{1}{M_1^2}\,L_2\right] \frac{(\lbr_{pr})^\ast(\lori_{st})}{M_1^2}\nonumber\\
			&-\left[4\left(1 + L_1\right)g^{\prime 2}(2Y_u^2-2Y_eY_d-Y_eY_{S_1}) \right.\nonumber\\
			&+\left. 2g^{\prime 2}Y_u(Y_d + 3Y_u) - 2g^{\prime 2}Y_e(3Y_d - Y_u)\right]\frac{(\lbr_{pr})^\ast (\lori_{st})}{4M_1^2}\nonumber\\
			&-2g^{\prime2}\frac{(\lbr_{ps})^\ast (\lori_{rt})}{M_1^2}\left(\frac{3}{2} + L_1\right)\left[Y_d(Y_e-Y_u) - Y_u(Y_e + Y_u)\right] \nonumber\\
			&-g_s^2\frac{1}{2}\left(\frac{7}{3} + L_1\right)\frac{(\lbr_{pr})^\ast (\lori_{st})}{M_1^2}\;.
			\label{eq:bnv4}
		\end{align}
In total, 109 out of 139 operators are generated in the Green basis which translates into 53 out of 59 operators in the Warsaw basis. The only operators not generated by the $S_1+\tilde{S}_2$-model 
in the Warsaw basis are the CP violating ones, namely $ Q_{3(\tilde{G},\tilde{W}), H(\tilde{B},\tilde{W},\tilde{G}), H\tilde{W}B} $, which are of course absent in the Green basis as well.
		
The UOLEA parameters $\tilde{f}_{11}-\tilde{f}_{19}$ appearing in Vector-Bosons-Scalar operators are given separately in  Appendix~\ref{app:UOLEAexps}. 	
The hypercharges of the SM chiral fermions and the Higgs are, 
\begin{equation}
Y_\ell = -\frac{1}{2}\;, \quad  Y_e=-1\;, \quad Y_q=\frac{1}{6}\;, \quad  Y_u=\frac{2}{3}\;, \quad Y_d=-\frac{1}{3}\;, \quad {\rm and}  \quad Y_H=\frac{1}{2} \;,
\label{eq:Y}
\end{equation}
respectively, while  $Y_{S_1}=1/3$ and $ Y_{\tilde{S}_2}=1/6 $ for leptoquarks. 
			

		\subsection{Theoretical Remarks}
		Further remarks on our findings for the complete set of $d\le 6$ Wilson coefficients  in Green basis at one-loop are in order.

\subsubsection{Evanescent Operators}
	Evanescent operators appear in 4-point functions involving fermions. Treating the integrals in $ d $-dimensions while using Fierz identities, that hold only in $ d=4 $ or encountering higher order of $ \gamma $-matrices products, give rise to evanescent operators that in general vanish in $ d=4 $. The scheme we will be using for this type of structures is the introduction of local counterterms $ a_{ev},\ldots ,f_{ev} $. For details the reader is referred to refs.~\cite{BURAS199066,Herrlich:1994kh,Dugan:1990df,Aebischer:2020dsw,Dekens:2019ept}.
	
	Although, strictly-speaking, not part of the actual matching calculation, in translating the raw results of the traces after substituting the $ U $ matrices, one needs to choose a definite scheme to reduce the $ \gamma $-matrix structure appearing in the subsequent equations and match it to a certain basis, such as the Green basis. In the model under consideration evanescent operators make their appearance in the Supertrace of (\ref{eq:2.35}) where both left and right projection operators appear in the same trace. We keep a general parameter $ a_{ev} $ not resorting to any particular scheme. The usual scheme choice for evanescent operators, however, is $ a_{ev}=\ldots=f_{ev}=1 $. The relevant Dirac-structures appearing in the model at hand are (in the NDR scheme $ d=4-\epsilon $),
		\begin{align}
			P_{L}\gamma_{\mu}\gamma_{\nu}P_{L}\otimes P_{L}\gamma^{\mu}\gamma^{\nu}P_{L} &= 4\left(1 - \frac{\epsilon}{4}\right)\,P_{L}\otimes P_{L} - P_{L}\sigma_{\mu\nu}P_{L}\otimes P_{L}\sigma^{\mu\nu}P_{L}\;,\\
			P_{L}\gamma_{\mu}\gamma_{\nu}P_{L}\otimes P_{L}\gamma^{\nu}\gamma^{\mu}P_{L} &= 4\left(1 - \frac{\epsilon}{4}\right)\,P_{L}\otimes P_{L} + P_{L}\sigma_{\mu\nu}P_{L}\otimes P_{L}\sigma^{\mu\nu}P_{L}\;,\\
			P_{L}\gamma^{\mu}\gamma^{\nu}P_{L}\otimes P_{R}\gamma^{\mu}\gamma^{\nu}P_{R} &= 4\left(1 + a_{ev}\frac{\epsilon}{2}\right)\,P_{L}\otimes P_{R} + E^{(2)}_{LR}\;,\\
			P_{L}\gamma^{\mu}\gamma^{\nu}P_{L}\otimes P_{R}\gamma^{\nu}\gamma^{\mu}P_{R} &= 4\left[1 - \frac{\epsilon}{2}\left(1+a_{ev}\right)\right]\,P_{L}\otimes P_{R} + E^{(2)}_{LR}\;.
		\end{align}
	Plugging in a specific value for the coefficient $ a_{ev} $, in our case, defines the evanescent operators $ E^{(2)}_{LR} $.
	
\subsubsection{RGE checks}
	As a further cross check of our results for Wilson coefficients we have calculated the Renormalization Group Equations (RGEs) for a certain set. For purely one-loop generated operators one has to just take the derivative with respect to the renormalization scale $ \mu $ and extract the relevant $ \beta $-function. For instance explicitly taking the derivative with respect to $ \mu $ on $ [C_{eW}]_{pr} $ we find,
		\begin{align}
			\frac{d [C_{eW}]_{pr}}{d \ln\mu} = -N_c \ \frac{g}{4} \ \frac{(\XLU)_{pr}}{M_1^2}\;. 
		\end{align} 
	Comparing with the $ \beta $-functions from \cite{Jenkins:2013zja,Jenkins:2013wua,Alonso:2013hga,Celis:2017hod},
		\begin{align}
			[\beta_{eW}]_{pr} = 6 \ g \ (y_U)^\ast_{st} \ \left[C_{\ell equ}^{(3)}\right]_{prst}\;,
		\end{align}
	after plugging in the value of the relevant Wc  [eq.~(\ref{eq:3.14})], we get,
		\begin{align}
			[\beta_{eW}]_{pr} = -\frac{3g}{4} \ \frac{(\XLU)_{pr}}{M_1^2}\;,
		\end{align}
	which is in complete agreement, for $ N_c = 3 $, with the direct application of the derivative. The same procedure has been followed for every other purely one-loop generated Wcs and we have found no discrepancies in the comparison.
	
	For operators generated at tree-level as well the picture is a bit different because at 
	tree-level the coupling of the respective Wcs must be considered as running parameters due to shifts of the corresponding fields. However these exact shifts will cancel with the RG running due to wavefunction renormalization. For example, to bring back the lepton and quark doublet kinetic term we must make the following shift,
		\begin{align}
			\ell_{p} \ &\longrightarrow \ \ell_{p} - \frac{1}{2} \ (\delta Z_{\ell})_{pp_1}\ \ell_{p_1}\;,\\
			q_{p} \  &\longrightarrow \ q_{p} - \frac{1}{2} \ (\delta Z_{q})_{pp_1} \ q_{p_1}\;,
		\end{align}
	where $ \delta Z_{(\ell,q)} $ correspond to eqs.~\eqref{eq:3.43} and~\eqref{eq:3.45}, respectively. We have also suppressed all other indices apart from generation indices.
	In turn this produces a shift in the coupling $ \lol_{pr} $. To absorb this shift we redefine the coupling as,
		\begin{align}
			(\lol_{pr})^{\text{eff.}} = \lol_{pr} - \frac{1}{2} \lol_{pw} \ (\delta Z_{\ell})_{wr} - \frac{1}{2} \lol_{wr} \ (\delta Z_{q})_{wp}\;.
		\end{align}
	Taking the derivative w.r.t $ \mu $-parameter we find the $ \beta $-function for the effective running coupling,
		\begin{align}
			[\beta_{\lol}]_{pr} = \ &\frac{N_c}{2} \ \left[\lol(\lol)^\dagger\lol + \lol(\ltil)^\dagger\ltil\right]_{pr}\nonumber\\
			+&\frac{1}{2} \ \left[(\lol)^T(\lol)^\ast(\lol)^T - 8(\lol)^T \lbl(\lbl)^\dagger\right]_{pr}\;.
		\end{align}
	The same of course can be done for the complex conjugate coupling, 
		\begin{align}
			[\beta_{(\lol)^\dagger}]_{pr} = &\  [\beta_{\lol}]^\dagger_{pr}\;.
		\end{align}
	These two will contribute, for instance, in the RG evolution of the operator $ C_{\ell q}^{(1)} $. Ignoring finite parts, and for the clarity of the cancellation considering only the part produced by the shifts in the fermion fields, we find
		\begin{align}
			[C^{(1)}_{\ell q}]_{prst} \propto [C^{(1)}_{\ell q}]^{(0)}_{prst} - \frac{1}{2}&\left\{(\delta Z_{\ell})_{pw} \ [C^{(1)}_{\ell q}]^{(0)}_{wrst} + (\delta Z_{\ell})_{rw} \ [C^{(1)}_{\ell q}]^{(0)}_{pwst}\right.\nonumber\\
			&\left. +(\delta Z_{q})_{sw} \ [C^{(1)}_{\ell q}]^{(0)}_{prwt} + (\delta Z_{q})_{tw} \ [C^{(1)}_{\ell q}]^{(0)}_{prsw}\right\}\nonumber\\
			+& \left[c_1 \ L_1 + \frac{\tilde{M}_2^2}{2M_1^2} \ L_2 \ (3\coottilde + \ctotilde) + 8|A^\prime|^2 \ \frac{1}{M_1^2} \ L_2 \right] \frac{(\lol_{ps})^\dagger (\lol_{tr})}{M_1^2}\;.
		\end{align}
	Schematically for the operator above we have,
		\begin{align}
			\frac{d [C^{(1)}_{\ell q}]_{prst}}{d \ln\mu} \propto \frac{d [C^{(1)}_{\ell q}]_{prst}^{(0)}}{d \ln\mu} - \frac{1}{2}\frac{d}{d\ln\mu} &\left\{(\delta Z_{\ell})_{pw} \ [C^{(1)}_{\ell q}]^{(0)}_{wrst} + (\delta Z_{\ell})_{rw} \ [C^{(1)}_{\ell q}]^{(0)}_{pwst}\right.\nonumber\\
			&\left. +(\delta Z_{q})_{sw} \ [C^{(1)}_{\ell q}]^{(0)}_{prwt} + (\delta Z_{q})_{tw} \ [C^{(1)}_{\ell q}]^{(0)}_{prsw}\right\}\nonumber\\
			+& \left[2 c_1 + \frac{\tilde{M}_2^2}{M_1^2} \ (3\coottilde + \ctotilde) + 16|A^\prime|^2 \ \frac{1}{M_1^2}\right] \frac{(\lol_{ps})^\dagger (\lol_{tr})}{M_1^2}\;.
			\label{eq:3.183}
		\end{align}
	On the other hand the derivative of the tree-level operator reads,
		\begin{align}
			\frac{d[C^{(1)}_{\ell q}]^{(0)}_{prst}}{d \ln\mu} = \frac{(\lol_{ps})^\dagger}{4M_1^2} \ (\beta_{\lol})_{tr} + \frac{(\lol)_{tr}}{4M_1^2} \ (\beta_{(\lol)^\dagger})_{ps} - \frac{(\lol_{ps})^\dagger (\lol)_{tr}}{4M_1^2} \ \gamma_{M_1^2}\;,
			\label{eq:3.184}
		\end{align}
	where $ \gamma_{M_1^2} = d \ln M_1^2/d\ln\mu $ is the anomalous dimension of the leptoquark mass. The last term in \eqref{eq:3.184} will cancel with the last line of \eqref{eq:3.183}. In contrast to the coupling, this cancellation is not captured by the matching procedure, where we have assumed no heavy-external-field legs.
	
	The remaining terms inside the curly brackets in \eqref{eq:3.183}, due to the redefintion of the fermion fields, exactly cancel the contributions to the $ \beta $-function from the tree-level result (\ref{eq:3.184}). Cancellations aside, by taking the explicit derivative of the whole set of Wcs we can cross-check the logarithmic part of our results. Comparing the $ \beta $-functions produced by our model, with the relevant parts of \cite{Jenkins:2013zja,Jenkins:2013wua,Alonso:2013hga,Celis:2017hod}, we find complete agreement.	
	\subsection{Phenomenological Aspects}
		
In this section we present some interesting phenomenological aspects 
arising from the $S_1+\tilde{S}_2$ LQ-model whose one-loop effective Lagrangian 
derived previously in this section.

\subsubsection{Lepton magnetic and electric dipole moments}
	There is a recent excitement about the muon anomalous magnetic moment. 
	FNAL experiment~\cite{Abi:2021gix} confirmed previous results by 
	BNL experiment~\cite{Bennett:2006fi}
	and   found a 4.2$\sigma$ excess w.r.t the SM, 
	$\Delta a_\mu = a_\mu^{\rm exp} - a_\mu^{\rm SM}= (251\pm 59)\times 10^{-11}$~\cite{Abi:2021gix}. 
	As a demonstration of our one-loop effective Lagrangian 
	we work out the contribution to $\Delta a_\mu$ from the decoupling of
	$S_1+ \tilde{S}_2$-leptoquarks, and compare it with the fixed order one-loop calculation.
	
	Within  functional approach, contributions to magnetic moments of fermions arise from eqs.~\eqref{g-2a} and \eqref{g-2b}. The corresponding functional supertrace diagram and its expression is displayed in \eqref{Strg-2b}. The above contributions along with the insertion of the tree-level operator $ [\mathcal{O}^{(3)}_{\ell equ}] $, associated with $ [C^{(3)}_{\ell equ}] $ given in eq.~(\ref{eq:3.14}), in a one-loop diagram computed for example in ref.~\cite{Aebischer:2021uvt} constitute the full EFT formula in this model. 
	The dominant new physics contributions to the anomalous magnetic moment of the $\ell$-generation lepton are thus
	\begin{equation}
		\Delta a^\ell \ = \ \frac{4 m_\ell v}{\sqrt{2}} \left [ \frac{1}{g'} \Re e [\mathcal{C}_{eB}] \ - \ \frac{1}{g} \Re e [\mathcal{C}_{eW}] \right ]_{\ell\ell}  + \frac{2 N_c}{3\pi^2}\,\sum_{q} m_{\ell} m_{q}\,\log\left(\frac{\mu^2}{m_q^2}\right)\, \Re e \left\{[\mathcal{C}^{(3)}_{\ell equ}]_{\ell\ell qq}\right\}\;,
		\label{eq:Dal}
	\end{equation}
	where $v$ is the vev, $m_{\ell}$ is the $\ell=e,\mu,\tau$ lepton mass and $ m_q $ is the quark mass running in the loop with $ q = t, c, u  $. We note for later, that when evolving down to the top quark mass, $ m_t $, the dominant part of the last term vanishes, hence we can neglect all other sub-leading terms in the sum. Therefore, we are left with the first two terms in the square bracket. The coefficients $\mathcal{C}_{eB}$ and $\mathcal{C}_{eW}$ are defined at low energies in mass basis of ref.~\cite{Dedes:2017zog}. They are related to the Warsaw gauge basis coefficients, $C_{eB}$ and $C_{eW}$, through the expressions,
	\begin{equation}
		\mathcal{C}_{eB} = U_{e_L}^\dagger C_{eB} U_{e_R}\;, \qquad \mathcal{C}_{eW} = U_{e_L}^\dagger C_{eW} U_{e_R} \;,
		\label{eq:mCeB}
	\end{equation}
	where the unitary matrices $U_{e_{L,R}}$ diagonalize the lepton mass matrices, $U_{e_L}^\dagger y_E U_{e_R} = \hat{y}_E= diag(y_e, y_\mu, y_\tau)$.
	Since our results are given in Green basis we need a translation from Green to Warsaw basis. This translation is nicely given in ref.~\cite{Gherardi:2020det}. After a little 
	algebra, we find  the coefficients at renormalization scale $\mu$ (still 
	gauge basis in $3\times 3$ matrix notation) to be
	\begin{align}
		[C_{eB}]^{(1)}(\mu) \ &= \ \frac{g' N_c}{16 \pi^2} \left \{ \frac{5}{24} \left [\log\left (\frac{\mu^2}{M_1^2}\right ) + \frac{19}{10} \right ] \frac{Y_{1U}^{1L}}{M_1^2}
		 - \frac{1}{24} \frac{y_E  \cdot \Lambda_e}{M_1^2}
		- \frac{1}{48} \frac{\tilde{\Lambda}_\ell \cdot y_E}{\tilde{M}_2^2}  \right \} \;, \label{eq:CeB} \\[2mm]
		[C_{eW}]^{(1)}(\mu) \  &= \ \frac{g N_c}{16 \pi^2} \left \{ -\frac{1}{8} \left [\log\left (\frac{\mu^2}{M_1^2}\right ) + \frac{3}{2} \right ] \frac{Y_{1U}^{1L}}{M_1^2} + \frac{1}{24} \frac{\Lambda_\ell \cdot y_E}{M_1^2}
		- \frac{1}{48} \frac{\tilde{\Lambda}_\ell \cdot y_E}{\tilde{M}_2^2}  \right \} \;, \label{eq:CeW}
	\end{align}
	where $``\cdot "$ means matrix multiplication. These results are in agreement with ref.~\cite{Aebischer:2021uvt}. 
	The parameters $Y_{1U}^{1L}, \Lambda_e,\tilde{\Lambda}_\ell$ and  ${\Lambda}_\ell$ are defined in eqs.~\eqref{eq:Y1U1L},\eqref{eq:Lambdae} and \eqref{eq:Lambdaell}, respectively.
	We then use the RGE running of  the coefficients $C_{eB}, C_{eW}$ 
	from the heavy leptoquark mass scale $M_1$ down to the top-quark mass scale $m_t$
	and plug the result into \eqref{eq:Dal} to  find at leading-log approximation (for $N_c=3$):\footnote{To  leading-log order the result is the same by setting $\mu=m_t$
	in \eqref{eq:CeB} and \eqref{eq:CeW} and then take the difference in  \eqref{eq:Dal}, and neglecting the contribution from light quark masses.}
	\\	
	\begin{align}
		\Delta a^{(S_1+\tilde{S}_2)}_\ell  &=   \sum_{q=u,c,t} \frac{m_\ell}{4\pi^2} \: \frac{m_q}{M_1^2} \: \left [ \log\left (\frac{m_t^2}{M_1^2}\right ) + \frac{7}{4} \right ] 
		\Re e (\hat{\lambda}^{1L*}_{q\ell} \hat{\lambda}^{1R}_{q\ell}) \nonumber \\[2mm]
		&-\frac{m_\ell^2}{32 \pi^2 M_1^2} \: \left ( \hat{\lambda}^{1L*}_{q\ell} \hat{\lambda}^{1L}_{q\ell} + \hat{\lambda}^{1R *}_{q\ell} \hat{\lambda}^{1R}_{q\ell} \right )\;,\label{eq:DaS1S2}
	\end{align}
	where all parameters and masses are to be evaluated at $m_t$ and the ``hatted'' couplings are defined in mass basis as,
	\begin{equation}
		\hat{\lambda}^{1L} = U_{u_L}^T {\lambda}^{1L} U_{e_L}\;,  \quad \hat{\lambda}^{1R} = U_{u_R}^T {\lambda}^{1R} U_{e_R}\;, \quad \hat{\tilde{\lambda}} = U^{\dagger}_{d_R}\tilde{\lambda}\,U_{eL}\;, \label{eqs:lhat}
	\end{equation}
	with $U_{u_L}^\dagger y_U U_{u_R} = \hat{y}_U= \mathrm{diag}(y_u, y_s, y_t)$ being the up-quark  fermion Yukawa couplings. 
	Note that, as it should, the one-loop expression \eqref{eq:DaS1S2}   agrees with the fixed order 
	calculation of ref.~\cite{Bauer:2015knc} for the $S_1$-leptoquark decoupling.
	Also obvious from \eqref{eq:DaS1S2} is a natural enhancement of $O(m_t/m_\ell)$
	due to $S_1$-decoupling, while there is no effect from the $ \tilde{S}_{2} $-particle decoupling.
	However, a similar enhancement is shown in
	 eq.~\eqref{eq:Ymu},  for  the one-loop corrections to the
	Yukawa coupling of the leptons, and subsequently to the lepton mass itself.
	

Moreover, a bound on electron Electric Dipole Moment (eEDM), $|d_e|<1.1 \times 10^{-29}~\mathrm{e\cdot cm}$ at 90\% CL,  anounced by ACME collaboration~\cite{ACME:2018yjb} almost three years ago.
Our complete 1-loop functional matching LQs  renders the calculation of eEDM
very easy. As mentioned previously, in the model at hand ($S_1+\tilde{S}_2$), bosonic operators
of the form $H^2 F \tilde{F}$ and $F^3$ are not induced, therefore the only effect at one-loop
arises from the Warsaw-basis operators $Q_{eB}$, $Q_{eW}$ and $ Q^{(3)}_{\ell equ} $ as before. Basically, the calculation is the same with the lepton magnetic moments we performed above. Again, neglecting the contribution from $ C^{(3)}_{\ell equ} $ since the evaluation is at $ \mu = m_t $, the eEDM reads,
	\begin{equation}
		\frac{d_e}{e}(m_t) \ = \ \sqrt{2}\; v \left [ \frac{1}{g'} \Im m [\mathcal{C}_{eB}(m_t)] \ - \ \frac{1}{g} \Im m [\mathcal{C}_{eW}(m_t)] \right ]_{11} \;.
		\label{eq:de}
	\end{equation}
Plugging into \eqref{eq:de} the imaginary parts of eqs.~\eqref{eq:CeB} and \eqref{eq:CeW} after substituting $ \mu = m_t $,
we find 
\begin{eqnarray}
\frac{d_e}{e} \ = \  \frac{1}{8\pi^2} \, \frac{m_t}{M_1^2}\, \left [
\log \left (\frac{m_t^2}{M_1^2} \right ) + \frac{7}{4} \right ] \Im m (\hat{\lambda}^{1L\, *}_{31}
\hat{\lambda}^{1R}_{31} )\;.
\label{eq:deS1S2}
\end{eqnarray}
This result agrees with the fixed order 1-loop calculation, up to $\mathcal{O}(m_t^2/M_1^4)$-terms,
obtained by applying the general one-loop formula for lepton EDMs
 from ref.~\cite{Dedes:2007ef} onto the particular $S_1+\tilde{S}_2$ LQ-model. The leading-log
term of \eqref{eq:deS1S2} also agrees with the one obtained in refs.~\cite{Dorsner:2016wpm,Panico:2018hal}.

The reader should note that this is a SMEFT calculation, i.e. we have chosen $ \mu = m_t $. Ideally it's more appropriate to match onto Low Energy Effective Field Theory (LEFT) and perform the calculation of dipole moments at lower scales, which is beyond the scope of this paper. For a thorough EFT analysis of leptonic magnetic and electric dipole moments along these lines the reader is referred to refs.~\cite{Dekens:2018bci,Aebischer:2021uvt}.

\subsubsection{Radiative Neutrino masses}
	A nice feature of the $S_1+\tilde{S}_2$ model is that  neutrino masses are induced radiatively at one-loop. The single dimension-5 operator $Q_{\nu\nu}$, defined in Appendix~\ref{app:GreenOps}, 
	arises in the effective Lagrangian from the supertrace functional diagram \eqref{eq:2.34} which after calculation provides us the associated Wilson coefficient $G_{\nu\nu}=C_{\nu\nu}$
	 \eqref{eq:WeinbergOp} in both Green and  Warsaw basis. The result is finite and agrees with ref.~\cite{Crivellin:2020mjs}. 
	 Then going to the mass basis SMEFT Lagrangian of ref.~\cite{Dedes:2017zog} we have for the 
	 diagonal neutrino mass matrix
	\begin{equation}
		m_{\nu} \ = \ - v^2 \: U_{\nu_L}^T \, C_{\nu\nu} \, U_{\nu_L} \;.
	\end{equation}
	 By using the 
	 tree level definitions of CKM and MNS matrices\footnote{There is no "pollution" to CKM or MNS matrices from other tree level operators in this model.}, 
	 as $K_{\rm CKM}= U_{u_L}^\dagger U_{d_L}$ and $U_{\rm MNS} = U_{e_L}^\dagger U_{\nu_L}$ and  eq.~\eqref{eqs:lhat} we obtain
	\begin{equation}
		m_\nu \ = \ - \frac{\sqrt{2}}{16\pi^2} \frac{v A_{\tilde{2} 1}}{M_1^2-\tilde{M}_2^2} \: \left [U_{\rm MNS}^T\, (\hat{\lambda}^{1L})^T \, K_{\rm CKM} \, m_d \hat{\tilde{\lambda}}\, U_{\rm MNS} \right ]
		\log \left ( \frac{M_1^2}{\tilde{M}_2^2} \right ) \;,
	\end{equation}
	where $m_d$ is the diagonal down quark mass matrix. The neutrino masses clearly follow a ``down-quark''  mass  hierarchy with 
	the couplings   $\hat{\lambda}^{1L}$,  $\hat{\tilde{\lambda}}$ and the CKM matrix defining off-diagonal transitions. Obviously, depending 
	on the value of the parameter $A_{\tilde{2}1}$ that mixes both leptoquarks with the Higgs boson, we can probe two different mass
	scales: (i) $ A_{\tilde{2}1}/max(M_1,M_2) \simeq 1$, and (ii) $A_{\tilde{2}1}/max(M_1, M_2) \to 0$. In case (i) correct order of magnitude of neutrino 
	mass, $m_\nu\lesssim 0.1$ eV, and $\lambda \sim O(1)$ requires $M_1\approx M_2 \gtrsim 10^{13}$ GeV and this is currently consistent with proton decay bounds (see below), 
	whereas in case (ii) $M_1\approx M_2 \simeq 1$ TeV requires $A_{\tilde{2}1} \approx 10^{-7}$ GeV which is technically natural in terms of a $\mathbb{Z}_3$ (or $\mathbb{Z}_4$)
	softly broken discrete symmetry discussed in  section~\ref{sec:3} (but still baryon number violating couplings, $\lbl$ and $\lbr$, are allowed and have to be set to  zero by another symmetry in 
	order to avoid fast proton decay).

\subsubsection{Proton decay}

Baryon number is violated in the $S_1+\tilde{S}_2$ model. \emph{All} $d=6$ 
baryon number
violating (BNV) operators appear already at tree level in the effective Lagrangian [eqs.~\eqref{eq:dim6BNV1},\eqref{eq:dim6BNV2}], 
after the decoupling of $S_1$-field. From these expressions and Table~\ref{tab:BL}
we easily see they have $\Delta B=\Delta L=1$ consistent with  eqs.~\eqref{eq:dmin}
and \eqref{eq:deven}. Rotating fermion fields into the mass basis of 
ref.~\cite{Dedes:2017zog}
we find for the tree-level Wilson-coefficients 
\begin{eqnarray}
\left[\mathcal{C}^{qqu} \right]^{(0)}_{f_1f_2f_3f_4} \ &=& \ \frac{1}{M_1^2} \: [K_{\rm CKM}^T \, (\hat{\lbl})^* ]_{f_1f_2} \, (\hat{\lambda}^{1R})_{f_3f_4} \;, \label{eq:p1}\\
\left[\mathcal{C}^{duq}\right ]^{(0)}_{f_1f_2f_3f_4} \ &=& \ 
\frac{1}{M_1^2} \:
(\hat{\lbr})^*_{f_1f_2} \, [ K_{\rm CKM}^T\,  \hat{\lambda}^{1L}]_{f_3f_4}\;, \\
\left[\mathcal{C}^{duu}\right ]^{(0)}_{f_1f_2f_3f_4} \ &=& \ \frac{1}{M_1^2} \: 
(\hat{\lbr})^*_{f_1f_2} \: (\hat{\lambda}^{1R})_{f_3f_4} \;, \\
\left[\mathcal{C}^{qqq}\right]^{(0)}_{f_1f_2f_3f_4} \ &=& \ -\frac{2}{M_1^2} \: 
[K_{\rm CKM}^T \: (\hat{\lbl})^*]_{f_1f_2} \: [K^T_{\rm CKM} \hat{\lambda}^{1L}]_{f_3f_4} \;,
\label{eq:p4}
\end{eqnarray}
where the ``hatted" couplings are given in eq.~\eqref{eqs:lhat}  in addition to
\begin{equation}
\hat{\lbl} \ = \ U^\dagger_{u_L} \lbl U_{d_L}^* \;, \qquad 
\hat{\lbr} \ = \ U^\dagger_{d_R} \lbr U_{u_R}^* \;.
\end{equation}
Obviously, due to lepton and baryon quantum numbers 
arranged in Table~\ref{tab:BL}, only one BNV-coupling, $\lbl$ or $\lbr$, appear for $\Delta L=\Delta B=1$.
Plugging these into the Feynman Rules of ref.~\cite{Dedes:2017zog} we derive 
decay rates for general nucleon decay processes.

To date, proton decay has not been observed  and Super Kamiokande has 
increased the proton lifetime limits up to  $\sim 10^{34}$ years with bounds~\cite{Super-Kamiokande:2016exg,Super-Kamiokande:2014otb}
\begin{equation}
\tau (p \rightarrow e^+ \pi^0) > 1.6 \times 10^{34}~\mathrm{yrs}\;, \qquad 
\tau (p \rightarrow \bar{\nu} K^+) > 0.59 \times 10^{34}~\mathrm{yrs} \;,
\label{eq:pdecaylimits}
\end{equation}
being the most sensitive ones to BSM physics~\cite{Nath:2006ut}.
Based on the first of these bounds
we derive constraints on the following products of couplings:
\begin{equation}
2(\hat{\lbr})^*_{11} \hat{\lambda}_{11}^{1R} \;, \quad 
2 (\hat{\lbr})^*_{11} \hat{\lambda}_{11}^{1L}\;, \quad 
2 (\hat{\lbl})^*_{11} \hat{\lambda}_{11}^{1R}\;, \quad
4 (\hat{\lbl})^*_{11} \hat{\lambda}_{11}^{1L} \ \lesssim \ 10^{-6} \, \left ( 
\frac{M_1}{3\times 10^{12}} \right )^2 \;.
\label{eq:bnvbounds}
\end{equation}
Note that although the CKM-matrix, $K_{\rm CKM}$,
appears explicitly in tree expressions \eqref{eq:p1}-\eqref{eq:p4}, it disappears completely from 
the relevant BNV vertices of ref.~\cite{Dedes:2017zog} for physical fields.
In summary, $M_1 \gtrsim 3\times 10^{12}~{\rm GeV}$ and first generation
$\hat{\lambda}^{1L,1R}\sim O(1)$ and  $\lbl$ and $\lbr$ of the order of electron yukawa coupling is, currently, a safe combination.

At tree level, the proton decay constraints in eq.~\eqref{eq:bnvbounds} 
 apply on the 11 entries of $\lambda$-matrices. This changes by going to higher orders where in principle all flavour structure of those matrices are getting involved through the 
 CKM-matrix. One loop contributions to BNV Wilson coefficients are given in eqs.~\eqref{eq:bnv1}-\eqref{eq:bnv4}. 
For example, by just looking at $[G_{qqq}]^{(1)}$ and the term proportional to 
the strong QCD coupling, $g_s^2 (\lbl_{ps})^*\, (\lol_{st})$ we find a contribution
to the $p\to e^+ \pi^0$ amplitude of the form $g_s^2 (\hat{\lbl} K_{\rm CKM}^\dagger)_{11}
\, (K_{CKM}^T \hat{\lambda}^{1L})_{11}$ which, although CKM-suppressed,  displays a certain sensitivity in the off-diagonal $\lambda_{1i}$ entries under the strong proton decay bounds of \eqref{eq:pdecaylimits}. A partial list of 
dedicated studies on proton decay in leptoquark-like
 models are given in refs.~\cite{Dorsner:2012nq, Baldes:2011mh, Hambye:2017qix, Helo:2019yqp} and related reviews in refs.~\cite{Nath:2006ut,Heeck:2019kgr}.

In summary, unless there is a symmetry to prohibit BNV-couplings, the mass $M_1$ 
must be bigger than the ``intermediate" scale $\sim10^{12}$ GeV.
 As we discuss in the next paragraph, 
this will bring every other leptoquark masses e.g. $\tilde{M}_2$, at around that scale 
unless unnatural fine tuning is called upon in  the Higgs sector.

\subsubsection{Perturbativity and Fine Tuning}
	An interesting interplay between the LQ masses can be explored in some operators that contain the ratio of the two masses. We can thus put some bounds on the ratio $ M_1/\tilde{M}_2 $ so that perturbation theory is not violated. For example, for the operator $ G_{\ell d} $ adding the tree, eq.\eqref{eq:Gldtree}, and one-loop level, eq.\eqref{eq:Gldloop} that depends on the LQ mass ratio, we have,
		\begin{align}
			\left[G_{\ell d}\right]_{prst} \propto -\frac{\ltil^\ast_{tp} \ltil_{sr}}{2\tilde{M}_2^2}\,\left(1 + \frac{1}{16\pi^2}\,\frac{M_1^2}{\tilde{M}_2^2}(N_c \coottilde + \ctotilde)(1+L_1)\right)\;.
		\end{align}
	Other operators depend on the inverse mass ratio $ \tilde{M}_2/M_1 $. For instance, adding eq.\eqref{eq:Gqqtree} and eq.\eqref{eq:Gqqloop}, we have,
		\begin{align}
			\left[G_{\ell q}^{(1)}\right]_{prst} \propto \frac{(\lol)_{sp}^\ast (\lol_{tr})}{4M_1^2}\,\left(1 + \frac{2}{16\pi^2}\,\frac{\tilde{M}_2^2}{M_1^2}(N_c\, \coottilde + \ctotilde)(1 + L_2)\right)\;.
		\end{align}
	For perturbation theory to work, loop level contributions have to be way smaller than tree level ones. Then we can get a combined constrain for the ratio of the masses,
		\begin{align}
			2\frac{N_c \coottilde + \ctotilde}{16\pi^2}\,(1+L_2) < \frac{M_1^2}{\tilde{M}_2^2} < \frac{16\pi^2}{(1+L_1)(N_c\, \coottilde + \ctotilde)}\;,
		\end{align}
	which depends explicitly on parameters of the self interactions of the leptoquarks. Next, we define the following quantities,
		\begin{align}
			r^2 \equiv \frac{M_1^2}{\tilde{M}_2^2}\,,\qquad\qquad \alpha \equiv \frac{N_c\,\coottilde + \ctotilde}{16\pi^2}\;.\label{eq:alpha}
		\end{align}
	The inequality then becomes,
		\begin{align}
			2\alpha (1+L_2) < r^2 < \frac{1}{\alpha (1+L_1)}\;.
		\end{align}
	Which can be written in the following form,
		\begin{align}
			\frac{\mu^2}{\tilde{M}_2^2}\,\exp\left(1 - \frac{1}{\alpha r^2}\right) < r^2 < \frac{M_1^2}{\mu^2}\,\exp\left(\frac{r^2}{2\alpha} - 1\right)\;.
		\end{align}
	From here on we will pick up a renormalization scale, first we will pick $ \mu = M_1 $ and then $ \mu = \tilde{M}_2 $. As for the parameters $ c_{1\tilde{2}}^{(1,2)} $ we will take them first of order unity and second to a fine tuned value of the order of $ 10^{-4} $. Plugging these numbers into eq.\eqref{eq:alpha} we get $ \alpha = 1/4\pi^2 $ and $ \alpha = 10^{-4}/4\pi^2 $ for each respective value of c's. Both these values will be used for both cases of the renormalization scale and extract some bounds for the masses.
		\begin{itemize}
			\item For the first case, $ \mu = M_1 $. In this case the combined inequality reads, for $ N_c =3 $,
				\begin{align}
					r^2\,\exp\left(1 - \frac{1}{\alpha r^2}\right) < r^2 < \exp\left(\frac{r^2}{2\alpha} - 1\right)\;.
				\end{align}
			The right inequality is always satisfied, while from the left one we can get that,
				\begin{align}
					r^2 < \frac{1}{\alpha}\;.
				\end{align}
			Considering the two distinct values of $ \alpha $, we can get $ M_1 < (2\pi,\,2\pi\times 10^2)\, \tilde{M}_2$.
			\\[.2cm]
			\item For the second case $ \mu = \tilde{M}_2 $. The inequality becomes,
				\begin{align}
				\exp\left(1 - \frac{1}{\alpha r^2}\right) < r^2 < r^2\,\exp\left(\frac{r^2}{2\alpha} - 1\right)\;.
				\end{align}
				The right part gives, $ r^2 > 2\alpha $, while the left part is always true.
		\end{itemize}
	If we combine the two results we can get the following bound,
		\begin{align}
			2\alpha < r^2 < \frac{1}{\alpha}\;.
		\end{align}
	Plugging the values of $ \alpha $ we get,
		\begin{align}
			\frac{1}{\sqrt{2}\pi}\,(1,\,10^{-2}) < \frac{M_1}{\tilde{M}_2} < 2\pi\,(1,\,10^{2})\;.
		\end{align}
	Therefore, we can conclude that when couplings $ c_{1\tilde{2}}^{(1,2)} $ are of $ \mathcal{O}(1) $ the masses can be taken to be of similar magnitude without violating perturbation theory. If we fine tune the couplings to even smaller values we can further increase the mass ratio giving us more room to tune the numerical values of the masses.
	

Noteworthy, the situation we just described for the Wcs, $G_{\ell d}^{(1)}$ and $G_{\ell q}^{(1)}$,
 share the same characteristics with
 the Higgs mass hierarchy problem which is of course evident in LQ-models. 
Indeed, by performing the matching procedure we have assumed that the Higgs mass $m$ is zero i.e., 
the Higgs field  is part of the low energy EFT. Therefore, 
one-loop contributions to the Higgs mass found in \eqref{eq:dm2} 
have to be of the order of the EW scale. For this to happen there are two cases \textit{(i)} 
LQ-masses are of the order of the TeV-scale and Higgs couplings naturally 
of order $\mathcal{O}(1)$ or \textit{(ii)} 
LQ-masses are heavier but Higgs couplings e.g. $\lambda_{H1}, \tilde{\lambda}_{H2}, \lambda_{\tilde{2}\tilde{2}}$ together with $A_{\tilde{2}1}/M_1$ are small enough, although there is 
no known symmetry to naturally accommodate all these limits.

	\section{Conclusions}
		
	The resurgence of functional techniques to matching has led to a fair amount of universal results and compact formulae over the last few years. In this work we explored the matching of all scalar leptoquark representations that can be constructed under the SM gauge group. We have extracted through the use of \textit{Supertrace functional techniques}~\cite{Cohen:2020fcu} a universal formula, eq.~\eqref{eq:treeEFT} for tree level and eq.~\eqref{eq:LEFT1L} for one-loop matching plus all 
$\mathbf{X}$-matrices in Appendix~\ref{app:X}, for the decoupling of all scalar leptoquarks and put it to use in two distinct models. First we cross-tested it with the Feynman diagrammatic approach of the $ S_1 + S_3 $ model~\cite{Gherardi:2020det}. Then we applied it to the, phenomenologically richer, $ S_1 + \tilde{S}_2 $ model taking also baryon number violating couplings into account. In total, the latter model generates the single dimension-5 Weinberg operator at one loop, which gives rise to radiative neutrino masses. At dimension-6, 109 operators are generated in the Green basis while in the translation to Warsaw basis we are left with 53 (out of 59) operators thus covering almost the entire spectrum of dimension-6 operators, the only exception being the set of 6 bosonic CP-violating ones. Our results of the given Wilson coefficients, derived in section~\ref{sec:3},
have been also cross checked with the one loop RGEs finding complete agreement.

On the phenomenological side, we have briefly explored several distinct observables. First, we studied the implications to the lepton magnetic and electric dipole moments where there has been  recent experimental advances. With this example we demonstrated the use of the matching 
in arriving at  know results from fixed order calculations. 
	Secondly, we have investigated possible regions of leptoquark masses, at the scale of a few TeV and at a high scale, as well as their coupling with the Higgs field, to generate radiatively the order of magnitude of neutrino masses through the Weinberg operator. Furthermore, we were able to put certain bounds in the combinations of BNV and non-BNV couplings, through the investigation of proton decay at tree level and one-loop. Last but not least, we have constructed a combined inequality for the ratio of the two LQ masses so that perturbation theory is not violated and briefly discussed the hierarchy problem in LQ-models. 
	
	As a concluding remark we would like to point out that models covering almost the entirety of the given operator basis spectrum, can serve as excellent \textit{benchmarks} for various codes that will perform the matching automatically. The main reason for this argument is that within matching one needs to apply a fair amount of identities ranging from group theoretic, to Fierz identities and also accounting for evanescent operators arising from higher number of $ \gamma $-matrix structures. All of the above have been encountered in the models that have been investigated in this work, ultimately, adding up to the number of fully worked out examples of one loop matching.


	\section*{Acknowledgments}
			We would like to thank Valerio Gherardi for useful insights upon their work while comparing our results of the Wilson coefficients to theirs and for helping us catching several sign mistakes. We also thank Xiaochuan Lu for taking the time in writing useful comments on the use of STrEAM and clarifying some of its functions. We would like to thank the authors of Refs.~\cite{Coy:2021hyr} and \cite{Aebischer:2021uvt} each helping us to find a sign error in two Wilson coefficients. KM acknowledges funding from University of Ioannina Research Committee.

\newpage	
	\appendix
	
	\section{Lagrangian and $\mathbf{X}$-matrices for scalar leptoquarks}
	\label{app:X}	
  				We append here the general Lagrangian for all scalar leptoquarks and the relevant $\mathbf{X}$-matrices needed to construct the EFT Lagrangian at one-loop. 
	It can be split into three parts, in eqs.~\eqref{eq:lagSf},\eqref{eq:S-H} and \eqref{eq:S}. The notation for LQs is given in Table~\ref{tab:allcharges}.
	
	The LQ-SM fermion interactions are,
		\begin{align}
			\mathcal{L}_{\text{LQ-f}} &= \left[
			\left(\lol_{pr}\right) \bar{q}^{c}_{pi}\cdot\epsilon\cdot\ell_{r} + \left(\lori_{pr}\right) \bar{u}_{i}^{c}\,e_{r} 
			\right]S_{1i} +
			\text{h.c.}\nonumber\\
			&+ \left[(\lbl_{pr})\,\epsilon^{ijk}\,\bar{q}_{pj}\cdot\epsilon\cdot q^{c}_{rk} +(\lbr_{pr})\,\epsilon^{ijk}\bar{d}_{pj}\,u^{c}_{rk}\right]S_{1i} + \text{h.c.}\nonumber\\
			&+\left[(\ltilone_{pr})\,\bar{d}^{c}_{pi}\,e_{r} + (\ltilbsl_{pr})\,\epsilon^{ijk}\,\bar{u}_{pj}\,u^{c}_{rk}\right]\,\tilde{S}_{1i} + \text{h.c.}\nonumber\\
			&+\left[(\lLR_{pr})\,\bar{q}_{pi\alpha}e_{r} - (\lRL_{pr})\,\bar{u}_{pi}\,\ell_{r\beta}\,\epsilon^{\beta\alpha}\right]\,S_{2i\alpha} + \text{h.c.} \nonumber\\
			&+ (\ltil_{pr})\,\bar{d}_{pi}\tilde{S}^{T}_{2i}\cdot\epsilon\cdot\ell_{r} + \text{h.c.}\nonumber\\
			&+ \left[(\lLLL_{pr})\,\bar{q}^{c}_{pi}\cdot\epsilon\cdot\sigma^{I}\cdot\ell_{r} + (\lBBB_{pr})\,\epsilon^{ijk}\,\bar{q}_{pj}\cdot\sigma^{I}\cdot\epsilon\cdot q^{c}_{rk}\right]S_{3i}^{I} + \text{h.c.}\;.
		\end{align} 

	 The LQ-Higgs interactions read,
		\begin{align}
			\mathcal{L}_{\text{LQ-H}} &= -\sum_{n}\,\left(M_{n}^{2} + \lambda_{Hn}\;|S_{n}|^{2}\right)\,|H|^{2} + \sum_{n=2,\tilde{2}}\,\lambda_{nn}\,(S^{\dagger}_{ni}\cdot H)\,(H^{\dagger}\cdot S_{ni})\nonumber\\
			&\left[-A_{\tilde{2}1}\,S^{\dagger}_{1i}\,(\tilde{S}^{\dagger}_{2i}\cdot H) + A_{\tilde{2}3}\,S^{I\dagger}_{3i}\,(\tilde{S}^{\dagger}_{2i}\cdot\sigma^{I}\cdot H)\right.\nonumber\\
			&+ \left. \lambda_{2\tilde{2}}\,(S^{\dagger}_{2i}\cdot H)\,(H^{T}\cdot\epsilon\cdot\tilde{S}_{2i}) + \lambda_{3\tilde{1}}\,\tilde{S}^{\dagger}_{1i}\,(H^{T}\cdot\epsilon\cdot\sigma^{I}\cdot H)S^{I}_{3i}\right.\nonumber\\
			&+ \left. \lambda_{H13}\,(H^{\dagger}\cdot\sigma^{I}\cdot H)\,S^{I\dagger}_{3i}\,S_{1i} + \text{h.c.}\right]\nonumber\\
			&- i\lambda_{\epsilon H3}\,\epsilon^{IJK}\,(H^\dagger\cdot\sigma^{I}\cdot H)S^{J\dagger}_{3i}\,S^{K\dagger}_{3i}\;.
		\end{align}
	Where the index $ n $ runs through, $ n=1,\tilde{1},2,\tilde{2},3 $ in one to one correspondence to the sets $ \left\{S_{1},\tilde{S}_{1},S_{2},\tilde{S}_{2},S_{3}\right\} $, $ \left\{M_1,\tilde{M}_1,M_2,\tilde{M}_2,M_3\right\} $ and $ \left\{\lambda_{H1}, \tilde{\lambda}_{H1}, \lambda_{H2}, \tilde{\lambda}_{H2}, \lambda_{H3}\right\} $. 
	
	Finally, self-interactions among scalar leptoquarks are 
\begin{equation}
\mathcal{L}_{\rm S} \ = \ - V(S) \;,
\end{equation}
where $V(S)$ is the tree-level potential that is built from gauge group invariant combinations among LQ fields. 
In general $V(S)$ for all LQs  is quite lengthy. For combinations $S_1 + \tilde{S}_2$, the potential $V(S_1, \tilde{S}_2)$ can be
read from \eqref{eq:sonly}, for $S_1 + S_3$ from eq.~(2.3) of ref.~\cite{Gherardi:2020det} while the most general one
in eqs.~(46) and (49) of ref.~\cite{Crivellin:2021tmz}. Note that $X(U)$-matrices are constructed solely from simple second field
derivatives of $V(S)$ and there is no need to be written down explicitly.

		In what follows we reserve letters $ i,\alpha,p,\mu, A $ and $ I $ to denote the respective field indices in the \emph{left} hand side multiplet $\bar{\varphi} = \left(\bar{\varphi}_{S},\,\bar{\varphi}_{L}\right)$, while the letters $j,\beta,r,\nu,B,J$ are used for the \emph{right} hand side multiplet $\varphi =\left(\varphi_{S},\,\varphi_{L}\right)$ and we suppress spinor indices. Each letter represents the respective gauge group representation given in Table~\ref{tab:allcharges}. Additionally, the chirality projection operators regarding the Weyl to Dirac conversion of fermions mentioned in the main text, are left implicit.
		\subsection{$ \mathbf{X_{SS}} $}
			\begin{center}
		$\underline{\mathbf{U_{S_1S_1}}}$
	\end{center}
		\begin{align}
			U_{S_1^\dagger S_1} &= \lambda_{H1} |H|^2\delta_{ij} + \frac{\partial^2 V}{\partial S^{\dagger}_{1i}\,\partial S_{1j}}\;,\\
			U_{S_1 S_1^\dagger} &= \lambda_{H1} |H|^2\delta_{ij} + \frac{\partial^2 V}{\partial S_{1i}\,\partial S^{\dagger}_{1j}}\;.
		\end{align}
		
	\begin{center}
		$\underline{\mathbf{U_{S_1\tilde{S}_2}}}$
	\end{center}
		\begin{align}
			U_{S_1\tilde{S}_2} &= A_{\tilde{2}1}^{\ast} H_{\beta}^{\ast} \delta_{ij} + \frac{\partial^2 V}{\partial S_{1i}\,\partial \tilde{S}_{2j\beta}}\;,\\
			U_{S_1^\dagger \tilde{S}_2^\ast} &= A_{\tilde{2}1} H_{\beta}\delta_{ij} + \frac{\partial^2 V}{\partial S^{\dagger}_{1i}\,\partial \tilde{S}^{\ast}_{2j\beta}}\;.
		\end{align}
		
	\begin{center}
		$\underline{\mathbf{U_{S_1 S_3}}}$
	\end{center}
		\begin{align}
			U_{S^\dagger_1 S_3} &= \lambda_{H13}^{\ast}\,\delta_{ij}\,(H^\dagger\sigma^{J}H) + \frac{\partial^2 V}{\partial S^{\dagger}_{1i}\,\partial S^{J}_{3j}}\;,\\
			U_{S_1 S^\ast_3} &= \lambda_{H13}\,\delta_{ij}\,(H^\dagger\sigma^{J}H) + \frac{\partial^2 V}{\partial S_{1i}\,\partial S^{I\ast}_{3j}}\;.
		\end{align}
	
	\begin{center}
		$\underline{\mathbf{U_{\tilde{S}_1 \tilde{S}_1}}}$
	\end{center}
		\begin{align}
			U_{\tilde{S}_1^\dagger \tilde{S}_1} &= \tilde{\lambda}_{H1} |H|^2\delta_{ij} + \frac{\partial^2 V}{\partial \tilde{S}^{\dagger}_{1i}\,\partial \tilde{S}_{1j}}\;,\\
			U_{\tilde{S}_1 \tilde{S}_1^\dagger} &= \tilde{\lambda}_{H1} |H|^2\delta_{ij} + \frac{\partial^2 V}{\partial \tilde{S}_{1i}\,\partial \tilde{S}^{\dagger}_{1j}}\;.
		\end{align}

	\begin{center}
		$\underline{\mathbf{U_{\tilde{S}_1 S_3}}}$
	\end{center}
		\begin{align}
			U_{\tilde{S}_1^\dagger S_3} &= -\lambda_{3\tilde{1}}\,\delta_{ij}\,(H^T\cdot\epsilon\cdot\sigma^{J}\cdot H) + \frac{\partial^2 V}{\partial \tilde{S}^{\dagger}_{1i}\,\partial S^{J}_{3j}}\;,\\
			U_{\tilde{S}_1 S_3^{\ast}} &= \lambda_{3\tilde{1}}^{\ast}\,\delta_{ij}\,(H^\dagger\cdot\sigma^{J}\cdot\epsilon\cdot H^{\ast}) + \frac{\partial^2 V}{\partial \tilde{S}_{1i}\,\partial S^{J\ast}_{3j}}\;.
		\end{align}
		
	\begin{center}
		$\underline{\mathbf{U_{S_2 S_2}}}$
	\end{center}
		\begin{align}
			U_{S_2^\dagger S_2} &= \delta_{ij}\delta_{\alpha\beta} \lambda_{H2} |H|^2 - \lambda_{22}\,\delta_{ij}\,H_{\alpha}\,H^{\ast}_{\beta} + \frac{\partial^{2} V}{\partial S^{\dagger}_{2i\alpha}\,\partial S_{2j\beta}}\;,\\
			U_{S_2^T S_2^\ast} &= \delta_{ij}\delta_{\alpha\beta} \lambda_{H2} |H|^{2} - \lambda_{22}\,\delta_{ij}\,H^{\ast}_{\alpha}\,H_{\beta} + \frac{\partial^{2} V}{\partial S^{T}_{2i\alpha}\,\partial S^{\ast}_{2j\beta}}\;.
		\end{align}

	\begin{center}
		$\underline{\mathbf{U_{S_2 \tilde{S}_2}}}$
	\end{center}
		\begin{align}
			U_{S_2^\dagger \tilde{S}_2} &= \lambda_{2\tilde{2}}\,\delta_{ij}\,H_{\alpha}\,(\epsilon\cdot H)^{T}_{\beta} + \frac{\partial^{2} V}{\partial S^{\dagger}_{2i\alpha}\,\partial \tilde{S}_{2j\beta}}\;,\\
			U_{S_2^T \tilde{S}_2^\ast} &= \lambda_{2\tilde{2}}\,\delta_{ij}\,H^{\ast}_{\alpha}\,(\epsilon\cdot H^{\ast})_{\beta} + \frac{\partial^{2} V}{\partial S^{T}_{2i\alpha}\,\partial \tilde{S}^{\ast}_{2j\beta}}\;.
		\end{align}
			
	\begin{center}
		$\underline{\mathbf{U_{\tilde{S}_2\tilde{S}_2}}}$
	\end{center}
		\begin{align}
			U_{\tilde{S}_2^\dagger \tilde{S}_2} &= \delta_{ij}\delta_{\alpha\beta} \tilde{\lambda}_{H2} |H|^2 - \lambda_{\tilde{2}\tilde{2}}\,\delta_{ij}\,H_{\alpha}\,H^{\ast}_{\beta} + \frac{\partial^{2} V}{\partial \tilde{S}^{\dagger}_{2i\alpha}\,\partial \tilde{S}_{2j\beta}}\;,\\
			U_{\tilde{S}_2^T \tilde{S}_2} &= \frac{2}{3}\lambda_{5}\epsilon^{ijk}\left[\epsilon^{\alpha\alpha_1} \tilde{S}_{2j\alpha_1} H_{\beta}^{\ast} + \tilde{S}_{2k\alpha_1}\epsilon^{\alpha_1\beta} H_{\alpha}^{\ast} - \epsilon^{\alpha\beta}(H^{\dagger}\cdot\tilde{S}_{2k})\right] + \frac{\partial^{2} V}{\partial \tilde{S}^{T}_{2i\alpha}\,\partial \tilde{S}_{2j\beta}}\;,\\
			U_{\tilde{S}_2^\dagger \tilde{S}_2^\ast} &= \frac{2}{3}\lambda_{5}\epsilon^{ijk} \left[\epsilon^{\alpha\alpha_1} \tilde{S}_{2k\alpha_1} H_{\beta} + \tilde{S}_{2k\alpha_1} \epsilon^{\alpha_1\beta} H_{\alpha} - \epsilon^{\alpha\beta} (\tilde{S}_{2k}^{\dagger}\cdot H)\right] + \frac{\partial^{2} V}{\partial \tilde{S}^{\dagger}_{2i\alpha}\,\partial \tilde{S}^{\ast}_{2j\beta}}\;,\\
			U_{\tilde{S}_2^T \tilde{S}_2^\ast} &= \delta_{ij}\delta_{\alpha\beta}\tilde{\lambda}_{H2} |H|^{2} - \lambda_{\tilde{2}\tilde{2}}\,\delta_{ij}\,H^{\ast}_{\alpha}\,H_{\beta} + \frac{\partial^{2} V}{\partial \tilde{S}^{T}_{2i\alpha}\,\partial \tilde{S}^{\ast}_{2j\beta}}\;.
		\end{align}
		
		\begin{center}
			$\underline{\mathbf{U_{\tilde{S}_2S_1}}}$
		\end{center}
		\begin{align}
			U_{\tilde{S}_2^T S_1} &= A^{\ast}_{\tilde{2}1} H_{\alpha}^{\ast}\delta_{ij} + \frac{\partial^2 V}{\partial \tilde{S}^{T}_{2i\alpha}\,\partial S_{1j}}\;,\\
			U_{\tilde{S}_2^\dagger S_1^\dagger} &= A_{\tilde{2}1} H_{\alpha} \delta_{ij} + \frac{\partial^2 V}{\partial \tilde{S}^{\dagger}_{2i\alpha}\,\partial S^{\dagger}_{1j}}\;.
		\end{align}
	
	\begin{center}
		$\underline{\mathbf{U_{\tilde{S}_2S_2}}}$
	\end{center}
		\begin{align}
			U_{\tilde{S}_2^\dagger S_2} &= \lambda_{2\tilde{2}}\,\delta_{ij}\,H_{\beta}^{\ast}\,(\epsilon\cdot H^{\ast})_{\alpha} + \frac{\partial^2 V}{\partial \tilde{S}^{\ast}_{2i\alpha}\,\partial S_{2j\beta}}\;,\\
			U_{\tilde{S}_2^T S_2^\ast} &= \lambda_{2\tilde{2}}\,\delta_{ij}\,H_{\beta}\,(\epsilon\cdot H)^{T}_{\alpha} + \frac{\partial^2 V}{\partial \tilde{S}_{2i\alpha}\,\partial S^{\ast}_{2j\beta}}\;.
		\end{align}
			
	\begin{center}
		$\underline{\mathbf{U_{\tilde{S}_2S_3}}}$
	\end{center}
		\begin{align}
			U_{\tilde{S}_2^\dagger S_3^{\ast}} &= A_{\tilde{2}3}\,\delta_{ij}\,(\sigma^{J}\cdot H)_{\alpha} + \frac{\partial^2 V}{\partial \tilde{S}_{2i\alpha}^{\ast}\,\partial S^{J\ast}_{3j}}\;,\\
			U_{\tilde{S}_2^T S_3} &= A^{\ast}_{\tilde{2}3}\,\delta_{ij}\,(\sigma^{J}\cdot H^{\ast})^{T}_{\alpha} + \frac{\partial^2 V}{\partial \tilde{S}_{2i\alpha}\,\partial S^{J}_{3j}}\;.
		\end{align}
		
	\begin{center}
		$\underline{\mathbf{U_{S_3 S_3}}}$
	\end{center}
		\begin{align}
			U_{S_3^\dagger S_3} &=\lambda_{H3}\,\delta_{ij}\,\delta^{IJ}\,|H|^2 + i\lambda_{\epsilon H3}\,\epsilon^{IJK}\,(H^\dagger\sigma^{K}H)\,\delta_{ij} + \frac{\partial^2 V}{\partial S_{3i}^{I\ast}\,\partial S^{J}_{3j}}\;,\\
			U_{S_3^T S_3} &= \lambda_{H3}\,\delta_{ij}\,\delta^{IJ}\,|H|^2 - i\lambda_{\epsilon H3}\,\epsilon^{IJK}\,(H^\dagger\sigma^{K}H)\,\delta_{ij} + \frac{\partial^2 V}{\partial S_{3i}^{I}\,\partial S^{J\ast}_{3j}}\;.
		\end{align}
		
	\begin{center}
		$\underline{\mathbf{U_{S_3 S_1}}}$
	\end{center}
		\begin{align}
			U_{S_3^\dagger S_1} &=\lambda_{H13}\,\delta_{ij}\,(H^{\dagger}\sigma^{I}H) + \frac{\partial^2 V}{\partial S_{3i}^{I\ast}\,\partial S_{1j}}\;,\\
			U_{S_3^T S_1^{\ast}} &= \lambda_{H13}^{\ast}\,\delta_{ij}\,(H^{\dagger}\sigma^{I}H) + \frac{\partial^2 V}{\partial S_{3i}^{I}\,\partial S^{\ast}_{1j}}\;.
		\end{align}
		
	\begin{center}
		$\underline{\mathbf{U_{S_3 \tilde{S}_1}}}$
	\end{center}
		\begin{align}
			U_{S_3^\dagger \tilde{S}_1} &=\lambda_{3\tilde{1}}^{\ast}\,\delta_{ij}\,(H^\dagger\cdot\sigma^{I}\cdot\epsilon\cdot H^{\ast}) + \frac{\partial^2 V}{\partial S_{3i}^{I\ast}\,\partial \tilde{S}_{1j}}\;,\\
			U_{S_3^T \tilde{S}_1^{\ast}} &= - \lambda_{3\tilde{1}}\,\delta_{ij}\,(H^T\cdot\epsilon\cdot\sigma^{I}\cdot H^{\ast}) + \frac{\partial^2 V}{\partial S_{3i}^{I}\,\partial \tilde{S}^{\ast}_{1j}}\;.
		\end{align}
		
	\begin{center}
		$\underline{\mathbf{U_{S_3 \tilde{S}_2}}}$
	\end{center}
		\begin{align}
			U_{S_3^\dagger \tilde{S}_2^{\ast}} &= -A_{\tilde{2}3}\,\delta_{ij}\,(\sigma^{I}\cdot H)_{\beta} + \frac{\partial^2 V}{\partial S_{3i}^{I\ast}\,\partial \tilde{S}^{\ast}_{2j\beta}}\;,\\
			U_{S_3^T \tilde{S}_2} &= -A_{\tilde{2}3}^{\ast}\,\delta_{ij}\,(\sigma^{I}\cdot H^{\ast})^{T}_{\beta} + \frac{\partial^2 V}{\partial S_{3i}^{I}\,\partial \tilde{S}_{2j\beta}}\;.
		\end{align}
		\\[.1cm]
		\begin{center}
			\underline{\textbf{Matrix Structure}}
		\end{center}
		\begin{align}
			&\mathbf{U_{S_{n}S_{m}}} = 
			\begin{pmatrix}
				U_{S^{\dagger}_{n}S_{m}} & U_{S^{\dagger}_{n}S^{\ast}_{m}}\\
				U_{S^{T}_{n}S_{m}} & U_{S^{T}_{n}S^{\ast}_{m}}
			\end{pmatrix}\;,
		\end{align}
		with $ n,m = 1,\tilde{1},2,\tilde{2},3 $. All combinations make up the whole matrix structure of the heavy-only $ \mathbf{U_{SS}} $. Here we have listed all terms involving the Higgs field as well. There are also terms coming from the potential of all leptoquarks which are found by the general formula,
		\\
		\begin{align}
			U_{S_n^\dagger S_m} = \frac{\partial V}{\partial S_n^\dagger\,\partial S_m}\;,\quad
			U_{S_n^\dagger S_m^\ast} = \frac{\partial V}{\partial S_n^\dagger\,\partial S_m^\ast}\;,\quad
			U_{S_n^T S_m} = \frac{\partial V}{\partial S_n^T\,\partial S_m}\;,\qquad
			U_{S_n^T S_m^\ast} = \frac{\partial V}{\partial S_n^T\,\partial S_m^\ast} \;.
		\end{align} 
		
		\subsection{$ \mathbf{X_{SL}} $}
		  		\begin{center}
			\underline{$\mathbf{U_{S_nf}}$}
		\end{center}
		\begin{align}
			&U_{S_1\ell} = -(\lol_{pr})\,\bar{q}^{c}_{pi\alpha}\epsilon^{\alpha\beta}\;, \hfill 
			&&U_{S_1^\dagger \ell^c} =- (\lol_{pr})^{\ast}\,\bar{q}_{pi\alpha}\epsilon^{\alpha\beta}\;,\\
			&U_{S_1^\dagger q} = 2(\lbl)^{\ast} \bar{q}^{c}_{pk\alpha}\epsilon^{\alpha\beta}\epsilon^{ijk}\;,
			\hfill
			&&U_{S_1^\dagger q^c} = (\lol_{pr})^{\dagger}\delta_{ij}\bar{\ell}_{p\alpha}\epsilon^{\alpha\beta}\;,\\
			&U_{S_1 q} = (\lol_{pr})^{T}\,\delta_{ij}\,\bar{\ell}^{c}_{p\alpha}\epsilon^{\alpha\beta}\;,
			\hfill
			&&U_{S_1 q^c} = 2(\lbl_{pr}) \epsilon^{ijk}\bar{q}_{pk\alpha}\epsilon^{\alpha\beta}\;,\\
			&U_{S_1^\dagger u} = (\lbr_{pr})^{\ast}\,\bar{d}^{c}_{pk}\,\epsilon^{ijk}\;,
			\hfill
			&&U_{S_1^\dagger u^c} = -(\lori_{pr})^{\dagger}\,\bar{e}_{p}\,\delta_{ij}\;,\\
			&U_{S_1 u} = -(\lori_{pr})^{T} \bar{e}^{c}_{p}\,\delta_{ij}\;,
			\hfill
			&&U_{S_1 u^c} = (\lbr_{pr})\epsilon^{ijk}\bar{d}_{pk}\;,\\
			&U_{S_1 e} = -(\lori_{pr})\bar{u}^{c}_{pi}\;,
			\hfill
			&&U_{S_1^\dagger e^c} = -(\lori_{pr})^{\ast}\,\bar{u}_{pi}\;,\\
			&U_{S_1^\dagger d} = -(\lbr_{pr})^{\dagger}\,\bar{u}^{c}_{pk}\epsilon^{ijk}\;,
			\hfill
			&&U_{S_1 d^c} = -(\lbr_{pr})^{T}\,\epsilon^{ijk} \bar{u}_{pk}\;,\\
			\nonumber\\
			&U_{\tilde{S}_1^\dagger u} = -2(\ltilbsl_{pr})^{\dagger}\,\bar{u}^{c}_{pk}\,\epsilon^{ijk}\;,
			&&U_{\tilde{S}_{1} u^c} = 2(\ltilbsl_{pr})\,\bar{u}_{pk}\,\epsilon^{ijk}\;,\\
			&U_{\tilde{S}_1 e} = -(\ltilone_{pr})\,\bar{d}^{c}_{pi}\;,
			&&U_{\tilde{S}_1^\dagger e^c} = -(\ltilone_{pr})^{\ast}\,\bar{d}_{pi}\;,\\
			&U_{\tilde{S}_1 d} = -(\ltilone_{pr})^{T}\,\bar{e}_{p}^{c}\,\delta_{ij}\;,
			&&U_{\tilde{S}_1^\dagger d^c} = -(\ltilone_{pr})^{\dagger}\,\bar{e}_{p}\,\delta_{ij}\;,\\
			\nonumber\\
			&U_{S_2^T \ell} = -(\lRL_{pr})\,\bar{u}_{pi}\,\epsilon^{\alpha\beta}\;,
			&&U_{S_2^\dagger \ell^c} = -(\lRL_{pr})^{\ast}\,\epsilon^{\alpha\beta}\,\bar{u}^{c}_{pi}\;,\\
			&U_{S_2^\dagger q} = -(\lLR_{pr})^{\dagger}\,\delta_{\alpha\beta}\delta_{ij}\,\bar{e}_{p}\;,
			&&U_{S_2^T q^c} = -(\lLR_{pr})^{T}\,\delta_{\alpha\beta}\delta_{ij}\,\bar{e}^{c}_{p}\;,\\
			&U_{S_2^\dagger u} = (\lRL_{pr})^{\dagger}\,\delta_{ij}\,\bar{\ell}_{p\beta}\epsilon^{\beta\alpha}\;,
			&&U_{S_2^T u^c} = (\lRL_{pr})^{T}\,\delta_{ij}\,\bar{\ell}^{c}_{p\beta}\,\epsilon^{\beta\alpha}\;,\\
			&U_{S_2^T e} = -(\lLR_{pr})\,\bar{q}_{pi\alpha}\;,
			&&U_{S_2^\dagger e^c} = -(\lLR_{pr})^{\ast}\,\bar{q}^{c}_{pi\alpha}\;,\\
			\nonumber\\
			&U_{\tilde{S}_2^T \ell} = -(\ltil_{pr})\,\bar{d}_{pi}\,\epsilon^{\alpha\beta}\;,
			\hfill
			&&U_{\tilde{S}_2^\dagger \ell^c} = - (\ltil_{pr})^{\ast}\,\bar{d}^{c}_{pi}\,\epsilon^{\alpha\beta}\;,\\
			&U_{\tilde{S}_2^\dagger d} = (\ltil_{pr})^{\dagger}\,\delta_{ij}\bar{\ell}_{p\beta}\,\epsilon^{\beta\alpha}\;,
			\hfill
			&&U_{\tilde{S}_2^T d^c} = (\ltil_{pr})^{T} \delta_{ij}\,\bar{\ell}^{c}_{p\beta}\,\epsilon^{\beta\alpha}\;,\\
			\nonumber\\
			&U_{S_3^T \ell} = -(\lLLL_{pr})\,\bar{q}^{c}_{pi\alpha}\epsilon^{\alpha\gamma}\,\sigma^{I}_{\gamma\beta}\;,
			&&U_{S_3^\dagger \ell^c} = (\lLLL_{pr})^{\ast}\,\bar{q}_{pi\alpha}\epsilon^{\gamma\alpha}\,\sigma^{I}_{\beta\gamma}\;,\\
			&U_{S_3^T q} = -(\lLLL_{pr})^{T}\,\bar{\ell}^{c}_{p\alpha}\,\sigma^{I}_{\gamma\alpha}\,\epsilon^{\beta\gamma}\delta_{ij}\;,
			&&U_{S_3^\dagger q^c} = (\lLLL_{pr})^{\dagger}\,\bar{\ell}_{p\alpha}\,\sigma^{I}_{\alpha\gamma}\,\epsilon^{\gamma\beta}\delta_{ij}\;,\\
			&U_{S_3^\dagger q} = 2(\lBBB_{pr})^\dagger\,\epsilon^{ijk}\,(\bar{q}^c_{pk}\cdot\epsilon\cdot\sigma^{I})_{\beta}\;,
			&&U_{S_3^T q^c} = 2(\lBBB_{pr})\,\epsilon^{ijk}\,(\bar{q}_{pk}\cdot\sigma^{I}\cdot\epsilon)_{\beta}\;.
		\end{align}
		\\[.1cm]
		\begin{center}
			\underline{\textbf{Matrix Structure}}
		\end{center}
		\begin{align}
			\mathbf{U_{S_1 \ell}} =
			\begin{pmatrix}
				0 & U_{S_1^\dagger \ell^c}\\
				U_{S_1 \ell} & 0
			\end{pmatrix}\;,
			\qquad\qquad
			\mathbf{U_{S_1 q}} = 
			\begin{pmatrix}
				U_{S_1^\dagger q} & U_{S_1^\dagger q^c}\\
				U_{S_1 q}	& U_{S_1 q^c}
			\end{pmatrix}\;,
			\qquad\qquad\quad
			\\
			\mathbf{U_{S_1 u}} =
			\begin{pmatrix}
				U_{S_1^\dagger u} & U_{S_1^\dagger u^c}\\
				U_{S_1 u} & U_{S_1 u^c}
			\end{pmatrix}\;,
			\quad
			\mathbf{U_{S_1 d}} = 
			\begin{pmatrix}
				U_{S_1^\dagger d} & 0\\
				0 & U_{S_1 d^c}
			\end{pmatrix}\;,
			\quad
			\mathbf{U_{S_1e}} = 
			\begin{pmatrix}
				0 & U_{S_1^\dagger e^c}\\
				U_{S_1 e} & 0
			\end{pmatrix}\;,
			\qquad
			\\
			\nonumber\\
			\mathbf{U_{\tilde{S}_1 u}} =
			\begin{pmatrix}
			U_{\tilde{S}_1^\dagger u} & 0\\
			0 & U_{\tilde{S}_1 u^c}
			\end{pmatrix}\;,
			\qquad
			\mathbf{U_{\tilde{S}_1 e}} =
			\begin{pmatrix}
			0 & U_{\tilde{S}_1^\dagger e^c}\\
			U_{\tilde{S}_1 e} & 0
			\end{pmatrix}\;.
			\qquad
			\mathbf{U_{\tilde{S}_1 d}} =
			\begin{pmatrix}
			0 & U_{\tilde{S}_1^\dagger d^c}\\
			U_{\tilde{S}_1 d} & 
			\end{pmatrix}\;,
			\qquad
			\\
			\nonumber\\
			\mathbf{U_{S_2 \ell}} =
			\begin{pmatrix}
			0 & U_{S_2^\dagger \ell^c}\\
			U_{S_2^T \ell} & 0
			\end{pmatrix}\;,
			\qquad\qquad
			\mathbf{U_{S_2 d}} =
			\begin{pmatrix}
			U_{S_2^\dagger q} & 0\\
			0 & U_{S_2^T q^c}
			\end{pmatrix}\;.
			\qquad\qquad
			\\
			\qquad\qquad
			\mathbf{U_{S_2 u}} =
			\begin{pmatrix}
			U_{S_2^\dagger u} & 0\\
			0 & U_{S_2^T u^c}
			\end{pmatrix}\;.
			\qquad\qquad
			\mathbf{U_{S_2 e}} =
			\begin{pmatrix}
			0 & U_{S_2^\dagger e^c}\\
			U_{S_2^T e} & 0
			\end{pmatrix}\;,
			\qquad\qquad
			\\
			\nonumber\\
			\mathbf{U_{\tilde{S}_2 \ell}} =
			\begin{pmatrix}
			0 & U_{\tilde{S}_2^\dagger \ell^c}\\
			U_{\tilde{S}_2^T \ell} & 0
			\end{pmatrix}\;,
			\qquad\qquad
			\mathbf{U_{\tilde{S}_2 d}} =
			\begin{pmatrix}
			U_{\tilde{S}_2^\dagger d} & 0\\
			0 & U_{\tilde{S}_2^T d^c}
			\end{pmatrix}\;,
			\qquad\qquad
			\\
			\nonumber\\
			\mathbf{U_{S_3 \ell}} =
			\begin{pmatrix}
			0 & U_{S_3^\dagger \ell^c}\\
			U_{S_3^T \ell} & 0
			\end{pmatrix}\;,
			\qquad\qquad
			\mathbf{U_{S_3 q}} =
			\begin{pmatrix}
			U_{S_3^\dagger q} & U_{S_3^\dagger q^c}\\
			U_{S_3^T q} & U_{S_3^T q^c}
			\end{pmatrix}\;.
			\qquad\qquad
		\end{align}
		\\
		\begin{center}
			\underline{$\mathbf{U_{S_nH}}$}
		\end{center}
		\begin{align}
			U_{S_1^\dagger H} &= \lambda_{H1} H_{\beta}^{\ast}S_{1i} + A_{\tilde{2}1}\,\tilde{S}_{2i\beta}^{\ast} + \lambda_{H13}^{\ast}\,S_{3i}^{I}\,(H^\dagger\cdot\sigma^{I})_{\beta}\;,\\
			U_{S_1^\dagger H^\ast} &= \lambda_{H1}H_{\beta}S_{1i}^{\dagger} + \lambda_{H13}^{\ast}\,S_{3i}^{I}\,(\sigma^{I}\cdot H)^T_{\beta}\;,\\
			U_{S_1 H} &= \lambda_{H1} H_{\beta}^{\ast} S_{1i} + \lambda_{H13}\,S_{3i}^{I\dagger}\,(H^\dagger\cdot\sigma^{I})_{\beta}\;,\\
			U_{S_1 H^\ast} &= \lambda_{H1}H_{\beta}S_{1i}^{\dagger} + A_{\tilde{2}1}^{\ast} \tilde{S}_{2i\beta} + \lambda_{H13}\,S_{3i}^{I\dagger}\,(\sigma^{I}\cdot H)^{T}_{\beta}\;.
			\end{align}
		\begin{align}
			U_{\tilde{S}_1^{\dagger} H} &= \lambda_{H\tilde{1}}\,H_{\beta}^{\ast}\,\tilde{S}_{1i} - \lambda_{3\tilde{1}}\left[(H^T\cdot\epsilon\cdot\sigma^{I})_{\beta} - (\epsilon\cdot\sigma^{I}\cdot H)^{T}_{\beta}\right]S_{3i}^{I}\;,\\
			U_{\tilde{S}_1^\dagger H^\ast} &= \lambda_{H\tilde{1}}\,H_{\beta}\,\tilde{S}_{1i}\;,\\
			U_{\tilde{S}_1 H} &= \lambda_{H\tilde{1}}\,H_{\beta}^{\ast}\,\tilde{S}_{1i}^{\dagger}\;,\\
			U_{\tilde{S}_1 H^\ast} &= \lambda_{H\tilde{1}}\,H_{\beta}\,\tilde{S}_{1i}^{\dagger} - \lambda_{3\tilde{1}}^{\ast}\,S_{3i}^{I\dagger}\left[(\sigma^{I}\cdot\epsilon\cdot H^{\ast})^{T}_{\beta} - (H^\dagger\cdot\sigma^{I}\cdot\epsilon)_{\beta}\right]\;.
		\end{align}
		\begin{align}
			U_{S_2^\dagger H} &= \lambda_{H2}\,H_{\beta}^{\ast}\,S_{2i\alpha}^{\ast} - \lambda_{2\tilde{2}}\,\delta_{\alpha\beta}(H^{T}\cdot\epsilon\cdot\tilde{S}_{2i}^{\ast}) - \lambda_{2\tilde{2}} H_{\alpha}(\epsilon\cdot\tilde{S}^{\ast}_{2i})^{T}_{\beta} - \lambda_{22}\,(\epsilon\cdot S_{2i}^{\ast})^{T}_{\beta} (\epsilon\cdot H^{\ast})^{T}_{\alpha}\;,\\
			U_{S_2^\dagger H^\ast} &= \lambda_{H2}\,H_{\beta}\,S_{2i\alpha}^{\ast} - \lambda_{22}\,(H^T\cdot\epsilon\cdot S_{2i}^{\ast})\,\epsilon^{\alpha\beta}\;,\\
			U_{S_2^T H} &= \lambda_{H2} H_{\beta}^{\ast}\,S_{2i\alpha} - \lambda_{22}\,\epsilon^{\beta\alpha}\,(S_{2i}^{T}\cdot\epsilon\cdot H^{\ast})\;,\\
			U_{S_2^T H^\ast} &= \lambda_{H2}\,H_{\beta}\,S_{2i\alpha} + \lambda_{2\tilde{2}}^{\ast}\,\delta_{\alpha\beta}\,(\tilde{S}_{2i}\cdot\epsilon\cdot H^{\ast}) + \lambda_{2\tilde{2}}^{\ast}\,H_{\alpha}^{\ast}\,(\tilde{S}_{2i}^{T}\cdot\epsilon)_{\beta} - \lambda_{22}\,(H^{T}\cdot\epsilon)_{\alpha}\,(S_{2i}^{T}\cdot\epsilon)_{\beta}\;.
		\end{align}
		\begin{align}
			U_{\tilde{S}_2^\dagger H} &= \tilde{\lambda}_{H2}H_{\beta}^{\ast}\tilde{S}_{2i\alpha} + A_{\tilde{2}1}S_{1i}^{\dagger}\delta_{\alpha\beta} - A_{\tilde{2}3}\,S_{3i}^{I\dagger}\,\sigma^{I}_{\alpha\beta} - \lambda_{\tilde{2}\tilde{2}}\,\delta_{\alpha\beta}(H^\dagger\cdot\tilde{S}_{2i})\nonumber\\ &+\frac{1}{3}\lambda_{5}\epsilon^{ijk}\left(-2\epsilon^{\alpha\alpha_1}\tilde{S}_{2j\alpha_1}\tilde{S}_{2k\beta} + \tilde{S}_{2k}^{T}\cdot\epsilon\cdot\tilde{S}_{2j}\delta_{\alpha\beta}\right)\;,\\
			U_{\tilde{S}_2^\dagger H^\ast} &= \tilde{\lambda}_{H2}H_{\beta}\tilde{S}_{2i\alpha} + \lambda_{2\tilde{2}}^{\ast}\,\epsilon^{\alpha\beta}\,(H^\dagger\cdot S_{2i}^{\ast}) + \lambda_{2\tilde{2}}^{\ast}\,S_{2i\beta}^{\ast}\,(\epsilon\cdot H^{\ast})^{T}_{\alpha} - \lambda_{\tilde{2}\tilde{2}}\,\epsilon^{\alpha\beta}\,(H^{T}\cdot\epsilon\cdot\tilde{S}_{2i}^{\ast})\;,\\
			U_{\tilde{S}_2^T H} &= \tilde{\lambda}_{H2}H^{\ast}_{\beta}\tilde{S}_{2i\alpha}^{\ast} - \lambda_{2\tilde{2}}\,(S_{2i}^{T}\cdot H)\epsilon^{\beta\alpha} - \lambda_{2\tilde{2}}\,S_{2i\beta}\,(H^{T}\cdot\epsilon)_{\alpha} - \lambda_{\tilde{2}\tilde{2}}\,\epsilon^{\beta\alpha}\,(\tilde{S}_{2i}^{T}\cdot\epsilon\cdot H^{\ast})\;,\\
			U_{\tilde{S}_2^T H^\ast} &= \tilde{\lambda}_{H2}H_{\beta}\tilde{S}_{2i\alpha}^{\ast} + A^{\ast}_{\tilde{2}1}\delta_{\alpha\beta}S_{1i} - A_{\tilde{2}3}^{\ast}\,\sigma^{I}_{\beta\alpha}\,S_{3i}^{I} - \lambda_{\tilde{2}\tilde{2}}\,(\tilde{S}_{2i}^{\dagger}	\cdot H)\nonumber\\
			&+\frac{1}{3}\lambda_{5}\epsilon^{ijk}\left(-2\epsilon^{\alpha\alpha_1}\tilde{S}_{2j\alpha_1}^{\ast}\tilde{S}_{2k\beta}^{\ast} + \tilde{S}^{\dagger}_{2k}\cdot\epsilon\cdot\tilde{S}_{2j}^{\ast}\delta_{\alpha\beta}\right)\;.
		\end{align}
		\begin{align}
			U_{S_3^\dagger H} &= \lambda_{H3}\,H_{\beta}^{\ast}\,S_{3i}^{I} + \lambda_{H13}\,(H^\dagger\cdot\sigma^{I})_{\beta}\,S_{1i} - i\lambda_{\epsilon H3}\,\epsilon^{IJK}\,(H^\dagger\cdot\sigma^{J})_{\beta}\,S_{3i}^{K}\;,\\
			U_{S_3^\dagger H^\ast} &= \lambda_{H3}\,H_{\beta}\,S_{3i}^{I} + \lambda_{H13}\,(\sigma^{I}\cdot H)^{T}_{\beta}\,S_{1i} - i\lambda_{\epsilon H3}\,\epsilon^{IJK}\,(\sigma^{J}\cdot H)^{T}_{\beta}\,S_{3i}^{K}\nonumber\\
			&+\lambda_{3\tilde{1}}\,\tilde{S}_{1i}\left[(H^\dagger\cdot\sigma^{I}\cdot\epsilon) - (\sigma^{I}\cdot\epsilon\cdot H^{\ast})^{T}\right]_{\beta}\;,\\
			U_{S_3^T H} &= \lambda_{H3}\,H_{\beta}^{\ast}\,S_{3i}^{I\ast} + \lambda_{H13}^{\ast}\,(H^\dagger\cdot\sigma^{I})_{\beta}\,S_{1i}^{\dagger} - i\lambda_{\epsilon H3}\,\epsilon^{IJK}\,(H^\dagger\cdot\sigma^{K})_{\beta}\,S_{3i}^{J\ast}\nonumber\\
			&-\lambda_{3\tilde{1}}\,\tilde{S}_{1i}^{\dagger}\,\left[(H^T\cdot\epsilon\cdot\sigma^{I}) - (\epsilon\cdot\sigma^{I}\cdot H)^T\right]_{\beta}\;,\\
			U_{S_3^T H^\ast} &= \lambda_{H3}\,H_{\beta}\,S_{3i}^{I\ast} + \lambda_{H13}^{\ast}\,S_{1i}^{\dagger}\,(\sigma^{I}\cdot H)^{T}_{\beta} - i\lambda_{\epsilon H3}\,\epsilon^{IJK}\,(\sigma^{K}\cdot H)^{T}_{\beta}\,S_{3i}^{J\ast}
		\end{align}
		\\[.1cm]
		\begin{center}
			\underline{\textbf{Matrix Structure}}
		\end{center}
		\begin{align}
			\mathbf{U_{S_n H}} =
			\begin{pmatrix}
				U_{S_n^\dagger H} & U_{S_n^\dagger H^\ast}\\
				U_{S_n^T H} & U_{S_n^T H^\ast}
			\end{pmatrix}\;.
		\end{align}
		
		\subsection{$\mathbf{X_{LS}}$}
				\begin{center}
			\underline{$\mathbf{U_{fS_n}}$}
		\end{center}
		\begin{align}
			&U_{\bar{\ell}^c S_1} = (\lol_{pr})^T \epsilon^{\alpha\beta} q_{rj\beta}\;,
			&&U_{\bar{\ell} S_1^\dagger} = (\lol_{pr})^{\dagger} \epsilon^{\alpha\beta} q^c_{rj\beta}\;,\\
			&U_{\bar{q} S_1} = 2(\lbl_{pr}) \epsilon^{ijk} \epsilon^{\alpha\beta} q^c_{rj\beta}\;,
			&&U_{\bar{q} S_1^\dagger} = -(\lol_{pr})^{\ast} \epsilon^{\alpha\beta} \ell^{c}_{r\beta} \delta_{ij}\;,\\
			&U_{\bar{q}^c S_1} = -(\lol_{pr}) \epsilon^{\alpha\beta} \ell_{r\beta} \delta_{ij}\;,
			&&U_{\bar{q}^c S_1^\dagger} = 2(\lbl_{pr})^{\ast} \epsilon^{ijk} \epsilon^{\alpha\beta} q_{rk\beta}\;,\\
			&U_{\bar{u} S_1} = -(\lbr_{pr})^T \epsilon^{ijk} d^{c}_{rk}\;,
			&&U_{\bar{u} S_1^\dagger} = -(\lori_{pr})^{\ast} e^{c}_{r} \delta_{ij}\;,\\
			&U_{\bar{u}^c S_1} = -(\lori_{pr}) e_{r} \delta_{ij}\;,
			&&U_{\bar{u}^c S_1^\dagger} = -(\lbr_{pr})^{\dagger} \epsilon^{ijk} d_{rk}\;,\\
			&U_{\bar{e}^c S_1} = -(\lori_{pr})^T u_{rj}\;,
			&&U_{\bar{e} S_1^\dagger} = -(\lori_{pr})^{\dagger} u^{c}_{rj}\;,\\
			&U_{\bar{d} S_1} = (\lbr_{pr}) \epsilon^{ijk} u^{c}_{rk}\;,
			&&U_{\bar{d}^c S_1^\dagger} = (\lbr_{pr})^{\ast} \epsilon_{ijk} u_{rk}\;,\\
			\nonumber\\
			&U_{\bar{u} \tilde{S}_1} = 2(\ltilbsl_{pr})\,\epsilon^{ijk}\,u^c_{rk}\;,
			&&U_{\bar{u}^c \tilde{S}_1^\dagger} = -2(\ltilbsl_{pr})^{\dagger}\,\epsilon^{ijk}\,u_{rk}\;,\\
			&U_{\bar{e}^c \tilde{S}_1} = -(\ltilone_{pr})^{T}\,d_{rj}\;,
			&&U_{\bar{e} \tilde{S}_1^\dagger} = -(\ltilone_{pr})^{\dagger}\,d^c_{rj}\;,\\
			&U_{\bar{d}^c \tilde{S}_1} = -(\ltilone_{pr})\,e_r\,\delta_{ij}\;,
			&&U_{\bar{d} \tilde{S}_1^\dagger}= -(\ltilone_{pr})^{\ast}\,e^c_r\,\delta_{ij}\;,\\
			\nonumber\\
			&U_{\bar{\ell}^c S_2} = -(\lRL_{pr})^{T}\,\epsilon^{\beta\alpha}\,u^c_{rj}\;,
			&&U_{\bar{\ell} S_2^\ast} = (\lRL_{pr})^\dagger\,\epsilon^{\alpha\beta}\,u_{rj}\;,\\
			&U_{\bar{q} S_2} = -(\lLR_{pr})\,\delta_{\alpha\beta}\,\delta_{ij}\,e_{r}\;,
			&&U_{\bar{q}^c S_2^\ast} = -(\lLR_{pr})^{\ast}\,\delta_{\alpha\beta}\,\delta_{ij}\,e^c_{r}\;,\\
			&U_{\bar{u} S_2} = -(\lRL_{pr})\,\delta_{ij}\,\epsilon^{\beta\alpha}\,\ell_{r\alpha}\alpha\;,
			&&U_{\bar{u}^c S_2^\ast} = (\lRL_{pr})^{\ast}\,\delta_{ij}\,\ell^c_{r\alpha}\,\epsilon^{\alpha\beta}\;,\\
			&U_{\bar{e}^c S_2} = -(\lLR_{pr})^{T}\,q^{c}_{pj\beta}\;,
			&&U_{\bar{e} S_2^\ast} = -(\lLR_{pr})^{\dagger}\,q_{rj\beta}\;,\\
			\nonumber\\
			&U_{\bar{\ell}^c \tilde{S}_2} = (\ltil_{pr})^{T} d^{c}_{ri} \epsilon^{\alpha\beta}\;,
			&&U_{\bar{\ell} \tilde{S}_2^\ast} = (\ltil{pr})^{\dagger} d_{rj} \epsilon^{\alpha\beta}\;,\\
			&U_{\bar{d} \tilde{S}_2} = -(\ltil_{pr})\delta_{ij}\epsilon^{\beta\alpha} \ell_{r\alpha}\;,
			&&U_{\bar{d}^c \tilde{S}_2^\ast} = (\ltil_{pr})^{\ast} \delta_{ij} \ell^{c}_{r\alpha} \epsilon^{\alpha\beta}\;,\\
			\nonumber\\
			&U_{\bar{\ell}^c S_3} = -(\lLLL_{pr})^{T}\,(q_{rj}\cdot\epsilon\cdot\sigma^{J})_{\alpha}\;,
			&&U_{\bar{\ell} S_3^\ast} = (\lLLL_{pr})^{\dagger}\,\sigma^{J}_{\alpha\alpha_1}\,\epsilon^{\alpha_1\alpha_2}\,q^c_{rj\alpha_2}\;,\\
			&U_{\bar{q}^c S_3} = -(\lLLL_{pr})\,\delta_{ij}\,\epsilon^{\alpha\alpha_1}\sigma^{I}_{\alpha_1\alpha_2}\ell_{r\alpha_2}\;,
			&&U_{\bar{q} S_3^\ast} = (\lLLL_{pr})^{\ast}\,(\ell^c_{r}\cdot\sigma^{J}\cdot\epsilon)_{\alpha}\;,\\
			&U_{\bar{q} S_3} = 2(\lBBB_{pr})\,\epsilon^{ijk}\,\sigma^{J}_{\alpha\alpha_1}\epsilon^{\alpha_1\alpha_2}\,q^c_{rk\alpha_2}\;,
			&&U_{\bar{q}^c S_3^\ast} = 2(\lBBB_{pr})^{\dagger}\,\epsilon^{ijk}\,\epsilon^{\alpha\alpha_1}\,\sigma^{J}_{\alpha_1\alpha_2}\,q_{rk\alpha_2}\;.
		\end{align}
		\\[.1cm]
		\begin{center}
			\underline{\textbf{Matrix Structure}}
		\end{center}
		\begin{align}
			\mathbf{U_{\ell S_1}} =
			\begin{pmatrix}
				0 & U_{\bar{\ell} S_1^\dagger} \\
				U_{\bar{\ell}^{c} S_1} & 0
			\end{pmatrix}\;,
			\qquad\qquad
			\mathbf{U_{q S_1}} =
			\begin{pmatrix}
				U_{\bar{q} S_1} & U_{\bar{q} S_1^\dagger}\\
				U_{\bar{q}^c S_1} & U_{\bar{q}^{c} S_1^\dagger}
			\end{pmatrix}\;,
			\qquad\qquad
			\\
			\mathbf{U_{u S_1}} =
			\begin{pmatrix}
				U_{\bar{u} S_1} & U_{\bar{u} S_1^\dagger}\\
				U_{\bar{u}^c S_1} & U_{\bar{u}^{c} S_1^\dagger}
			\end{pmatrix}\;,
			\qquad
			\mathbf{U_{d S_1}} =
			\begin{pmatrix}
				U_{\bar{d} S_1} & 0\\
				0 & U_{\bar{d}^{c} S_1^\dagger}
			\end{pmatrix}\;,
			\qquad
			\mathbf{U_{e S_1}} =
			\begin{pmatrix}
				0 & U_{\bar{e} S_1^\dagger}\\
				U_{\bar{e}^c S_1} & 0
			\end{pmatrix}\;,\\
			\mathbf{U_{u \tilde{S}_1}} =
			\begin{pmatrix}
			U_{\bar{u} \tilde{S}_1} & 0\\
			0 & U_{\bar{\ell}^c \tilde{S}_1^{\dagger}}
			\end{pmatrix}\;,
			\qquad
			\mathbf{U_{e \tilde{S}_1}} =
			\begin{pmatrix}
			0 & U_{\bar{e} \tilde{S}_1^\dagger}\\
			U_{\bar{e}^c \tilde{S}_1} & 0
			\end{pmatrix}\;,
			\qquad
			\mathbf{U_{d \tilde{S}_1}} =
			\begin{pmatrix}
			0 & U_{\bar{d} \tilde{S}_1^\dagger}\\
			U_{\bar{d}^c \tilde{S}_1} & 0
			\end{pmatrix}\;,
			\qquad
			\\
			\mathbf{U_{\ell S_2}} =
			\begin{pmatrix}
				0 & U_{\bar{\ell} S_2^\ast}\\
				U_{\bar{\ell}^c S_2} & 0
			\end{pmatrix}\;,
			\qquad\qquad
			\mathbf{U_{q S_2}} =
			\begin{pmatrix}
				U_{\bar{q} S_2} & 0\\
				0 & U_{\bar{q}^c S_2^\ast}
			\end{pmatrix}\;,
			\qquad\qquad
			\\
			\mathbf{U_{u S_2}} =
			\begin{pmatrix}
			U_{\bar{u} S_2} & 0\\
			0 & U_{\bar{u}^c S_2^\ast}
			\end{pmatrix}\;,
			\qquad\qquad
			\mathbf{U_{e S_2}} =
			\begin{pmatrix}
			0 & U_{\bar{e} S_2^\ast}\\
			U_{\bar{e}^c S_2} & 0
			\end{pmatrix}\;,
			\qquad\qquad
			\\
			\nonumber\\
			\mathbf{U_{\ell \tilde{S}_2}} =
			\begin{pmatrix}
			0 & U_{\bar{\ell} \tilde{S}_2^\ast}\\
			U_{\bar{\ell}^c \tilde{S}_2} & 0
			\end{pmatrix}\;,
			\qquad\qquad
			\mathbf{U_{d \tilde{S}_2}} =
			\begin{pmatrix}
			U_{\bar{d} \tilde{S}_2} & 0\\
			0 & U_{\bar{d}^c \tilde{S}_2^\ast}
			\end{pmatrix}\;,
			\qquad\qquad
			\\
			\nonumber\\
			\mathbf{U_{\ell S_3}} =
			\begin{pmatrix}
			0 & U_{\bar{\ell} S_3^\ast}\\
			U_{\bar{\ell}^c S_3} & 0
			\end{pmatrix}\;,
			\qquad\qquad
			\mathbf{U_{q S_3}} =
			\begin{pmatrix}
			U_{\bar{q} S_3} & U_{\bar{q} S_3^\ast}\\
			U_{\bar{q}^c S_3} & U_{\bar{q}^c S_3^\ast}
			\end{pmatrix}\;.
			\qquad\qquad
		\end{align}
		\\
		\begin{center}
			$\underline{\mathbf{U_{HS_n}}}$
		\end{center}
		\begin{align}
			U_{H^\dagger S_1} &= \lambda_{H1} H_{\alpha} S_{1j}^{\dagger} + A_{\tilde{2}1}^{\ast} \tilde{S}_{2j\alpha} + \lambda_{H13}\,S_{3j}^{J\dagger}\,(\sigma^{J}\cdot H)^{T}_{\alpha}\;,\\
			U_{H^\dagger S_1^\dagger} &= \lambda_{H1} H_{\alpha} S_{1j} + \lambda_{H13}^{\ast}\,S_{3j}^{J}\,(\sigma^{J}\cdot H)^T_{\alpha}\;,\\
			U_{H^T S_1} &= \lambda_{H1} H^{\ast}_{\alpha} S^{\dagger}_{1j} + \lambda_{H13}\,S_{3j}^{J\dagger}\,(H^\dagger\cdot\sigma^{J})_{\alpha}\;,\\
			U_{H^T S_1^\dagger} &= \lambda_{H1} H_{\alpha}^{\ast} S_{1j} + A_{\tilde{2}1} \tilde{S}_{2j\alpha}^{\ast} + \lambda_{H13}^{\ast}\,(H^\dagger\cdot\sigma^{J})\,S_{3j}^{J}\;,
		\end{align}
		\begin{align}
			U_{H^\dagger \tilde{S}_1} &= \lambda_{H\tilde{1}}\,H_{\alpha}\,\tilde{S}_{1j}^{\dagger} + \lambda_{3\tilde{1}}^{\ast}\left[(\sigma^{I}\cdot\epsilon\cdot H^\ast) - (H^\dagger\cdot\sigma^{I}\cdot\epsilon)^T\right]_{\alpha}\, S_{3j}^{I\dagger}\;,\\
			U_{H^\dagger \tilde{S}_1^\dagger} &= \lambda_{H\tilde{1}}\,H_{\alpha}\,\tilde{S}_{1j}\;,\\
			U_{H^T \tilde{S}_1} &= \lambda_{H\tilde{1}}\,H_{\alpha}^{\ast}\,\tilde{S}_{1j}^{\dagger}\;,\\
			U_{H^T \tilde{S}_1^\dagger} &= \lambda_{H\tilde{1}}\,H_{\alpha}^{\ast}\,\tilde{S}_{1j} - \lambda_{3\tilde{1}} \left[(H^T\cdot\epsilon\cdot\sigma^{I}) - (\epsilon\cdot\sigma^{I}\cdot H)^T\right]_{\alpha} S_{3j}^{I}\;.
		\end{align}
		\begin{align}
			U_{H^\dagger S_2} &= \lambda_{H2}\,H_{\alpha}\,S_{2j\beta} - \lambda_{2\tilde{2}}^{\ast}\,\left[H_{\beta}^{\ast}\,(\tilde{S}_{2j}^T\cdot\epsilon)_{\alpha} - \delta_{\alpha\beta}(\tilde{S}^{T}_{2j}\cdot\epsilon\cdot H^\ast)\right] - \lambda_{22}\,(S_{2j}\cdot\epsilon)^{T}_{\alpha}\,(H^T\cdot\epsilon)^{T}_{\beta}\;,\\
			U_{H^\dagger S_2^\ast} &= \lambda_{H2}\,H_{\alpha}\,S_{2j\beta}^{\ast} - \lambda_{22}\,\epsilon^{\beta\alpha}\,(H^T\cdot\epsilon\cdot S_{2j}^{\ast})\;,\\
			U_{H^T S_2} &= \lambda_{H2}\,H_{\alpha}^{\ast}\,S_{2j\beta} - \lambda_{22}\,\epsilon^{\alpha\beta}\,(S_{2j}^{T}\cdot\epsilon\cdot H^{\ast})\;,\\
			U_{H^T S_2^\ast} &= \lambda_{H2}\,H_{\alpha}\,S_{2j\beta}^{\ast} - \lambda_{2\tilde{2}}\,\left[H_{\beta}\,(\epsilon\cdot\tilde{S}_{2i}^{\ast})_{\alpha} + (H^T\cdot\epsilon\cdot\tilde{S}_{2j}^{\ast})\delta_{\alpha\beta}\right] - \lambda_{22}\,(\epsilon\cdot S_{2j}^{\ast})_{\alpha}\,(\epsilon\cdot H^\ast)_{\beta}\;.
		\end{align}
		\begin{align}
			U_{H^\dagger \tilde{S}_2} &= \tilde{\lambda}_{H2} H_{\alpha} \tilde{S}^{\ast}_{2j\beta} + A^{\ast}_{\tilde{2}1} \delta_{\alpha\beta} S_{1j} - A_{\tilde{2}3}^{\ast}\,S_{3j}^{I}\,\sigma^{I}_{\alpha\beta} - \lambda_{\tilde{2}\tilde{2}}\,(\tilde{S}^{\dagger}_{2i}\cdot H)\nonumber\\
			&+\frac{1}{3}\lambda_{5} \epsilon^{ijk} \left[ -2\tilde{S}_{2i\alpha_1} \epsilon^{\alpha_1\beta} \tilde{S}_{2k\alpha} + \left(\tilde{S}_{2i}\cdot\epsilon\cdot\tilde{S}_{2k}\right) \delta_{\alpha\beta} \right]\;,\\
			U_{H^\dagger \tilde{S}_2^\ast } &= \tilde{\lambda}_{H2} H_{\alpha} \tilde{S}_{2j\beta} + \lambda_{2\tilde{2}}^{\ast}\,\left[S_{2j\alpha}^{\ast}\,(\epsilon\cdot H^\ast)_{\beta} - \epsilon^{\alpha\beta}\,(H^\dagger\cdot S_{2j}^{\ast})\right] + \lambda_{\tilde{2}\tilde{2}}\,\epsilon^{\alpha\beta}\,(H^T\cdot\epsilon\cdot\tilde{S}_{2j}^{\ast})\;,\\
			U_{H^T \tilde{S}_2} &= \tilde{\lambda}_{H2} H_{\alpha}^{\ast} \tilde{S}_{2j\beta}^{\ast} - \lambda_{2\tilde{2}}\left[\epsilon^{\alpha\beta}\,(S_{2j}^{T}\cdot H) - S_{2j\alpha}\,(H^T\cdot\epsilon)^{T}_{\beta}\right] - \lambda_{\tilde{2}\tilde{2}}\,\epsilon^{\alpha\beta}\,(\tilde{S}_{2j}\cdot\epsilon\cdot H^\ast)\;,\\
			U_{H^T \tilde{S}_2^\ast} &= \tilde{\lambda}_{H2} H_{\alpha}^{\ast} \tilde{S}_{2i\beta} + A_{\tilde{2}1} S^{\dagger}_{1j} \delta_{\alpha\beta} - A_{\tilde{2}3}\,S_{3j}^{I\dagger}\,\sigma^{I}_{\beta\alpha} - \lambda_{\tilde{2}\tilde{2}}\,(H^\dagger \cdot \tilde{S}_{2i})\nonumber\\
			&+\frac{1}{3}\lambda_{5} \epsilon^{ijk} \left[-2 \tilde{S}_{2i\alpha_1}^{\ast} \epsilon^{\alpha_1\beta} \tilde{S}_{2k\alpha}^{\ast} +\left(\tilde{S}_{2i}^{\dagger}\cdot\epsilon\cdot\tilde{S}_{2k}^{\ast}\right) \delta_{\alpha\beta} \right]\;.
		\end{align}
		\begin{align}
			U_{H^\dagger S_3} &= \lambda_{H3}\,H_{\alpha}\,S_{3j}^{J\dagger} + \lambda_{H13}^{\ast}\,(\sigma^{J}\cdot H)_{\alpha}\,S_{1j}^{\dagger} - i\lambda_{\epsilon H3}\,\epsilon^{IJK}\,(H^\dagger\cdot\sigma^{I})^{T}_{\alpha}\,S_{3j}^{K\dagger}\;,\\
			U_{H^\dagger S_3^\ast} &= \lambda_{H3}\,H_{\alpha}\,S_{3j}^{J} + \lambda_{H13}\,(\sigma^{J}\cdot H)_{\alpha}\,S_{1j} + i\lambda_{\epsilon H3}\,\epsilon^{IJK}\,(\sigma^{I}\cdot H)_{\alpha}\,S_{3j}^{K}\nonumber\\
			&+\lambda_{3\tilde{1}}^{\ast}\,\tilde{S}_{1j}\left[(\sigma^{J}\cdot\epsilon\cdot H^{\ast}) - (H^\dagger\cdot\sigma^{J}\cdot\epsilon)^{T}\right]_{\alpha}\;,\\
			U_{H^T S_3} &= \lambda_{H3}\,H_{\alpha}^{\ast}\,S_{3j}^{J\dagger} + \lambda_{H13}^{\ast}\,(\sigma^{I}\cdot H^{\ast})_{\alpha}\,S_{1j}^{\dagger} - i\lambda_{\epsilon H3}\,\epsilon^{IJK}\,(H^\dagger\cdot\sigma^{I})^{T}_{\alpha}\, S_{3j}^{K\dagger}\nonumber\\
			&-\lambda_{3\tilde{1}}\,\tilde{S}_{1j}^{\dagger}\left[(\epsilon\cdot\sigma^{J}\cdot H) - (H^T\cdot\epsilon\cdot\sigma^{J})^{T}\right]_{\alpha}\;,\\
			U_{H^T S_3^\ast} &= \lambda_{H3}\,H_{\alpha}^{\ast}\,S_{3j}^{J} + \lambda_{H13}\,(H^\dagger\cdot\sigma^{J})^{T}_{\alpha}\,S_{1j} + i\lambda_{\epsilon H3}\,\epsilon^{IJK}\,(H^\dagger\cdot\sigma^{I})_{\alpha}\,S_{3j}^{K}\;.
		\end{align}
		\\[.1cm]
		\begin{center}
			\underline{\textbf{Matrix Structure}}
		\end{center}
		\begin{align}
			\mathbf{U_{H S_n}} = 
			\begin{pmatrix}
				U_{H^\dagger S_n} & U_{H^\dagger S_n^\ast}\\
				U_{H^T S_n} & U_{H^T S_n^\ast}
			\end{pmatrix}\;.
		\end{align}

		\subsection{$ \mathbf{X_{LL}} $}
				\begin{center}
			$\underline{\mathbf{U_{ff}}}$
		\end{center}
		\begin{align}
			&U_{\bar{\ell} q^c} = (\lol_{pr})^{\dagger} S_{1j}^{\dagger} \epsilon^{\alpha\beta} + (\lLLL_{pr})^{\dagger}\,S_{3j}^{J}\,\sigma^{J}_{\alpha\gamma}\,\epsilon^{\gamma\beta}\;,
			&&U_{\bar{\ell}^c q} = -(\lol_{pr})^{T} S_{1j} \epsilon^{\alpha\beta} + (\lLLL_{pr})^{T}\,S^{J}_{3j}\,\epsilon^{\beta\gamma}\,\sigma^{J}_{\gamma\alpha}\;,\\
			&U_{\bar{\ell} u} = (\lRL_{pr})^{\dagger}\,(\epsilon\cdot S^{\ast}_{2i})_{\alpha}\;,
			&&U_{\bar{\ell}^{c} u^{c}} = (\lRL_{pr})^{T}\,(\epsilon\cdot S_{2j})^{T}_{\alpha}\;,
			\\
			&U_{\bar{\ell} e} = (y_{E})_{pr} H_{\alpha}\;,
			&&U_{\bar{\ell}^c e^c} = (y_{E})_{pr}^{\ast} H_{\alpha}^{\ast}\;,\\
			&U_{\bar{\ell} d} = (\ltil_{pr})^{\dagger} \epsilon^{\alpha\beta} \tilde{S}^{\ast}_{2i\beta}\;,
			&&U_{\bar{\ell}^c d^c} = (\ltil_{pr})^{T} \epsilon^{\alpha\beta} \tilde{S}_{2j\beta}\;,\\
			\nonumber\\
			&U_{\bar{q} \ell^c} = -(\lol_{pr})^{\ast} S_{1i}^{\dagger} \epsilon^{\alpha\beta} + (\lLLL_{pr})^{\ast}S_{3i}^{I}\sigma^{I}_{\beta\gamma}\,\epsilon^{\gamma\alpha}\;,
			&&U_{\bar{q}^c \ell} = -(\lol_{pr}) \epsilon^{\alpha\beta} S_{1i} - (\lLLL_{pr})\,S_{3i}^{I}\,\sigma^{I}_{\alpha\gamma}\epsilon^{\gamma\beta}\;,\\
			&U_{\bar{q} q^c} = -2(\lbl_{pr}) \epsilon^{\alpha\beta}\epsilon^{ijk} S_{1k} \;,
			&&U_{\bar{q}^{c} \bar{q}} = -2(\lbl_{pr})^{\ast} \epsilon^{\alpha\beta} \epsilon^{ijk} S_{1k}^{\dagger}\;,\nonumber\\
			&-2(\lBBB_{pr})\,\epsilon^{ijk}S_{3k}^{K}\,\sigma^{K}_{\alpha\gamma}\epsilon^{\gamma\beta}\;,
			&& - (\lBBB_{pr})^{\dagger}\epsilon^{ijk}\,\epsilon^{\alpha\gamma}\sigma^{K}_{\gamma\beta}S_{3k}^{K\dagger}
			\\
			&U_{\bar{q} u} = (y_{U})_{pr} \delta_{ij} \epsilon^{\alpha\beta} H_{\beta}^{\ast}\;,
			&&U_{\bar{q}^{c} u^c} = (y_{U})^{\ast}_{pr} \delta_{ij}\epsilon^{\alpha\beta} H_{\beta}\;,\\
			&U_{\bar{q} e} = -(\lLR_{pr})\,S_{2i\alpha}\;,
			&&U_{\bar{q}^c e^c} = -(\lLR_{pr})^{\ast}\,S^{\ast}_{2i\alpha}\;,\\
			&U_{\bar{q} d} = (y_{D})_{pr} H_{\alpha} \delta_{ij}\;,
			&&U_{\bar{q}^c d^c} = (y_{D})^{\ast}_{pr} \delta_{ij} H_{\alpha}^{\ast}\;,\\
			\nonumber\\
			&U_{\bar{u} \ell} = -(\lRL_{pr})\,S_{2i\gamma}\epsilon^{\gamma\beta}\;,
			&&U_{\bar{u}^c \ell^c} = (\lRL_{pr})^{\ast}\,\epsilon^{\beta\gamma}\,S^{\ast}_{2i\gamma}\;,\\
			&U_{\bar{u} q} = -(y_{U})^{\dagger}_{pr} \epsilon^{\alpha\beta} H_{\alpha} \delta_{ij}\;,
			&&U_{\bar{u}^c \bar{q}^c} = -(y_{U})_{pr}^{T} \epsilon^{\beta\alpha} H_{\alpha}^{\ast} \delta_{ij}\;,\\
			&U_{\bar{u}u^c} = -2(\ltilbsl_{pr})\,\epsilon^{ijk}\,\tilde{S}_{1k}\;,
			&&U_{\bar{u}^c u} = 2(\ltilbsl_{pr})^{\dagger}\,\epsilon^{ijk}\,\tilde{S}^{\dagger}_{1k} \;,\\
			&U_{\bar{u} e^c} = -(\lori_{pr})^{\ast} S_{1i}^{\dagger} \;,
			&&U_{\bar{u}^c e} = -(\lori_{pr}) S_{1i}\;,\\
			&U_{\bar{u} d^c} = -(\lbr_{pr})^{T} \epsilon^{ijk} S_{1k}\;,
			&&U_{\bar{u}^c d}= -(\lbr_{pr})^{\dagger} \epsilon^{ijk} S^{\dagger}_{1k}\;,\\
			\nonumber\\
			&U_{\bar{e} \ell} = (y_{E})^{\dagger}_{pr} H_{\beta}^{\ast}\;,
			&&U_{\bar{e}^c \ell^c} = (y_{E})^{T}_{pr} H_{\beta}\;,\\
			&U_{\bar{e} q} = -(\lLR_{pr})^{\dagger}\,S^{\dagger}_{2j\alpha}\;,
			&&U_{\bar{e}^c q^c} =-(\lLR_{pr})^{T}\,S_{2j\beta}\;,\\
			&U_{\bar{e} u^c} = (\lori_{pr})^{\dagger} S_{1j}^{\dagger}\;,
			&&U_{\bar{e}^c u} = -(\lori_{pr})^{T} S_{1j}\;,\\
			&U_{\bar{e} d^c} = -(\ltil_{pr})^{\dagger}\,\tilde{S}_{1j}\;,
			&&U_{\bar{e}^c d} = -(\ltil_{pr})^{T}\,\tilde{S}_{1j}\;,\\
			\nonumber\\
			&U_{\bar{d} \ell} = -(\ltil_{pr}) \tilde{S}_{2i\alpha} \epsilon^{\alpha\beta}\;,
			&&U_{\bar{d}^c \ell^c} = (\ltil_{pr})^{\ast} \epsilon^{\beta\alpha} \tilde{S}_{2i\alpha}^{\ast}\;,\\
			&U_{\bar{d} q} = (y_{D})_{pr}^{\dagger} H_{\beta}^{\ast} \delta_{ij}\;,
			&&U_{\bar{d}^c q^c} = (y_{D})_{pr}^{T} H_{\alpha} \delta_{ij}\;,\\
			&U_{\bar{d} u^c} = (\lbr_{pr}) \epsilon^{ijk} S_{1k}\;,
			&&U_{\bar{d}^c u} = (\lbr_{pr})^{\ast} \epsilon^{ijk} S_{1k}^{\dagger}\;,\\
			&U_{\bar{d} e^c} = -(\ltil_{pr})^{\ast}\,\tilde{S}^{\dagger}_{1i}\;,
			&&U_{\bar{d}^c e} = -(\ltil_{pr})\,\tilde{S}_{1i}\;.
		\end{align}
		\\[3.9cm]
		\begin{center}
			\underline{\textbf{Matrix Structure}}
		\end{center}
		\begin{align}
			\mathbf{U_{\ell q}} = 
			\begin{pmatrix}
				0 & U_{\bar{\ell} q^c}\\
				U_{\bar{\ell}^c q} & 0
			\end{pmatrix}\;,
			\quad
			\mathbf{U_{\ell u}} =
			\begin{pmatrix}
			U_{\bar{\ell} u} & 0\\
			0 & U_{\bar{\ell}^c u^c}
			\end{pmatrix}\;,
			\quad
			\mathbf{U_{\ell e}} =
			\begin{pmatrix}
				U_{\bar{\ell} e} & 0\\
				0 & U_{\bar{\ell}^c e^c}
			\end{pmatrix}\;,
			\quad
			\mathbf{U_{\ell d}} = 
			\begin{pmatrix}
				U_{\bar{\ell} d} & 0\\
				0 & U_{\bar{\ell}^c d^c}
			\end{pmatrix}\;,
		\end{align}
		\begin{align}
			\mathbf{U_{q\ell}} = 
			\begin{pmatrix}
				0 & U_{\bar{q} \ell^c}\\
				U_{\bar{q}^c \ell} & 0
			\end{pmatrix}&\;,
			\qquad
			\mathbf{U_{qq}} =
			\begin{pmatrix}
				0 & U_{\bar{q} q^c}\\
				U_{\bar{q}^c q} & 0
			\end{pmatrix}\;,
			\qquad
			\mathbf{U_{qu}} = 
			\begin{pmatrix}
				U_{\bar{q} u} & 0\\
				0 & U_{\bar{q}^c u^c}
			\end{pmatrix}\;,
			\\
			&\mathbf{U_{qe}} =
			\begin{pmatrix}
				U_{\bar{q} e} & 0\\
				0 & U_{\bar{q}^c e^c}
			\end{pmatrix}\;,
			\qquad
			\mathbf{U_{qd}} =
			\begin{pmatrix}
			U_{\bar{q} d} & 0\\
			0 & U_{\bar{q}^c d^c}
			\end{pmatrix}\;,
		\end{align}
		\begin{align}
			\mathbf{U_{u\ell}} =
			\begin{pmatrix}
				U_{\bar{u} \ell} & 0\\
				0 & U_{\bar{u}^c \ell^c}
			\end{pmatrix}&\;,
			\qquad
			\mathbf{U_{uq}} =
			\begin{pmatrix}
				U_{\bar{u} q} & 0\\
				0 & U_{\bar{u}^c q^c}
			\end{pmatrix}\;,
			\qquad
			\mathbf{U_{uu}} =
			\begin{pmatrix}
				0 & U_{\bar{u} u^c}\\
				U_{\bar{u}^c u} & 0
			\end{pmatrix}\;,
			\qquad
			\\
			&\mathbf{U_{ue}} =
			\begin{pmatrix}
				0 & U_{\bar{u} e^c}\\
				U_{\bar{u}^c e} & 0
			\end{pmatrix}\;,
			\qquad
			\mathbf{U_{ud}} = 
			\begin{pmatrix}
				0 & U_{\bar{u} d^c}\\
				U_{\bar{u}^c d} & 0
			\end{pmatrix}\;,
		\end{align}
		\begin{align}
			\mathbf{U_{d \ell}} = 
			\begin{pmatrix}
				U_{\bar{d} \ell} & 0\\
				0 & U_{\bar{d}^c \ell^c} 
			\end{pmatrix}\;,
			\quad
			\mathbf{U_{dq}} = 
			\begin{pmatrix}
				U_{\bar{d} q} & 0\\
				0 & U_{\bar{d}^c q^c}
			\end{pmatrix}\;,
			\quad
			\mathbf{U_{du}} =
			\begin{pmatrix}
				0 & U_{\bar{d} u^c}\\
				U_{\bar{d}^c u} & 0
			\end{pmatrix}\;,
			\mathbf{U_{de}} =
			\begin{pmatrix}
			0 & U_{\bar{d} e^c}\\
			U_{\bar{d}^c e} & 0
			\end{pmatrix}\;,
		\end{align}
		\begin{align}
			\mathbf{U_{e\ell}} =
			\begin{pmatrix}
				U_{\bar{e} \ell} & 0\\
				0 & U_{\bar{e}^c \ell^c}
			\end{pmatrix}\;,
			\quad
			\mathbf{U_{eq}} =
			\begin{pmatrix}
				U_{\bar{e} q} & 0\\
				0 & U_{\bar{e}^c q^c}
			\end{pmatrix}\;,
			\quad
			\mathbf{U_{eu}} =
			\begin{pmatrix}
				0 & U_{\bar{e} u^c}\\
				U_{\bar{e}^c u} & 0
			\end{pmatrix}\;,
			\mathbf{U_{ed}} =
			\begin{pmatrix}
			0 & U_{\bar{e} d^c}\\
			U_{\bar{e}^c d} & 0
			\end{pmatrix}\;.
		\end{align}
		\begin{align}
			\mathbf{U_{\ell u}} = \mathbf{U_{\ell \ell}} = \mathbf{U_{q e}} = \mathbf{U_{u \ell}} = \mathbf{U_{u u}} = \mathbf{U_{dd}} = \mathbf{U_{de}} = \mathbf{U_{ee}} = \mathbf{U_{eq}} = \mathbf{U_{ed}} = \mathbf{0}
		\end{align}
		\\
		\begin{center}
			$\underline{\mathbf{U_{Hf}}}$
		\end{center}
		\begin{align}
			&U_{H^\dagger \ell} =(y_{E})_{pr}^{\dagger} \bar{e}_{p} \delta_{\alpha\beta}\;,
			&&U_{H^T \ell^c} = (y_{E})_{pr}^{T} \bar{e}^{c}_{p} \delta_{\alpha\beta}\;,\\
			&U_{H^\dagger q} = (y_{D})_{pr}^{\dagger} \bar{d}_{pj} \delta_{\alpha\beta}\;,
			&&U_{H^\dagger q^c} = (y_{U})_{pr}^{T} \bar{u}^{c}_{pj} \epsilon^{\beta\alpha}\;,\\
			&U_{H^T q} = (y_{U})_{pr}^{\dagger} \epsilon^{\alpha\beta} \bar{u}_{pj}\;,
			&&U_{H^T q^c} = (y_{D})^{T}_{pr} \bar{d}^{c}_{pj} \delta_{\alpha\beta}\;,\\
			&U_{H^\dagger u} =(y_{U})_{pr} \bar{q}_{pj\beta} \epsilon^{\beta\alpha}\;,
			&&U_{H^T u^c} = (y_U)^{\ast}_{pr} \bar{q}^{c}_{pj\beta} \epsilon^{\alpha\beta}\;,\\
			&U_{H^T e} = (y_E)_{pr} \bar{\ell}_{p\alpha}\;,
			&&U_{H^\dagger e^c} = (y_E)_{pr}^{\ast} \bar{\ell}^{c}_{p\alpha}\;,\\
			&U_{H^T d} = (y_D)_{pr} \bar{q}_{pj\alpha} \;,
			&&U_{H^\dagger d^c} = (y_D)_{pr}^{\ast} \bar{q}^{c}_{pj\alpha} \;.\\
		\end{align}
		\\[.1cm]
		\begin{center}
			\underline{\textbf{Matrix Structure}}
		\end{center}
		\begin{align}
			\mathbf{U_{H \ell}} =
			\begin{pmatrix}
				U_{H^\dagger \ell} & 0\\
				0 & U_{H^T \ell^c}
			\end{pmatrix}\;,
			\qquad\qquad
			\mathbf{U_{Hq}} =
			\begin{pmatrix}
				U_{H^\dagger q} & U_{H^\dagger q^c}\\
				U_{H^T q} & U_{H^T q^c}
			\end{pmatrix}\;,
			\qquad\qquad
			\\
			\mathbf{U_{Hu}} =
			\begin{pmatrix}
				U_{H^\dagger u} & 0\\
				0 & U_{H^T u^c}
			\end{pmatrix}\;,
			\qquad
			\mathbf{U_{Hd}} =
			\begin{pmatrix}
				0 & U_{H^\dagger d^c}\\
				U_{H^T d} & 0
			\end{pmatrix}\;,
			\qquad
			\mathbf{U_{He}} =
			\begin{pmatrix}
				0 & U_{H^\dagger e^c}\\
				U_{H^T e} & 0
			\end{pmatrix}\;.
		\end{align}
		\\
		\begin{center}
			$\underline{\mathbf{U_{fH}}}$
		\end{center}
		
		\begin{align}
			&U_{\bar{\ell} H} =(y_E)_{pr} e_{r} \delta_{\alpha\beta}\;,
			&&U_{\bar{\ell}^c H^\ast} = (y_E)_{pr}^{\ast} e^{c}_{r} \delta_{\alpha\beta}\;,\\
			&U_{\bar{q} H} = (y_D)_{pr} d_{ri} \delta_{\alpha\beta}\;,
			&&U_{\bar{q} H^\ast} = (y_U)_{pr} u_{ri} \epsilon^{\alpha\beta}\;,\\
			&U_{\bar{q}^c H} = (y_U)_{pr}^{\ast} \epsilon^{\alpha\beta} u^{c}_{ri}\;,
			&&U_{\bar{q}^c H^\ast} = (y_D)_{pr}^{\ast} d^{c}_{ri} \delta_{\alpha\beta}\;,\\
			&U_{\bar{u} H} = (y_U)_{pr}^{\dagger} \epsilon^{\beta\alpha} q_{ri\alpha}\;,
			&&U_{\bar{u}^c H^\ast} = (y_U)_{pr}^{T} \epsilon^{\alpha\beta} q^{c}_{ri\alpha}\;,\\
			&U_{\bar{d}^c H} = (y_D)_{pr}^{T} q^{c}_{ri\beta}\;,
			&&U_{\bar{d} H^\ast} = (y_D)_{pr}^{\dagger} q_{ri\beta}\;,\\
			&U_{\bar{e}^c H} = (y_E)_{pr}^{T} \ell^{c}_{r\beta}\;,
			&&U_{\bar{e} H^\ast} = (y_E)_{pr}^{\dagger} \ell_{r\beta}\;.
		\end{align}
		\\[.1cm]
		\begin{center}
			\underline{\textbf{Matrix Structure}}
		\end{center}
		\begin{align}
			\mathbf{U_{\ell H}} =
			\begin{pmatrix}
				U_{\bar{\ell} H} & 0\\
				0 &U_{\bar{\ell}^c H^\ast} 
			\end{pmatrix}\;,
			\qquad\qquad
			\mathbf{U_{q H}} =
			\begin{pmatrix}
				U_{\bar{q} H} & U_{\bar{q} H^\ast}\\
				U_{\bar{q}^c H} &U_{\bar{q}^c H^\ast} 
			\end{pmatrix}\;,
			\qquad\qquad\\
			\mathbf{U_{u H}} =
			\begin{pmatrix}
				U_{\bar{u} H} & 0\\
				0 &U_{\bar{u}^c H^\ast} 
			\end{pmatrix}\;,
			\qquad
			\mathbf{U_{d H}} =
			\begin{pmatrix}
				0 & U_{\bar{d} H^\ast}\\
				U_{\bar{d}^c H} & 0 
			\end{pmatrix}\;,´
			\qquad
			\mathbf{U_{e H}} =
			\begin{pmatrix}
				0 & U_{\bar{e} H^\ast}\\
				U_{\bar{e}^c H} & 0 
			\end{pmatrix}\;.
		\end{align}
		\\
		\begin{center}
			$\underline{\mathbf{U_{Vf}}}$
		\end{center}
		
		\begin{align}
			&U_{B\ell} = -\bar{\ell}_{r\beta} g^{\prime} Y_{\ell} \gamma^{\mu}\;,
			&&U_{B \ell^c} = \bar{\ell}^{c}_{r\beta} g^{\prime} Y_{\ell} \gamma^{\mu}\;,\\
			&U_{W\ell} = -\frac{g}{2} \bar{\ell}_{r\alpha_1} \sigma^{I}_{\alpha_1\beta} \gamma^{\mu}\;,
			&&U_{W \ell^c} = \frac{g}{2} \bar{\ell}_{r\alpha_1} \sigma^{I}_{\beta\alpha_1} \gamma^{\mu}\;,\\
			&U_{B q} = -\bar{q}_{rj\beta} g^{\prime} Y_{q} \gamma^{\mu}\;,
			&&U_{B q^c} = \bar{q}^{c}_{rj\beta} g^{\prime} Y_{q} \gamma^{\mu}\;,\\
			&U_{W q} = -\frac{g}{2} \bar{q}_{rj\alpha_1} \sigma^{I}_{\alpha_1\beta} \gamma^{\mu}\;,
			&&U_{W q^c} = \frac{g}{2} \bar{\ell}_{rj\alpha_1} \sigma^{I}_{\beta\alpha_1} \gamma^{\mu}\;,\\
			&U_{G q} = -g_s \bar{q}_{ri\beta} T^{A}_{ij} \gamma^{\mu}\;,
			&&U_{G q^c} = g_s \bar{q}^{c}_{ri\beta} T^{A}_{ji} \gamma^{\mu}\;,\\
			&U_{B u} = -\bar{u}_{rj} g^{\prime} Y_{u} \gamma^{\mu}\;,
			&&U_{B u^c} = \bar{u}^{c}_{ri} g^{\prime} Y_{u} \gamma^{\mu}\;,\\
			&U_{G u} = -g_s \bar{u}_{ri} T^{A}_{ij} \gamma^{\mu}\;,
			&&U_{G u^c} = g_s \bar{u}_{ri} T^{A}_{ji} \gamma^{\mu}\;,\\
			&U_{B d} = -\bar{d}_{rj} g^{\prime} Y_{d} \gamma^{\mu}\;,
			&&U_{B d^c} = \bar{d}^{c}_{rj} g^{\prime} Y_{d} \gamma^{\mu}\;,\\
			&U_{G d} = -g_s \bar{d}_{ri} T^{A}_{ij} \gamma^{\mu}\;,
			&&U_{G d^c} = g_s \bar{d}^{c}_{ri} T^{A}_{ji} \gamma^{\mu}\;,\\
			&U_{B e} = -\bar{e}_{r} g^{\prime} Y_{e} \gamma^{\mu}\;,
			&&U_{B e^c} = \bar{e}^{c}_{r} g^{\prime} Y_{e} \gamma^{\mu}\;.
		\end{align}
		\\[.1cm]
		\begin{center}
			\underline{\textbf{Matrix Structure}}
		\end{center}
		\begin{align}
			\mathbf{U_{V \ell}} =
			\begin{pmatrix}
				U_{B \ell} & U_{B \ell^c}\\
				U_{W \ell} & U_{W \ell^c}\\
				0 & 0
			\end{pmatrix}\;,
			\qquad\qquad
			\mathbf{U_{Vq}} =
			\begin{pmatrix}
				U_{B q} & U_{B q^c}\\
				U_{W q} & U_{W q^c}\\
				U_{G q} & U_{G q^c}
			\end{pmatrix}\;,
			\qquad\qquad
			\\
			\mathbf{U_{Vu}} =
			\begin{pmatrix}
				U_{B u} & U_{B u^c}\\
				0 & 0\\
				U_{G u} & U_{G u^c}
			\end{pmatrix}\;,
			\qquad
			\mathbf{U_{Vd}} =
			\begin{pmatrix}
				U_{B d} & U_{B d^c}\\
				0 & 0\\
				U_{G d} & U_{G d^c}
			\end{pmatrix}\;,
			\qquad
			\mathbf{U_{Ve}} =
			\begin{pmatrix}
				U_{B e} & U_{B e^c}\\
				0 & 0\\
				0 & 0
			\end{pmatrix}\;.
		\end{align}
		\\
		\begin{center}
			$\underline{\mathbf{U_{fV}}}$
		\end{center}
		
		\begin{align}
			&U_{\bar{\ell} B} = -g^{\prime} Y_{\ell} \gamma^{\nu} \ell_{p\alpha}\;,
			&&U_{\bar{\ell}^c B} = g^{\prime} Y_{\ell} \gamma^{\nu} \ell^{c}_{p\alpha}\;,\\
			&U_{\bar{\ell} W} = -\frac{g}{2} \sigma^{I}_{\alpha\alpha_1} \gamma^{\nu} \ell_{p\alpha_1}\;,
			&&U_{\bar{\ell}^c W} = \frac{g}{2} \sigma^{I}_{\alpha_1\alpha} \gamma^{\nu} \ell^{c}_{p\alpha_1}\;,\\
			&U_{\bar{q} B} = -g^{\prime} Y_{q} \gamma^{\nu} q_{pi\alpha}\;,
			&&U_{\bar{q}^c B} = g^{\prime} Y_{q} \gamma^{\nu} q^{c}_{pi\alpha}\;,\\
			&U_{\bar{q} W} = -\frac{g}{2} \sigma^{I}_{\alpha\alpha_1} \gamma^{\nu} q_{pi\alpha_1}\;,
			&&U_{\bar{q}^c W} = \frac{g}{2} \sigma^{I}_{\alpha_1\alpha} \gamma^{\nu} q_{pi\alpha_1}\;,\\
			&U_{\bar{q} G} = -g_s \gamma^{\mu} T^{B}_{ij} q_{pj\alpha}\;,
			&&U_{\bar{q}^c G} = g_s \gamma^{\mu} T^{B}_{ji} q^{c}_{pj\alpha}\;,\\
			&U_{\bar{u} B} = -g^{\prime} Y_{u} \gamma^{\nu} u_{pi}\;,
			&&U_{\bar{u}^c B} = g^{\prime} Y_{q} \gamma^{\nu} u^{c}_{pi}\;,\\
			&U_{\bar{u} G} = -g_s \gamma^{\mu} T^{B}_{ij} u_{pj}\;,
			&&U_{\bar{u}^c G} = g_s \gamma^{\mu} T^{B}_{ji} q^{c}_{pj}\;,\\
			&U_{\bar{d} B} = -g^{\prime} Y_{d} \gamma^{\nu} d_{pi}\;,
			&&U_{\bar{d}^c B} = g^{\prime} Y_{d} \gamma^{\nu} d^{c}_{pi}\;,\\
			&U_{\bar{d} G} = -g_s \gamma^{\mu} T^{B}_{ij} d_{pj}\;,
			&&U_{\bar{d}^c G} = g_s \gamma^{\mu} T^{B}_{ji} d^{c}_{pj}\;,\\
			&U_{\bar{e} B} = -g^{\prime} Y_{e} \gamma^{\nu} e_{p}\;,
			&&U_{\bar{e}^c B} = g^{\prime} Y_{e} \gamma^{\nu} e_{p}\;.
		\end{align}
		\\[.1cm]
		\begin{center}
			\underline{\textbf{Matrix Structure}}
		\end{center}
		\begin{align}
			\mathbf{U_{\ell V}} =
			\begin{pmatrix}
				U_{\bar{\ell} B} & U_{\bar{\ell} W} & 0\\
				U_{\bar{\ell}^c B} & U_{\bar{\ell}^c W} & 0 
			\end{pmatrix}\;,
			\qquad\qquad
			\mathbf{U_{q V}} =
			\begin{pmatrix}
				U_{\bar{q} B} & U_{\bar{q} W} & U_{\bar{q} G}\\
				U_{\bar{q}^c B} &U_{\bar{q}^c W} & U_{\bar{q}^c G}
			\end{pmatrix}\;,
			\qquad\qquad\\
			\mathbf{U_{u V}} =
			\begin{pmatrix}
				U_{\bar{u} B} & 0 & U_{\bar{u} G}\\
				U_{\bar{u} B} & 0 & U_{\bar{u}^c G} 
			\end{pmatrix}\;,
			\qquad
			\mathbf{U_{d V}} =
			\begin{pmatrix}
				U_{\bar{d} B} & 0 & U_{\bar{d} G}\\
				U_{\bar{d}^c B} & 0 & U_{\bar{d}^c G} 
			\end{pmatrix}\;,´
			\qquad
			\mathbf{U_{e V}} =
			\begin{pmatrix}
				U_{\bar{e} B} & 0 & 0\\
				U_{\bar{e}^c B} & 0 & 0 
			\end{pmatrix}\;.
		\end{align}
		\\
		\begin{center}
			$\underline{\mathbf{Z_{S_n V}}}$
		\end{center}
		
		\begin{align}
			&Z^{\rho\nu}_{S_{(1,\tilde{1})} B} = -g^{\rho\nu} g^\prime Y_{S_{(1,\tilde{1})}} S^{\dagger}_{(1,\tilde{1})i}\;,
			&&\bar{Z}^{\mu\kappa}_{B S_{(1,\tilde{1})}}= -g^{\mu\kappa} g^\prime Y_{S_{(1,\tilde{1})}} S^{\dagger}_{(1,\tilde{1})j}\;,\\
			&Z^{\rho\nu}_{S_{(1,\tilde{1})}^\dagger B} = -g^{\rho\nu} g^\prime Y_{S_{(1,\tilde{1})}} S_{(1,\tilde{1})i}\;,
			&&\bar{Z}^{\mu\kappa}_{B S_{(1,\tilde{1})}^\dagger}= -g^{\mu\kappa} g^\prime Y_{S_{(1,\tilde{1})}} S_{(1,\tilde{1})j}\;,\\
			&Z^{\rho\nu}_{S_{(1,\tilde{1})} G} = -g^{\rho\nu} g_s S^{\dagger}_{(1,\tilde{1})k} T^{B}_{ik}\;,
			&&\bar{Z}^{\mu\kappa}_{G S_{(1,\tilde{1})}}= -g^{\mu\kappa} g_s S^{\dagger}_{(1,\tilde{1})k} T^{B}_{kj} \;,\\
			&Z^{\rho\nu}_{S_{(1,\tilde{1})}^\dagger G} = -g^{\rho\nu} g_s T^{B}_{ik} S_{(1,\tilde{1})k}\;,
			&&\bar{Z}^{\mu\kappa}_{G S_{(1,\tilde{1})}^\dagger}= -g^{\mu\kappa} g_s T^{B}_{kj} S_{(1,\tilde{1})k}\;,\\
			\nonumber\\
			&Z^{\rho\nu}_{S_{(2,\tilde{2})}^T B} = -g^{\rho\nu} g^{\prime} Y_{S_{(2,\tilde{2})}} S^{\ast}_{(2,\tilde{2})i\alpha}\;,
			&&\bar{Z}^{\mu\kappa}_{B S_{(2,\tilde{2})}^T}= -g^{\mu\kappa} g^{\prime} Y_{S_{(2,\tilde{2})}} S_{(2,\tilde{2})j\beta}\;,\\
			&Z^{\rho\nu}_{S_{(2,\tilde{2})}^\dagger B} = -g^{\rho\nu} g^{\prime} Y_{S_{(2,\tilde{2})}} S_{(2,\tilde{2})i\alpha}\;,
			&&\bar{Z}^{\mu\kappa}_{B S_{(2,\tilde{2})}^\dagger}= -g^{\mu\kappa} g^{\prime} Y_{S_{(2,\tilde{2})}} S^{\ast}_{(2,\tilde{2})j\beta}\;,\\
			&Z^{\rho\nu}_{S_{(2,\tilde{2})}^T W} = -g^{\rho\nu} \frac{g}{2} \sigma^{J}_{\alpha\alpha_1} S_{(2,\tilde{2})i\alpha_1}^{\ast}\;,
			&&\bar{Z}^{\mu\kappa}_{W S_{(2,\tilde{2})}^T}= -g^{\mu\kappa} \frac{g}{2} \sigma^{I}_{\alpha_1\beta} S_{(2,\tilde{2})j\alpha_1}\;,\\
			&Z^{\rho\nu}_{S_{(2,\tilde{2})}^\dagger W} = -g^{\rho\nu} \frac{g}{2} \sigma^{J}_{\alpha\alpha_1} S_{(2,\tilde{2})i\alpha_1}\;,
			&&\bar{Z}^{\mu\kappa}_{W S_{(2,\tilde{2})}^\dagger}= -g^{\mu\kappa} \frac{g}{2} \sigma^{I}_{\alpha_1\beta} S_{(2,\tilde{2})j\alpha_1}^{\ast}\;,\\
			&Z^{\rho\nu}_{S_{(2,\tilde{2})}^T G} = -g^{\rho\nu} g_s T^{B}_{ik} S^{\ast}_{(2,\tilde{2})k\alpha}\;,
			&&\bar{Z}^{\mu\kappa}_{G S_{(2,\tilde{2})}^T}= -g^{\mu\kappa} g_s T^{B}_{kj} S_{(2,\tilde{2})k\beta}\;,\\
			&Z^{\mu\nu}_{S_{(2,\tilde{2})}^\dagger G} = -g^{\mu\nu} g_s T^{B}_{ik} S_{(2,\tilde{2})k\alpha}\;,
			&&\bar{Z}^{\mu\kappa}_{G S_{(2,\tilde{2})}^\dagger}= -g^{\mu\kappa} g_s T^{B}_{kj} S^{\ast}_{(2,\tilde{2})k\beta}\;,\\
			\nonumber\\
			&Z^{\rho\nu}_{S_{3}^T B} = -g^{\rho\nu} g^{\prime} Y_{S_{3}} S^{I\ast}_{3i}\;,
			&&\bar{Z}^{\mu\kappa}_{B S_{3}^T}= -g^{\mu\kappa} g^{\prime} Y_{S_{3}} S^{J}_{3j}\;,\\
			&Z^{\rho\nu}_{S_{3}^\dagger B} = -g^{\rho\nu} g^{\prime} Y_{S_{3}} S^{I}_{3i}\;,
			&&\bar{Z}^{\mu\kappa}_{B S_{3}^\dagger}= -g^{\mu\kappa} g^{\prime} Y_{S_{3}} S^{J\ast}_{3j}\;,\\
			&Z^{\rho\nu}_{S_{3}^T W} = ig^{\rho\nu} g \epsilon^{LKJ} S_{3l}^{K\ast}\;,
			&&\bar{Z}^{\mu\kappa}_{W S_{3}^T}= -ig^{\mu\kappa} g \epsilon^{LKI} S^{K}_{3l}\;,\\
			&Z^{\rho\nu}_{S_{3}^\dagger W} = ig^{\rho\nu} g\epsilon^{LKJ} S^{K}_{3l}\;,
			&&\bar{Z}^{\mu\kappa}_{W S_{3}^\dagger}= -ig^{\mu\kappa} g \epsilon^{LKI} S_{3j}^{K\ast}\;,\\
			&Z^{\rho\nu}_{S_{3}^T G} = -g^{\rho\nu} g_s T^{B}_{ik} S^{\ast}_{3k\alpha}\;,
			&&\bar{Z}^{\mu\kappa}_{G S_{3}^T}= -g^{\mu\kappa} g_s T^{B}_{kj} S_{3k\beta}\;,\\
			&Z^{\mu\nu}_{S_{3}^\dagger G} = -g^{\mu\nu} g_s T^{B}_{ik} S_{3k\alpha}\;,
			&&\bar{Z}^{\mu\kappa}_{G S_{3}^\dagger}= -g^{\mu\kappa} g_s T^{B}_{kj} S^{\ast}_{3k\beta}\;.
		\end{align}	
	
		\section{Supertraces vs covariant diagrams}	
		\label{app:a}
				The construction of  covariant diagrams precedes chronologically the method of Supertraces. The essential difference between these two techniques is the point at which the CDE is applied. In Supertraces one first makes superdiagrams from the distinct components of the interaction matrix $ \mathbf{X} $ and then applies the CDE in the resulting Supertrace. In the Covariant diagrams approach the components of the interaction matrix are split explicitly into, \textit{heavy-only}, \textit{heavy-light} and \textit{light-only} contributions. In the resulting trace the CDE is applied and the log-function is ultimately Taylor expanded. Finally, from the expanded formula one reads the components of the covariant diagrams in the same manner one would read of Feynman rules and the covariant diagrams are drawn. All in all Supertraces are a more compact way to present Covariant diagrams. One can also look at \cite{Cohen:2020fcu} for a more elaborate comparison between the two approaches.
	
	In Tables~\ref{tab:covdiagU} and~\ref{tab:covs_pu}  we present the relevant covariant diagrams for the Leptoquark Scalar Action. We have made combinations of matrices, and momentum insertions following the rules for constructing covariant diagrams in \cite{Zhang:2016pja}. We have classified them in terms of $ U $, $ P $ and $ Z $ insertions. In total we count 60 covariant diagrams.
	\\
	\begin{table}[h]
		\centering	
\begin{tabular}{|c|m{5.5cm}|l|m{5.5cm}|}
			\hline
			$ \mathcal{O}\left(U^{2}\right) $ & \vspace{5mm}\begin{equation*}\begin{fmffile}{OU2}
	\begin{fmfgraph*}(10,10)
		\fmfset{dash_len}{2mm}
		\fmfcurved
		\fmfsurroundn{v}{2}
		\fmfv{d.shape=circle, d.filled=empty,d.size=5}{v1,v2}
		\fmf{dashes,right=1,label=$ H $,label.side=right,label.dist=1.5mm}{v1,v2}
		\fmf{dbl_dashes,right=1,label=$ S_i $,label.side=right,label.dist=1.5mm}{v2,v1}
	\end{fmfgraph*}
\end{fmffile}\end{equation*} & $\mathcal{O}\left(U^{3}\right)$ & 
			\begin{center}
				\begin{fmffile}{OU3}
	\begin{fmfgraph*}(10,10)
		\fmfset{dash_len}{2mm}
		\fmfcurved
		\fmfsurroundn{v}{8}
		\fmfv{d.shape=circle, d.filled=empty,d.size=5}{v1,v7,v5}
		\fmf{plain,right=1,label=$ f $,label.side=right,label.dist=1.5mm}{v1,v5}
		\fmf{dbl_dashes,left=0.5,label=$ S_i $,label.side=left,label.dist=1.5mm}{v7,v5}
		\fmf{plain,right=0.5,label=$ f^\prime $,label.side=right,label.dist=.5mm}{v7,v1}
		\fmffreeze
		\fmf{dots,width=2,left=0.3}{v8,v3}
	\end{fmfgraph*}
\end{fmffile}
			\end{center}
			\\
\hline
		\end{tabular}
		\caption{$U$ and $Z$ only, heavy-light, covariant diagrams contributing to the construction of operators up to dimension 6. Where $ S_i = S_{1}, \tilde{S}_{2} $, $ f=\ell,q,u,e,d $ and $ V=B,W,G $. In total we count 13 diagrams.}
		\label{tab:covdiagU}
	\end{table}
	
	\newpage
		\begin{longtable}[h!]{|m{6.6cm}|m{6.6cm}|}
			\hline
			\begin{center}
				\vspace{-1cm}
				$ \mathcal{O}\left(U^{2}P\right) $ 
				\vspace{-5mm}
			\end{center}
			&
			\begin{center}
				\vspace{-1cm}
				$\mathcal{O}\left(U^{3}P\right)$
				\vspace{-5mm}
			\end{center}
			\\
\end{longtable}

	To summarize one can now clearly see the advantage of using the method of Supertraces. From 60 Covariant Diagrams we ended up with 15 Supertrace Diagrams calculated in Section~\ref{sec:1loop}.


		\section{Green Basis Operators}
		    \label{app:GreenOps}
		The dimension-5 operator has the same definition in both Green and Warsaw basis:
		\begin{table}[hb]
			\begin{center}
				\begin{tabular}{|c|c|}
					\hline
					\multicolumn{2}{|c|}{\textbf{\large{}$\boldsymbol{H^{2} \psi^2}$}}
					\\
					\hline 
					\textcolor{blue}{\cellcolor[gray]{0.92}}$\mathcal{O}_{\nu\nu}$ & 
					\textcolor{blue}{\cellcolor[gray]{0.92}}$\epsilon^{\alpha\beta}\epsilon^{\alpha_1 \beta_1} H^{\alpha} H^{\alpha_1}\,\bar{\ell}^{c}_{p\beta}\,\ell_{r\beta_1}$
						\\
					\hline
				\end{tabular}
			\end{center}
		\caption{\label{tab:G5} Single dimension-5 operator giving rise to neutrino masses.}
		\end{table}
			
		The dimension-6 operators in  Green basis are listed below. Shaded ones are included in  Warsaw basis too. 
		\begin{table}[hb]
\begin{centering}
\begin{tabular}{|c|c|c|c|c|c|}
\hline 
\multicolumn{2}{|c|}{\textbf{\large{}$\boldsymbol{X^{3}}$}} & \multicolumn{2}{c|}{{\large{}$\text{\ensuremath{\boldsymbol{X^{2}H^{2}}}}$}} & \multicolumn{2}{c|}{{\large{}$\boldsymbol{H^{2}D^{4}}$}}\tabularnewline
\hline 
\textcolor{blue}{\cellcolor[gray]{0.92}}$\mathcal{O}_{3G}$  & \textcolor{blue}{\cellcolor[gray]{0.92}}$f^{ABC}G_{\mu}^{A\nu}G_{\nu}^{B\rho}G_{\rho}^{C\mu}$  & \textcolor{blue}{\cellcolor[gray]{0.92}}$\mathcal{O}_{HG}$  & \textcolor{blue}{\cellcolor[gray]{0.92}}$G_{\mu\nu}^{A}G^{A\mu\nu}(H^{\dagger}H)$  & $\mathcal{O}_{DH}$  & $(D_{\mu}D^{\mu}H)^{\dagger}(D_{\nu}D^{\nu}H)$\tabularnewline
\cline{5-6} \cline{6-6} 
\textcolor{blue}{\cellcolor[gray]{0.92}}$\mathcal{O}_{\widetilde{3G}}$  & \textcolor{blue}{\cellcolor[gray]{0.92}}$f^{ABC}\widetilde{G}_{\mu}^{A\nu}G_{\nu}^{B\rho}G_{\rho}^{C\mu}$  & \textcolor{blue}{\cellcolor[gray]{0.92}}$\mathcal{O}_{H\widetilde{G}}$  & \textcolor{blue}{\cellcolor[gray]{0.92}}$\widetilde{G}_{\mu\nu}^{A}G^{A\mu\nu}(H^{\dagger}H)$  & \multicolumn{2}{c|}{{\large{}$\text{\ensuremath{\boldsymbol{H^{4}D^{2}}}}$}}\tabularnewline
\cline{5-6} \cline{6-6} 
\textcolor{blue}{\cellcolor[gray]{0.92}}$\mathcal{O}_{3W}$  & \textcolor{blue}{\cellcolor[gray]{0.92}}$\epsilon^{IJK}W_{\mu}^{I\nu}W_{\nu}^{J\rho}W_{\rho}^{K\mu}$  & \textcolor{blue}{\cellcolor[gray]{0.92}}$\mathcal{O}_{HW}$  & \textcolor{blue}{\cellcolor[gray]{0.92}}$W_{\mu\nu}^{I}W^{I\mu\nu}(H^{\dagger}H)$  & \textcolor{blue}{\cellcolor[gray]{0.92}}$\mathcal{O}_{H\square}$  & \textcolor{blue}{\cellcolor[gray]{0.92}}$(H^{\dagger}H)\square(H^{\dagger}H)$\tabularnewline
\hline 
\end{tabular}
\par\end{centering}
\caption{\label{tab:GreenBosonic} Bosonic operators in the Green's basis.}
\end{table}

\pagebreak{}
\begin{table}[ht]
\begin{centering}
\begin{tabular}{|c|c|c|c|c|c|}
\hline 
\multicolumn{2}{|c|}{{\large{}$\boldsymbol{\psi^{2}D^{3}}$}} & \multicolumn{2}{c|}{{\large{}$\boldsymbol{\psi^{2}XD}$}} & \multicolumn{2}{c|}{{\large{}$\boldsymbol{\psi^{2}DH^{2}}$}}\tabularnewline
\hline 
$\mathcal{O}_{qD}$  & $\frac{i}{2}\overline{q}\left\{ D_{\mu}D^{\mu},\slashed D\right\} q$  & $\mathcal{O}_{Gq}$  & $(\overline{q}T^{A}\gamma^{\mu}q)D^{\nu}G_{\mu\nu}^{A}$  & \textcolor{blue}{\cellcolor[gray]{0.92}}$\mathcal{O}_{Hq}^{(1)}$  & \textcolor{blue}{\cellcolor[gray]{0.92}}$(\overline{q}\gamma^{\mu}q)(H^{\dagger}i\overleftrightarrow{D}_{\mu}H)$\tabularnewline
$\mathcal{O}_{uD}$  & $\frac{i}{2}\overline{u}\left\{ D_{\mu}D^{\mu},\slashed D\right\} u$  & $\mathcal{O}_{Gq}^{\prime}$  & $\frac{1}{2}(\overline{q}T^{A}\gamma^{\mu}i\overleftrightarrow{D}^{\nu}q)G_{\mu\nu}^{A}$  & $\mathcal{O}_{Hq}^{\prime(1)}$  & $(\overline{q}i\overleftrightarrow{\slashed D}q)(H^{\dagger}H)$\tabularnewline
$\mathcal{O}_{dD}$  & $\frac{i}{2}\overline{d}\left\{ D_{\mu}D^{\mu},\slashed D\right\} d$  & $\mathcal{O}_{\widetilde{G}q}^{\prime}$  & $\frac{1}{2}(\overline{q}T^{A}\gamma^{\mu}i\overleftrightarrow{D}^{\nu}q)\widetilde{G}_{\mu\nu}^{A}$  & $\mathcal{O}_{Hq}^{\prime\prime(1)}$  & $(\overline{q}\gamma^{\mu}q)\partial_{\mu}(H^{\dagger}H)$\tabularnewline
$\mathcal{O}_{\ell D}$  & $\frac{i}{2}\overline{\ell}\left\{ D_{\mu}D^{\mu},\slashed D\right\} \ell$  & $\mathcal{O}_{Wq}$  & $(\overline{q}\sigma^{I}\gamma^{\mu}q)D^{\nu}W_{\mu\nu}^{I}$  & \textcolor{blue}{\cellcolor[gray]{0.92}}$\mathcal{O}_{Hq}^{(3)}$  & \textcolor{blue}{\cellcolor[gray]{0.92}}$(\overline{q}\sigma^{I}\gamma^{\mu}q)(H^{\dagger}i\overleftrightarrow{D}_{\mu}^{I}H)$\tabularnewline
\hline 
\end{tabular}
\par\end{centering}
\caption{\label{tab:GreenSingleFermion} Two-fermion operators in the Green's
basis. Generation indices are suppressed.}
\end{table}

\pagebreak{} 
\begin{table}[t]
\begin{centering}
\begin{tabular}{|c|c|c|c|c|c|}
\hline 
\multicolumn{2}{|c|}{{\large{}Four-quark}} & \multicolumn{2}{c|}{{\large{}Four-lepton}} & \multicolumn{2}{c|}{{\large{}Semileptonic}}\tabularnewline
\hline 
\textcolor{blue}{\cellcolor[gray]{0.92}}$\mathcal{O}_{qq}^{(1)}$  & \textcolor{blue}{\cellcolor[gray]{0.92}}$(\overline{q}\gamma^{\mu}q)(\overline{q}\gamma_{\mu}q)$  & \textcolor{blue}{\cellcolor[gray]{0.92}}$\mathcal{O}_{\ell\ell}$  & \textcolor{blue}{\cellcolor[gray]{0.92}}$(\overline{\ell}\gamma^{\mu}\ell)(\overline{\ell}\gamma_{\mu}\ell)$  & \textcolor{blue}{\cellcolor[gray]{0.92}}$\mathcal{O}_{\ell q}^{(1)}$  & \textcolor{blue}{\cellcolor[gray]{0.92}}$(\overline{\ell}\gamma^{\mu}\ell)(\overline{q}\gamma_{\mu}q)$\tabularnewline
\textcolor{blue}{\cellcolor[gray]{0.92}}$\mathcal{O}_{qq}^{(3)}$  & \textcolor{blue}{\cellcolor[gray]{0.92}}$(\overline{q}\gamma^{\mu}\sigma^{I}q)(\overline{q}\gamma_{\mu}\sigma^{I}q)$  & \textcolor{blue}{\cellcolor[gray]{0.92}}$\mathcal{O}_{ee}$  & \textcolor{blue}{\cellcolor[gray]{0.92}}$(\overline{e}\gamma^{\mu}e)(\overline{e}\gamma_{\mu}e)$  & \textcolor{blue}{\cellcolor[gray]{0.92}}$\mathcal{O}_{\ell q}^{(3)}$  & \textcolor{blue}{\cellcolor[gray]{0.92}}$(\overline{\ell}\gamma^{\mu}\sigma^{I}\ell)(\overline{q}\gamma_{\mu}\sigma^{I}q)$\tabularnewline
\textcolor{blue}{\cellcolor[gray]{0.92}}$\mathcal{O}_{uu}$  & \textcolor{blue}{\cellcolor[gray]{0.92}}$(\overline{u}\gamma^{\mu}u)(\overline{u}\gamma_{\mu}u)$  & \textcolor{blue}{\cellcolor[gray]{0.92}}$\mathcal{O}_{\ell e}$  & \textcolor{blue}{\cellcolor[gray]{0.92}}$(\overline{\ell}\gamma^{\mu}\ell)(\overline{e}\gamma_{\mu}e)$  & \textcolor{blue}{\cellcolor[gray]{0.92}}$\mathcal{O}_{eu}$  & \textcolor{blue}{\cellcolor[gray]{0.92}}$(\overline{e}\gamma^{\mu}e)(\overline{u}\gamma_{\mu}u)$\tabularnewline
\hline 
\end{tabular}
\par\end{centering}
\caption{\label{tab:4FermionBconserving} Four-fermion operators.
Generation indices are suppressed.}
\end{table}

\begin{table}[t]
\begin{centering}
\begin{tabular}{|c|c|}
\hline 
\multicolumn{2}{|c|}{{\large{}$B$ and $L$ violating}}\tabularnewline
\hline 
\textcolor{blue}{\cellcolor[gray]{0.92}}$\mathcal{O}_{duq}$  & \textcolor{blue}{\cellcolor[gray]{0.92}}$\varepsilon_{ijk}\epsilon_{\alpha\beta}\left[(d^{i})^{T}Cu^{j}\right]\left[(q^{k\alpha})^{T}C\ell^{\beta}\right]$\tabularnewline
\textcolor{blue}{\cellcolor[gray]{0.92}}$\mathcal{O}_{qqu}$  & \textcolor{blue}{\cellcolor[gray]{0.92}}$\varepsilon_{ijk}\epsilon_{\alpha\beta}\left[(q^{i\alpha})^{T}Cq^{j\beta}\right]\left[(u^{k})^{T}Ce\right]$\tabularnewline
\textcolor{blue}{\cellcolor[gray]{0.92}}$\mathcal{O}_{qqq}$  & \textcolor{blue}{\cellcolor[gray]{0.92}}$\varepsilon_{ijk}\epsilon_{\alpha\beta}\epsilon_{\gamma\delta}\left[(q^{i\alpha})^{T}Cq^{j\gamma}\right]\left[(q^{k\delta})^{T}C\ell^{\beta}\right]$\tabularnewline
\textcolor{blue}{\cellcolor[gray]{0.92}}$\mathcal{O}_{duu}$  & \textcolor{blue}{\cellcolor[gray]{0.92}}$\varepsilon_{ijk}\left[(d^{i})^{T}Cu^{j}\right]\left[(u^{k})^{T}Ce\right]$\tabularnewline
\hline 
\end{tabular}
\par\end{centering}
\caption{\label{tab:4FermionBviolating}Baryon and lepton number violating
four-fermion operators. Generation indices are suppressed and $C$ is the Dirac charge
conjugation matrix.}
\end{table}

\clearpage\newpage	
		\section{Auxiliary expressions from the UOLEA}
		\label{app:UOLEAexps}
		
		We append here expressions for the UOLEA coefficients that arise in case 
		of possible LQs mass degeneracies.
			 We split them as $ f_n = \frac{i}{16\pi^2}\tilde{f}_{n} $ listing only $ \tilde{f}_{n} $. We also adopt the notation $ \Delta^2_{ij} = M_i^2 - M_j^2 $ and wherever $ S_i=\{S_1,\tilde{S}_2\} $. The coefficients then read,
			\begin{align}
				\tilde{f}^{S_iS_jS_i}_{11} &=\frac{2M_i^6+M_j^6 + 3M_i^2M_j^2(M_i^2-2M_j^2) + M_i^4M_j^2\log M_j^2/M_i^2}{6M_i^2(\Delta_{ij})^{4}}\;,\\
				\tilde{f}^{S_iS_iS_j}_{11} &=\frac{M_i^4 + 4M_i^2M_j^2(1+\log M_j^2/M_i^2) + 2M_j^4(\log M_j^2/M_i^2 - 5/2)}{2(\Delta_{ij})^4}\;,\\
				\tilde{f}^{S_iS_j}_{12} &= \frac{M_i^4 + 10M_i^2M_j^2 + M_j^4}{12(\Delta_{ij}^2)^4}-\frac{M_i^2M_j^2(M_i^2 + M_j^2)\log M_i^2/M_j^2}{2(\Delta_{ij}^2)^5}\;,\\
				\tilde{f}_{13}^{S_iS_j} &=\frac{2M_i^4+5M_i^2M_j^2-M_j^4}{12M_i^2(\Delta^2_{ij})^{3}}-\frac{M_i^2M_j^2\log M_i^2/M_j^2}{(\Delta^2_{ij})^{5}}\;\\
				\tilde{f}_{14}^{S_iS_j} &=-\frac{M_i^4+10M_i^2M_j^2M_j^4}{6(\Delta^2_{ij})^{4}}+\frac{M_i^2M_j^2(M_i^2M_j^2)\log M_i^2/M_j^2}{(\Delta_{ij}^2)^5}\;,\\
				\tilde{f}^{S_iS_jS_iS_j}_{17} &= \tilde{f}^{S_iS_j}_{17} = \frac{18M_j^2(M_i^4 + M_i^2M_j^2)-2M_i^6 - 34M_j^6 + 12(M_j^4 + 3M_i^2M_j^2)\log M_j^2/M_i^2}{12M_j^2(\Delta_{ij}^2)^5 }\;,\\
				\tilde{f}_{18}^{S_iS_jS_iS_j} &= \tilde{f}_{18}^{S_iS_j} =\frac{M_i^2 + M_j^2}{6(\Delta_{ij}^{2})^{4}}+ \frac{M_j^8 -M_i^8}{12M_i^2M_j^2(\Delta_{ij}^2)^5}+\frac{M^2_iM^2_j\log M_j^2/M_i^2}{(\Delta_{ij}^2)^5}\;,\\
				\tilde{f}^{S_iS_jS_iS_jS_iS_j}_{19} &= \tilde{f}^{S_iS_j}_{19} = \frac{(M_i^6-M_j^6) + 9M_i^2M_j^2(\Delta_{ij}^2) + 6M^2_iM^2_j(M_i^2 + M_j^2)\log M_j^2/M_i^2}{12M_i^2M_j^2(\Delta_{ij}^2)^5} 
			\end{align}

			
			
						

	\clearpage\newpage

\bibliographystyle{JHEP}		
\bibliography{refs}
\end{document}